\documentclass[letterpaper,12pt,twoside]{report}
\usepackage{graphicx,epsf, epsfig, amssymb, multirow,bm,amsmath,anysize,setspace}% Include figure files
\usepackage[left=1.5in, right=1.5in, top=1in, bottom=1in]{geometry}
\usepackage[dvips,ps2pdf,bookmarks=true,breaklinks=true,linkbordercolor={0 0 1},bookmarksnumbered=true]{hyperref}

\usepackage{verbatim} 
\usepackage[font=small,format=plain,labelfont=bf,up,textfont=it,up]{caption}

\usepackage[square, comma, numbers, sort&compress]{natbib}
\usepackage{longtable}
\usepackage{chngcntr}

\usepackage{sectsty}
\chapterfont{\large\sc\centering}
\chaptertitlefont{\centering}
\subsubsectionfont{\centering}

\hypersetup{
	pdfauthor={Arunava Roy},
	pdftitle={Particle Phenomenology of Gravitational Events at the TeV Scale},
	colorlinks}

\def\missET{${E\mkern-11mu\slash}_T$}
\def\missPT{${P\mkern-11mu\slash}_T$}

\begin{document}
\begin{titlepage}
\begin{center}
\vspace*{1in}
{\LARGE Particle Phenomenology of Gravitational Events at the TeV Scale}
\par
\vspace{1.5in}
A Dissertation
\par
\vspace{0.1in}
Presented for the
\par
\vspace{0.1in}
Doctor of Philosophy
\par
\vspace{0.1in}
Degree
\par
\vspace{0.1in}
The University of Mississippi
\par
\vfill
{\large Arunava Roy}
\par
\vspace{0.5in}
June 2009
\end{center}
\end{titlepage}
\newpage
%\doublespacing   
\begin{center}
\vspace*{\fill}
 \textit{``Fill the brain with high thoughts, highest ideals, place them day and night before you, and out of that will come great work"}\\
Vivekananda, Philosopher, $12^{{\rm th}}$ January 1863 $-$ $4^{{\rm th}}$ July 1902
 \vspace*{\fill}
\end{center}
\newpage
\section*{\begin{center} Acknowledgments \end{center}} 
I am grateful to the many people who have helped and encouraged me. I am particularly thankful to: my dissertation advisor, Marco Cavagli\`a, and the rest of my dissertation committee: Lucien Cremaldi, Neil Manson and Don Summers.

Luis~A.~Anchordoqui, Alakabha Datta, Romulus Godang and David Sanders for fruitful discussions; Peter Skands and Xerxes Tata for help with simulation software.  

My parents, Tapan and Swapna; my brother and sister-in-law, Amitava and Debjani; and my nephew Aarush.

Finally, I would also like to thank my wife, Baishali,  for her constant encouragement and support.
\newpage
\section*{\begin{center} Abstract \end{center}} 
If the fundamental scale of gravity is of the order of 1 TeV, black holes might be produced at the Large Hadron Collider. This work presents simulations of black holes and other exotic models of physics beyond the Standard Model - supersymmetry, extra dimensional models and string theory. Isolated leptons with high transverse momenta can be used to distinguish black holes from supersymmetry and models of extra dimensions. $Z^0$ bosons and photons with high transverse momenta allow the discrimination of black holes and string resonances. The analysis of visible and missing energy/momenta and event shape variables complement these techniques. 
\newpage
\tableofcontents
\newpage
%\addcontentsline{toc}{chapter}{List of Tables}
\listoftables 
\newpage
%\addcontentsline{toc}{chapter}{List of Figures}
\listoffigures 
%\newpage
%
\chapter*{Introduction}
\addcontentsline{toc}{chapter}{Introduction}
\pagenumbering{arabic}
In 1964 the theory of quarks was formulated by M.~Gell-Mann and G.~Zweig \cite{GellMann1964214,Barnes:1964pd,Zweig:1981pd,Zweig:1964jf} and in 1979 S.~Glashow, A.~Salam, and S.~Weinberg shared the Nobel Prize in Physics for their discovery of the electroweak theory \cite{Sakurai:1979qy}. Quantum electrodynamics (QED), the theory of weak interactions, and Quantum chromodynamics (QCD), the theory of the quark interactions  form the basis of Standard Model (SM) of particle physics. Results from particle collider experiments, such as Large Electron Positron LEP (at CERN) and Fermilab Tevatron agree with the predictions of the SM to a very high precision. One of the major milestones in the experimental verification of the SM was the discovery of $W^{\pm}$ and $Z$ bosons by the SPS, UA1 and UA2 experiments at CERN \cite{Rubbia:1976um,Banner:1983jy,Bagnaia:1983zx,Arnison:1983rp,Arnison:1983mk}. 

The framework of the SM relies on the existence of an unobserved particle known as the Higgs boson, which was postulated in 1964 by R.~Brout, F.~Englert and P.~Higgs \cite{Englert:1964et,PhysRevLett.13.508}. In the minimal SM the existence of the  Higgs  particle is required to generate the masses of all other particles.
Many questions, however, remain unanswered. The SM does not explain why there are three quark generations, it does not address charge-parity (CP) violation, it does not provide a viable candidate for dark matter, and finally it does not incorporate the gravitational force. Various models such as extra dimensions (EDs) \cite{Appelquist:1987nr} and supersymmetry (SUSY) (see Ref.~\cite{Martin:1997ns,Peskin:2008nw} and references therein) have been proposed to overcome the drawbacks of the SM.

The center-of-mass (CM) energy of LEP was $\sim$ 200 GeV. Tevatron's $p\bar{p}$ beams reach a maximum energy of 1 TeV. The upcoming Large Hadron Collider (LHC) \cite{CERN:web}, with 600 million collisions per second and a CM energy of 14 TeV,  is supposed to provide an answer on some, if not all, of the open questions of the SM. 
\begin{figure*}[t]
\centerline{\null\hfill
    \includegraphics*[width=0.9\textwidth]{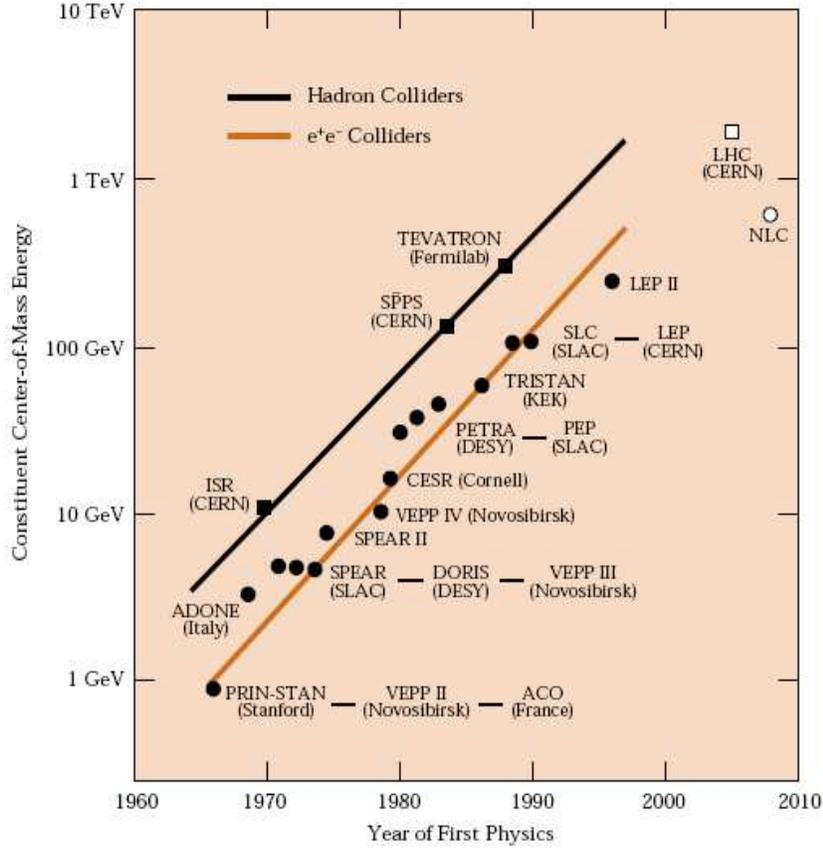}
    \hfill} 
\caption{The energy of electron-positron and hadron colliders constructed (filled circles and squares) or in the planning stage \cite{PANOFSKY}. 
%As the incident energy of proton energy is shared among its constituent quarks and gluons, the energy of hadron colliders has here been reduced by factors of 6–10 in accordance with the fact that . 
}    
\label{panofsky}
\end{figure*}

The most viable candidate of new physics beyond the SM is SUSY.  SUSY provides an explanation for the Higgs mass problem, a candidate for cold dark matter, and unification of low energy gauge couplings by introducing superpartners to SM fields. The lightest supersymmetric particle (LSP) in SUSY is an ideal dark matter candidate: a massive, weakly interacting and neutral particle. 

Models with large extra dimensions (LEDs) are an alternative to SUSY. LEDs provide a solution to the hierarchy problem. If they exist, gravity would become strong at the TeV scale. SUSY and LEDs are essential ingredients of string theory (ST). If ST happens to be the correct theory of physics at high energies, and LEDs do exist, the LHC could even start producing string resonances (SRs).

An interesting consequence of the presence of LEDs would be the production of micro black holes (BHs) in particle colliders \cite{Argyres:1998qn,Banks:1999gd,Dimopoulos:2001hw,Giddings:2001bu,Ahn:2002mj,Ahn:2002zn,Frolov:2002gf,Frolov:2002as,Cavaglia:2003qk,Chamblin:2003wg,Cavaglia:2004jw,Chamblin:2004zg,Harris:2004xt,Tanaka:2004xb,Webber:2005qa,Lonnblad:2005ah,Hewett:2005iw,Nayak:2006vf,Stoecker:2006we,Alberghi:2006qr,Koch:2007um,Feng:2001ib,Ringwald:2001vk,Anchordoqui:2001cg,Ahn:2003qn,Illana:2005pu,Cardoso:2004zi,Ahn:2003cz,Cafarella:2004hg,Ahn:2005bi,Cavaglia:2007bk,Casadio:2009ri} (for reviews, see Refs.~\cite{Cavaglia:2002si,Emparan:2003xu,Hossenfelder:2004af,Kanti:2004nr,Landsberg:2006mm}) and Earth's atmosphere by cosmic rays  \cite{Feng:2001ib,Ringwald:2001vk,Anchordoqui:2001cg,Ahn:2003qn,Illana:2005pu,Cardoso:2004zi,Ahn:2003cz,Cafarella:2004hg,Ahn:2005bi,Cavaglia:2007bk}. 
Production of  Kaluza-Klein (KK) excitations of gravitons and SM \cite{Giudice:1998ck,Mirabelli:1998rt,Cullen:1999hc} particles would also be consequences of EDs.

SUSY, LED models, KK excitations and ST lead to different physical signatures at the TeV scale. It is thus worthwhile to look into the means of comparing these models  \cite{Buescher:2006jm}. Comparisons of SUSY and universal EDs/little Higgs models in colliders have been investigated in Refs.~\cite{Rizzo:2001sd,Macesanu:2002db,Datta:2005zs,Battaglia:2005ma,Battaglia:2005zf,Konar:2005bd}. However, the literature lacks comprehensive and quantitative comparisons of the various models of  physics beyond the TeV scale.

The aim of this thesis is to show, in a quantitative way, how to distinguish BH events at the LHC from SUSY, graviton events and SRs. The original work of this thesis is based on Refs.~\cite{Roy:2008we,Roy:2007fx,Roy:2009pc}. SUSY, KK excitations and SRs can be distinguished from BHs by the use of kinematic and dynamical quantities. These methods will be discussed in detail in  Chapter~\ref{simanaly}, where the original work of this thesis is included.

This thesis is organised as follows. The next chapter reviews the most important features of the SM and the various models of new physics above the TeV scale, known problems with the SM, and how new physics beyond the SM is supposed to overcome these problems. The physics of particle collisions is discussed in Chapter~\ref{partcoll}. Simulation techniques are presented in Chapter~\ref{simtech}. Analysis and results are presented in Chapter~\ref{simanaly}. Finally, the last chapter provides some conclusions and discusses open issues. Notations, units, acronyms and symbols are included in the appendices.
\chapter{Physics of the  Standard Model and Beyond\label{theo}}
%What is the SM and why do we need new physics.
%
In this chapter we briefly review the basics of the SM and of models of physics beyond the SM. The purpose of this chapter is to introduce concepts and notations for our analysis of new physics above the TeV scale of Chapter~\ref{simanaly}. The content of this chapter is by no means exhaustive. For a more comprehensive discussion of the SM and physics beyond the SM the reader is referred to \cite{Donoghue:1992dd,Mohapatra:1990jc,Allanach:2006fy} and references therein.
\section{The Standard Model\label{SModel}}
The SM of particle physics describes the elementary particles and their interactions. The known
elementary particles can be grouped into two categories: matter particles (fermions, half spin) and force
carriers (bosons, integer spin). The particle content of the SM is shown in the figure below.
\begin{figure*}[htbp]
\centerline{\null\hfill
    \includegraphics*[width=0.7\textwidth]{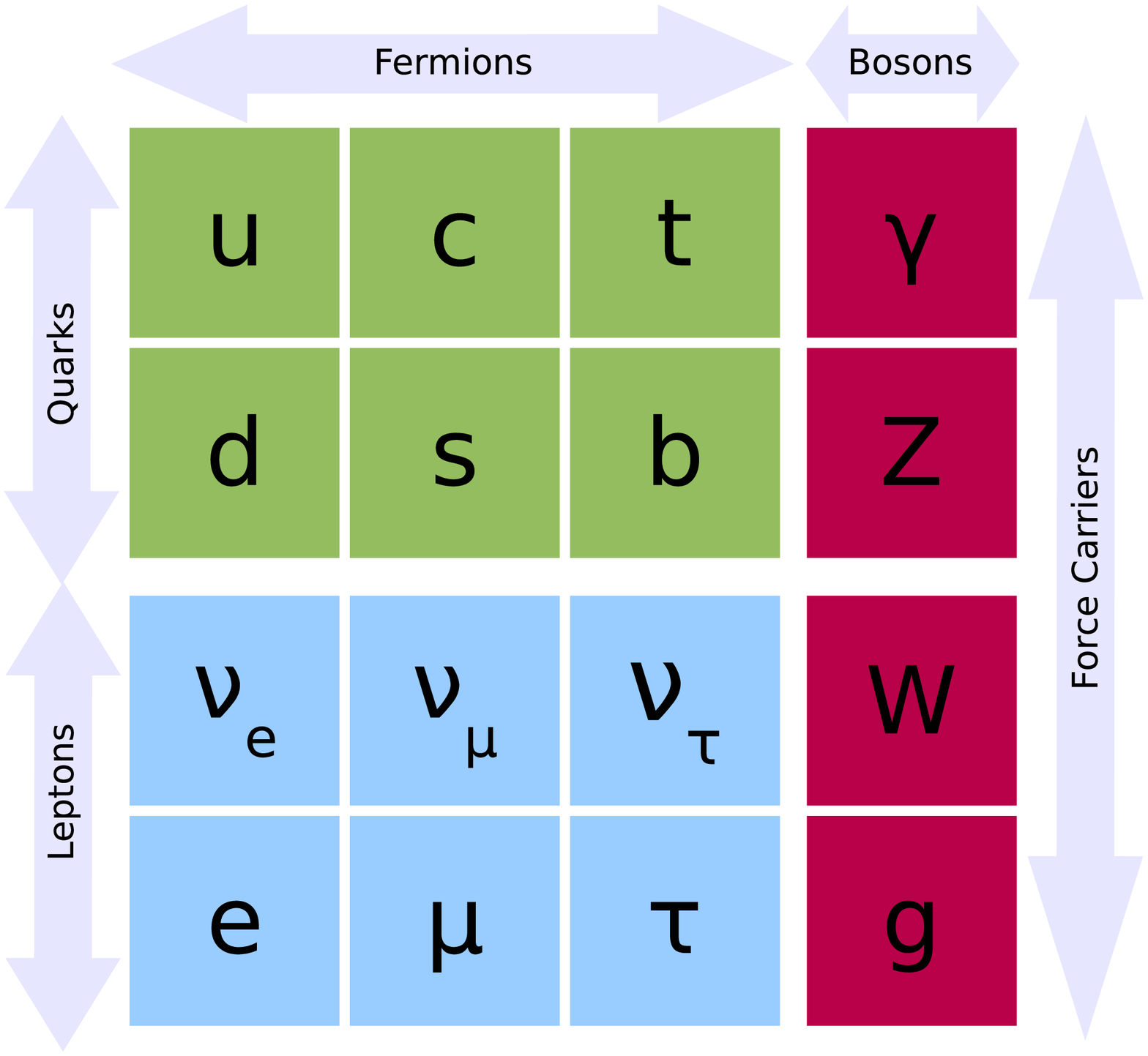}
    \hfill} 
\label{sm}
\end{figure*}

The matter particles can be further grouped into quarks and leptons which occur in three generations. The quarks are: up ($u$) and down ($d$) in the first generation, charm ($c$) and strange ($s$) in the second generation and top ($t$) and bottom ($b$) in the third generation. Quark's charges and masses are listed in Table~\ref{smquark}. The leptons are the electron ($e$), muon ($\mu$), tau ($\tau$) and the corresponding neutrinos ($\nu_i$), where $i$ denotes $e,\mu~\hbox{or}~\tau$. The masses of leptons are listed in Table~\ref{smlep}. The neutrinos are chargeless and have negligible mass. The electron and neutrinos are stable. Muons and taus have a lifetime of $\sim 10^{-6}$ and $10^{-15}$ seconds, respectively. Each matter particle in the SM is also accompanied by its corresponding antiparticle: antiparticles have equal mass and opposite charge of the corresponding particles. The bosons include the photon ($\gamma$), the $W$ and $Z^0$ weak bosons, and eight massless gluons; these particles are the force mediators. The photon is the mediator of the electromagnetic force, the $W,Z$ bosons mediate the weak nuclear force and the gluons carry the strong nuclear force (See Table~\ref{smbosons}.). 
\begin{table}[htbp]
\caption{Charges and masses of the SM quarks \cite{PDG}. Due to the asymptotic freedom structure of QCD, quarks and gluons are not capable of independent existence \cite{Greiner:2002ui,Gross:1973id,Gross:1973ju,Gross:1974cs} (and references therein).  Therefore, quark masses are measured indirectly through their influence on hadrons. The computational scheme used to calculate the quark mass from experimental observations is the source of the uncertainty in their mass.}
\begin{center}
\begin{tabular*}{0.4\textwidth}{@{\extracolsep{\fill}}c|ccc}
\hline\hline
& charge & mass (GeV) &\\
\hline
$u$ & 2/3  &  0.002 to 0.008 &\\
\hline
$d$ & -1/3 &  0.005 to 0.0015 &\\
\hline\hline
$c$ & 2/3  &  3.38 to 3.48 &\\
\hline 
$s$ & -1/3 &  0.1 to 0.106 &\\
\hline\hline
$t$ & 2/3  &  170 to  174 &\\
\hline
$b$ & -1/3 &  4.1 to 4.5 &\\
\hline
\end{tabular*}
\end{center}
\label{smquark}
\end{table}
\begin{table}[htbp]
\caption{Charges and masses of the SM leptons \cite{PDG}.}
\begin{center}
\begin{tabular*}{0.36\textwidth}{@{\extracolsep{\fill}}c|cc}
\hline\hline
& mass (MeV) &\\
\hline
$e^{\pm}$  &  0.511$\pm$0.000000013 &\\
\hline
$\mu^{\pm}$ &  105.6$\pm$0.000004 MeV &\\
\hline
$\tau^{\pm}$  &  1777$\pm$0.017 &\\
\hline 
\end{tabular*}
\end{center}
\label{smlep}
\end{table}
\begin{table}[htbp]
\caption{Charges and masses of the SM bosons\cite{PDG}.}
\begin{center}
\begin{tabular*}{0.66\textwidth}{@{\extracolsep{\fill}}c|cccc}
\hline\hline
& charge & mass (GeV) & force &\\
\hline
$\gamma$ & 0 & 0 & electromagnetic &\\
\hline
$W^{\pm}$ & $\pm$1 & 80.39$\pm$0.025 & weak &\\
\hline
$Z$ & 0 & 91.18$\pm$0.0021 & weak &\\
\hline 
$g$ & 0 & 0 & strong &\\
\hline
Higgs & 0 & $>$ 114.4 & &\\
\hline
\end{tabular*}
\end{center}
\label{smbosons}
\end{table}

Gluons bind quarks together and are responsible for the production of hadrons, which are divided in baryons (quark triplets) and mesons
(quark-antiquark pair). Mesons are unstable and quickly decay via the strong force or weak force into lighter particles. For example, the $\pi$ and the $K$ mesons decay through the weak interaction and are relatively long lived with lifetimes of the order of $10^{-8}$ seconds. The $J/\Psi$ meson decays via the strong interaction and has a lifetime of $\sim 10^{-21}$ seconds. 

The only massless particles of the SM are the photon and the gluons. The masses of other SM particles are generated through the Higgs mechanism of electroweak symmetry breaking. This model assumes the existence of a scalar field, called the Higgs field. Dedicated searches for the Higgs by CERN LEP and Fermilab Tevatron have been conducted in recent years. While the Higgs particle remains elusive, these experiments have set limits on its mass. According to the most recent data \cite{Phenomena:2009pt} the mass of the Higgs is constrained to 114.4 $\le m_{Higgs} \le$ 160 GeV or $m_{Higgs} >$ 170 GeV. 
\section{Physics beyond the Standard Model\label{BSM}}
There are strong reasons to believe that the SM is not a complete theory. Firstly, it describes the electromagnetic, strong and weak forces but not gravity. In natural units, the strength of the gravitational force is determined by the Planck mass. Max Planck introduced the Planck mass  $M_{PL}=\sqrt{\hbar c/G_4} \sim$ $10^{19}$ GeV from three fundamental constants: Newton's gravitational constant $G_4$, Planck's constant $\hbar$ and the speed of light $c$. At energy scales of the order of the Planck scale, quantum gravitational effects become important. This suggests that the SM may be a low energy approximation of some theory which is valid at higher energies. The SM is also plagued by the hierarchy problem \cite{Arkani-Hamed:1998rs,Randall:1999vf}: the fundamental scale of gravity is $\sim 10^{16}$ times higher than the electroweak scale. Why is gravity so much weaker than the other forces?

The hierarchy problem can also be understood by considering corrections to the Higgs mass. The Higgs mass receives divergent corrections, e.g. from production of a quark-antiquark pair. The observed Higgs mass $m_{Higgs}$ is related to the bare Higgs mass $m_{bare}$ by \cite{Schmaltz200340}
\begin{equation}
m_{Higgs}^2\sim m_{bare}^2-\Lambda^2,
\label{higgseqn}
\end{equation}
where $\Lambda$ is the cutoff introduced to regulate the divergence. Assuming that the SM is valid up to $M_{PL}$, a natural choice for the cutoff is of the order of the Planck scale. This would cause the Higgs mass to diverge, unless a fine tuning of the order of 1 out of $10^{16}$ is imposed to remove the quantum corrections in Eq.~(\ref{higgseqn}):
\begin{equation}
1\sim \frac{m_{bare}^2}{\Lambda^2}-\frac{m_{Higgs}^2}{\Lambda^2}.
\end{equation}

\begin{figure*}[htbp]
\centerline{\null\hfill
     \includegraphics*[width=0.5\textwidth]{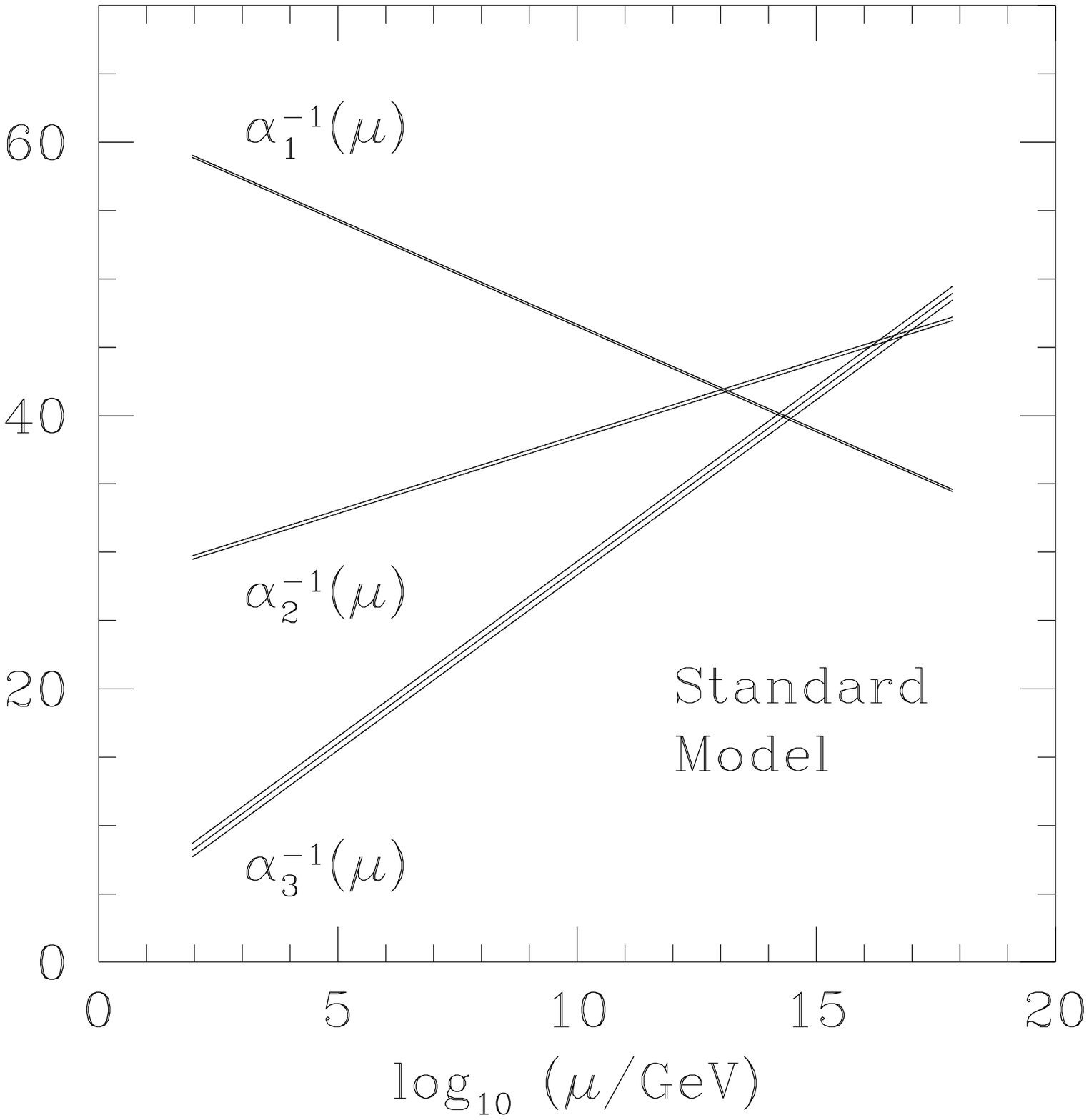}
     \null\hfill
    \includegraphics*[width=0.5\textwidth]{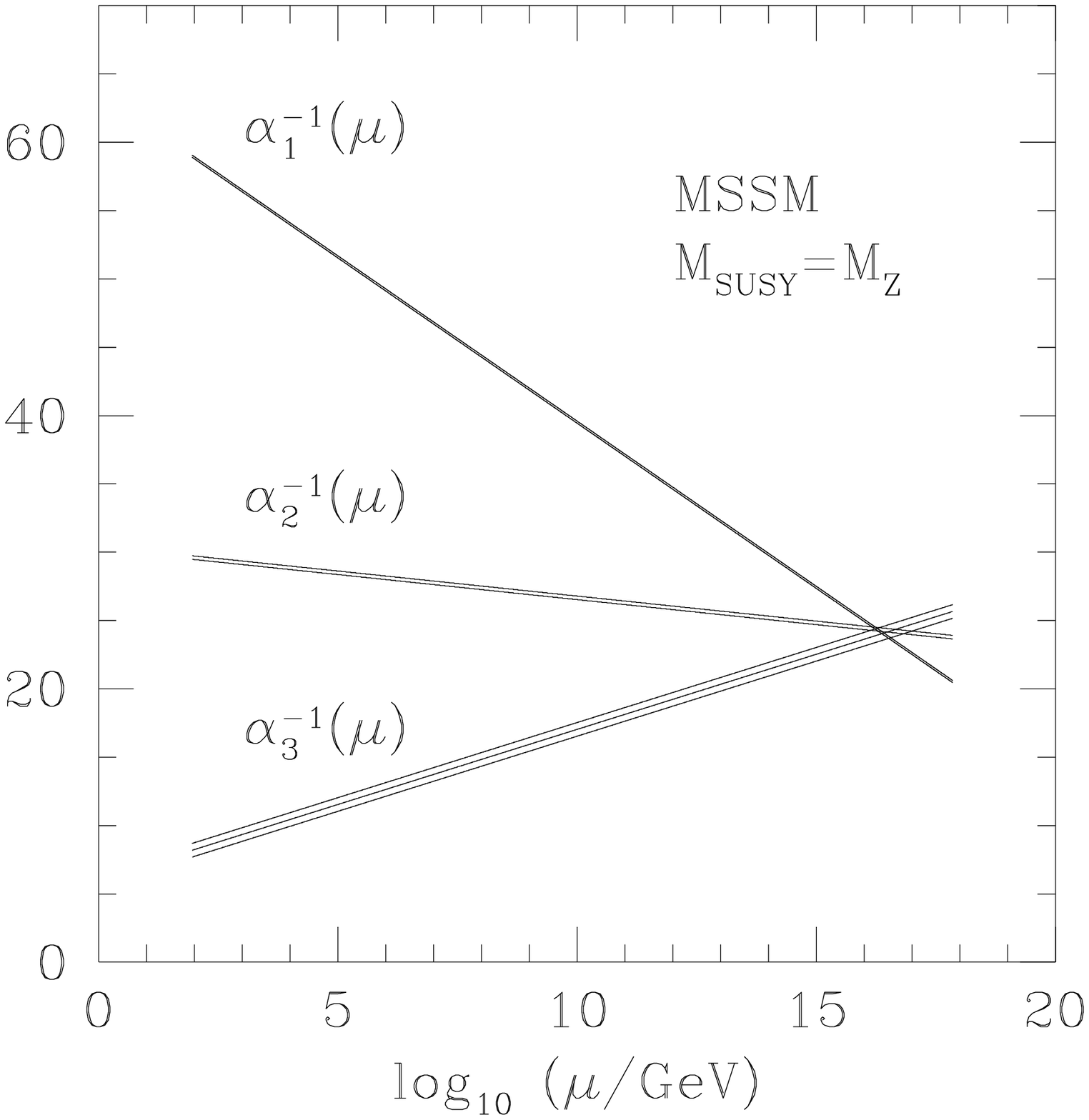}
    \null\hfill} 
\caption{Gauge couplings as a function of the energy in SM (left panel) and SUSY (right panel). The relative width of each line reflects current experimental uncertainties. Reproduced from \cite{Dienes:1996du}.}
\label{couplingfig}
\end{figure*}

The SM also does not address unification of fundamental interactions. Gauge coupling constants determine the strength of an interaction, for example the fine structure constant $\alpha$ determines the strength of the electromagnetic force. Coupling constants are not strictly ``constants" since they vary as the momentum transfer changes in a scattering process. If the SM is indeed embedded in some theory valid at higher energies, then at those higher energies all the three forces should have the same strength. The SM coupling constants, $\alpha_i$, where $i$ denotes electromagnetic ($i$=1) , weak ($i$=2) and strong  ($i$=3) forces, satisfy \cite{Kane:1987gb}
\begin{equation}
\frac{1}{\alpha_i(X^2)}=\frac{1}{\alpha_i(\mu^2)}+\frac{b_i}{4\pi}\ln\frac{X^2}{\mu^2},
\label{coupling}
\end{equation}
where, $\alpha_i(X^2)$ are the coupling constants at momentum transfer scale $X$, $\alpha_i(\mu^2)$ are the coupling constants at energy scale $\mu$ (for example the $Z^0$ mass scale $\sim$ 92 GeV), and $b_i$ are numerical factors which depends on the particle content of the SM: $b_1$=7, $b_2$=-3, $b_3$=-7 \cite{Dienes:1996du}. It is observed that the coupling constants do not meet at any scale when extrapolated to higher energies using Eq~(\ref{coupling}) (See left panel of Fig.\ref{couplingfig}).

\section[Supersymmetry]{Supersymmetry\label{SUSY}}
The most studied model of new physics beyond the SM is SUSY. In SUSY, each boson (fermion) has a fermionic (bosonic) superpartner. Quantum corrections to the Higgs mass due to boson and fermion loops have opposite signs. If the SM particles and their supersymmetric partners are degenerate in mass, their contributions to the Higgs mass cancel. Thus a Higgs boson with mass $\sim$ 100 GeV is possible \cite{Ellis:2007wa}\footnote{Note, however, that SUSY is a broken symmetry at low energy and therefore the degeneracy in the 
particle and sparticle masses is not exact.}.  The presence of superpartners leads to a modification of the $b_i$ constants of Eq.~(\ref{coupling}). Therefore, in SUSY the running gauge couplings when extrapolated to high energies meet at the Grand Unified Theory (or $GUT$) scale $10^{16}$ GeV (see right panel of Fig.~\ref{couplingfig}.). 

In SUSY, superpartners have identical properties as their SM counterparts (masses, charges and quantum numbers), differing only in their spin. The supersymmetric partners of the gluons (spin $1$) are gluinos (spin $\frac{1}{2}$), the superpartners of quarks (leptons) with spin $\frac{1}{2}$ are squarks (sleptons) with spin $0$,  the partner of the Higgs (spin $0$) is the Higgsino (spin $\frac{1}{2}$) and the partners of the of $W$,$Z$ and $\gamma$ (spin $1$) are gauginos with spin $\frac{1}{2}$. The gaugino is a common name for the supersymmetric partners of the gauge bosons. There are two charginos and four neutralinos that occur as a result of mixing between the Higgsinos and gauginos. 

\begin{figure*}[htbp]
\centerline{\null\hfill
    \includegraphics*[width=0.75\textwidth]{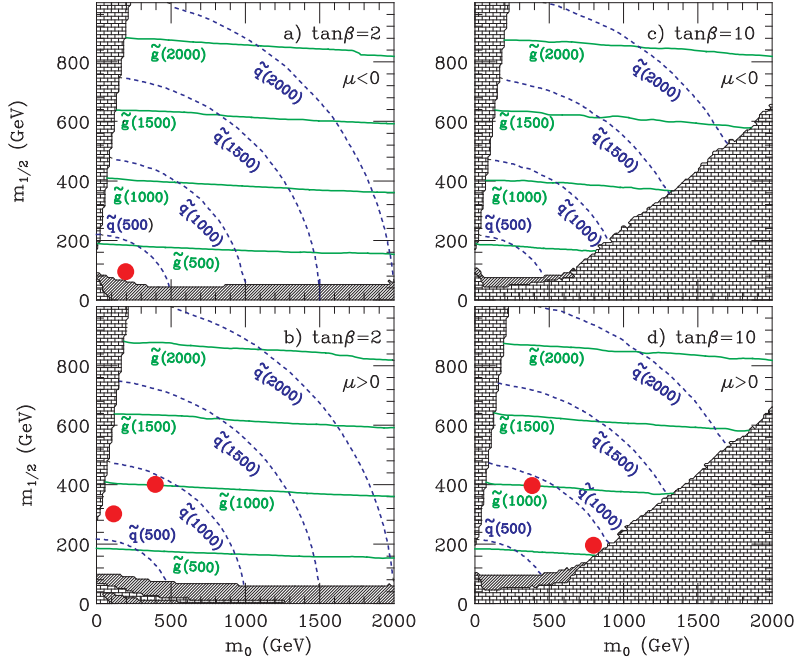}
    \hfill}
\caption{Five ATLAS points (large dots) shown in the $m_0 - m_{1/2}$ plane along with lines of constant squark and gluon mass. The shaded regions are excluded because of theoretical/ex\-pe\-ri\-men\-tal constraints. Reproduced from \cite{Paige:1997xb}.} 
\label{susy_plot}
\end{figure*}

As we have mentioned, the Minimal Supersymmetric extension of the Standard Model (MSSM) \cite{Wess:1973kz,Weinberg:2000cr} allows for the unification of electromagnetic, weak and strong forces at $M_{GUT}\sim 10^{16}$ GeV. Since superpartners of SM particles are not observed at low energies, SUSY must be a broken symmetry. The SUSY breaking
scale is generally assumed to be around 1 TeV. A method of SUSY breaking which is mediated by gravitational
interactions is supergravity (SUGRA). The gravitino is the supersymmetric counterpart of the graviton. In its minimal version, the physics of mSUGRA is determined by a point in the five-dimensional
moduli space with parameters: 
\begin{itemize}
\item $m_0$, the common mass of scalar particles (squarks and sleptons) at $M_{GUT}$,
\item $m_{1/2}$, the common gaugino and Higgsino mass at $M_{GUT}$,
\item $A_0$ the common trilinear coupling at $M_{GUT}$,
\item $\tan~\beta$, the ratio of the vacuum expectation values of the two Higgs fields,
\item $\mu$, the sign of the Higgsino mass parameter.
\end{itemize}
In the MSSM, the Higgs has five states denoted by $H_0,h_0, A_0, H^{\pm}$. The masses of these states are related to the masses of the massive $W^{\pm}$ and $Z^0$ bosons by the relations \cite{RichterWas:1996ak}
\begin{eqnarray}
m_{H_0,h_0}^2&=&\frac{1}{2}\left[m_{A_0}^2+m_Z^2\pm\sqrt{(m_{A_0}^2+m_z^2)^2-4m_{A_0}^2m_Z^2\cos^22\beta}\right],\\
m_{H\pm}&=&m_W^2+m_{A_0}^2.
\label{higgsmass}
\end{eqnarray}
Typical mSUGRA parameters which are relevant for LHC processes (LHC points) are given in Table~\ref{table1} \cite{Bartl:1996dr} and Fig.~\ref{susy_plot}. 
\begin{table}[thbp]
\caption{MSSM Parameters for typical LHC points. The first five points refer to ATLAS and LM1
refers to CMS (see Chapter~\ref{partcoll}). The scalar mass and the gaugino mass are given in GeV \cite{Allanach:2002nj}.}
\begin{center}
\begin{tabular*}{0.65\textwidth}{@{\extracolsep{\fill}}c|cccccc}
\hline\hline
~LHC point~ & $m_0$ & $m_{1/2}$ & $A_0$ & tan $\beta$ & $\mu$ &\\
\hline
A & 100 & 300 & 300 & 2.1 & + &\\
\hline
B & 400 & 400 & 0 & 2 & +     &\\
\hline
C & 400 & 400 & 0 & 10 & +    &\\
\hline 
D & 200 & 100 & 0 & 2 & -     &\\
\hline
E & 800 & 200 & 0 & 10 & +    &\\
\hline
LM1 & 60 & 250 & 0 & 10 & +    &\\
\hline
\end{tabular*}
\end{center}
\label{table1}
\end{table}

Neutralino $\tilde{\chi}_{i}^{0}$, gluino $\tilde{g}$ and squark $\tilde{q}$ masses are determined by $m_0$ and
$m_{1/2}$ through the relations $\tilde{\chi}_{1}^{0}\sim m_{1/2}/2$, $\tilde{\chi}_{2}^{0} \sim \tilde{\chi}_{1}^{\pm} \sim m_{1/2}$,
$\tilde{g} \sim 3m_{1/2}$ and $m(\tilde{q}) \sim ({m_0}^2+6 m_{1/2}^2)^{1/2}$ \cite{Dittmar:1998rb}.  The gluino and the
first two generations of squarks masses do not depend on $A_0$. The masses of the bosons depend on $m_0$ and the masses of the fermions depend on $m_{\frac{1}{2}}$ and $A_0$ \cite{Stark:2005mp}. 

A symmetry of the MSSM is $R$-parity \cite{Wess:1973kz,Weinberg:2000cr}. The  $R$-parity of a particle is defined as:
\begin{equation}
P_{R}=(-1)^{3B+L+2\textbf{s}}\,,
\end{equation}
where $L$ ($B$) is the lepton~(baryon) number and $\textbf{s}$ is the particle spin. Superpartners have
$P_R=-1$ and SM particles have $P_R=1$. As a consequence, $R$-parity implies that SUSY particles are always pair produced in the decay of SM particles. $R$-parity conservation ensures that a SUSY process at the LHC ends in a state with SM particles and two lightest stable SUSY
particles (LSPs), which are generally neutralinos. Being colorless and chargeless the LSPs escape the detector and are
the source of missing transverse momentum \missPT\, an important signature of SUSY events\footnote{If $R$-parity is not
conserved, the missing transverse energy is reduced by the LSP decay.}. Throughout this thesis we will assume that
$R$-parity is conserved, in agreement with the mSUGRA scenario. The mass spectrum for the typical LHC points of Table~\ref{table1} is given in Table~\ref{susymasstable}.
\begin{table}[t]
\begin{center}
\begin{tabular*}{0.65\textwidth}{@{\extracolsep{\fill}}c|ccccccc}
\hline\hline
					&	A	&	B	&	C	&	D	&	E	&	LM1	&\\
\hline													q
$\tilde{d_L}, \tilde{s_L}$		&	669	&	947	&	947	&	315	&	906	&	565	&\\
\hline													
$\tilde{u_L}, \tilde{c_L}$		&	665	&	944	&	943	&	309	&	903	&	559	&\\
\hline													
$\tilde{d_R}, \tilde{s_R}$		&	643	&	916	&	916	&	307	&	900	&	542	&\\
\hline													
$\tilde{u_R}, \tilde{c_R}$		&	644	&	918	&	917	&	306	&	900	&	542	&\\
\hline													
$\tilde{b_1}$				&	616	&	846	&	855	&	270	&	759	&	517	&\\
\hline													
$\tilde{t_1}$				&	472	&	630	&	688	&	253	&	575	&	405	&\\
\hline													
$\tilde{b_2}$				&	642	&	913	&	910	&	306	&	891	&	542	&\\
\hline													
$\tilde{t_2}$				&	677	&	895	&	895	&	316	&	778	&	581	&\\
\hline													
$\tilde{e_L^-}, \tilde{\mu_L^-}$	&	230	&	483	&	484	&	214	&	810	&	187	&\\
\hline													
$\tilde{e_R^-}, \tilde{\mu_R^-}$	&	152	&	427	&	428	&	206	&	804	&	118	&\\
\hline													
$\tilde{\nu_{e_L}}, \tilde{\nu_{{\mu}_L}}$	&	220	&	478	&	476	&	204	&	806	&	168	&\\
\hline													
$\tilde{\tau_1^-}$			&	154	&	427	&	423	&	205	&	795	&	111	&\\
\hline													
$\tilde{\nu_{{\tau}_L}}$		&	218	&	477	&	474	&	204	&	803	&	165	&\\
\hline													
$\tilde{\tau_2^-}$			&	229	&	483	&	483	&	214	&	808	&	189	&\\
\hline													
$\tilde{g}$				&	719	&	951	&	951	&	277	&	541	&	607	&\\
\hline													
$\tilde{\chi}_1^0$				&	117	&	163	&	163	&	43	&	78	&	97	&\\
\hline													
$\tilde{\chi}_2^0$				&	218	&	310	&	306	&	93	&	145	&	178	&\\
\hline													
$\tilde{\chi}_1^{\pm}$			&	217	&	309	&	307	&	93	&	145	&	178	&\\
\hline													
$\tilde{\chi}_2^{\pm}$			&	511	&	765	&	538	&	271	&	330	&	363	&\\
\hline													
$h_0$					&	89	&	92	&	114	&	66	&	110	&	110	&\\
\hline													
$H_0$					&	613	&	1021	&	703	&	376	&	856	&	377	&\\
\hline													
$A_0$					&	604	&	1012	&	698	&	366	&	851	&	374	&\\
\hline													
$H^+$					&	613	&	1021	&	707	&	376	&	860	&	385	&\\
\hline														
\end{tabular*}
\end{center}
\caption{SUSY mass spectrum in GeV. The first two generations of squarks and sleptons are degenerate. Masses are computed using \texttt{ISAJET}~(ver.\ 7.75) \cite{Paige:2003mg}.}
\label{susymasstable}
\end{table}

\begin{figure*}[htbp]
\centerline{\null\hfill
    \includegraphics*[width=0.6\textwidth]{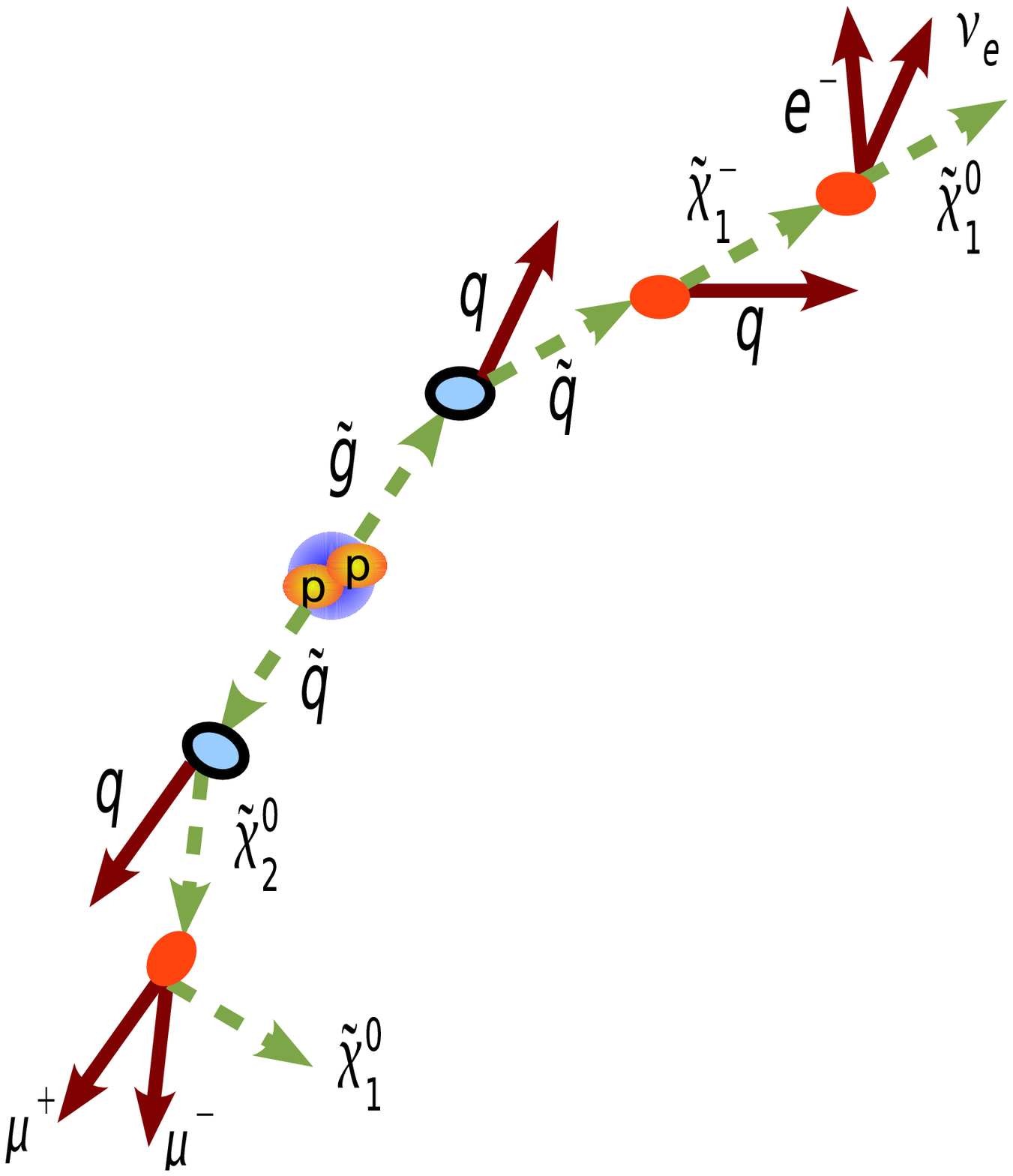}
    \hfill}
\caption{Typical SUSY cascade decay in a hadronic process ($pp$).} 
\label{cascade}
\end{figure*}

Hadronic SUSY processes are dominated by $\tilde{g}\tilde{g}$, $\tilde{g}\tilde{q}$ and $\tilde{q}\tilde{q}$ channels.
Table~\ref{susycross} shows the SUSY cross section for the LHC points of Table~\ref{table1}. Fig.~\ref{cascade} shows a typical SUSY decay chain. This example illustrates the three important characteristics of SUSY processes; production of jets, leptons and missing energy. The jets are produced at the beginning of the decay chain from the hadronization of quarks and gluons.
The end phases of the decay are dominated by leptons produced by the decay of charginos and neutralinos. The missing
energy is due to the LSPs. 
\begin{table}[htbp]
\caption{Cross section in fb for production of SUSY particles for the LHC points listed in Table~\ref{table1} \cite{ATLAS:1999,Drozdetsky:2007zza,Kcira:2007ty, Ball:2007zza}.}
\begin{center}
\begin{tabular*}{0.7\textwidth}{@{\extracolsep{\fill}}c|ccccccc}
\hline\hline
 & A & B & C & D & E & LM1 &\\
\hline
$\tilde{g}\tilde{g}$ 		& 1751 & 258 & 259 & 437189 & 10877 & 10550 &\\
\hline
$\tilde{q}\tilde{\bar{q}}$ 	& 2379 & 363 & 337 & 103059 & 455   & 8851 &\\
\hline
$\tilde{q}\tilde{q}$ 		& 2820 & 686 & 672 & 73769 & 909    & 6901 &\\
\hline 
$\tilde{g}\tilde{q}$ 		& 8306 & 1486 & 1444 & 642765 & 8259 & 28560 &\\
\hline
\end{tabular*}
\end{center}
\label{susycross}
\end{table}
%\clearpage
%
%

\section[Extra-dimensional Models]{Extra-dimensional Models\label{ed}}
%\section{Introduction\label{EDM}}
%
Alternatives to SUSY are extra-dimensional models such as LEDs \cite{Arkani-Hamed:1998rs,Antoniadis:1998ig,Arkani-Hamed:1998nn}, warped braneworlds \cite{Randall:1999ee,Randall:1999vf} and universal extra dimensions \cite{Appelquist:2000nn,Cembranos:2006gt}. In these scenarios, the fundamental scale of gravity $M_\star$ is lowered to $\sim$ 1 TeV by the presence of $n$ extra spatial dimensions or by a warping of the metric. 

The idea of extra spatial dimensions was first proposed by Kaluza and Klein in the 1920's in an attempt to unify gravity and electromagnetism. The recent interest in extra-dimensional models arises from the possibility that these models may provide a solution to the hierarchy problem.

The ArkaniHamed-Dimopoulos-Dvali (ADD) \cite{Arkani-Hamed:1998rs,Antoniadis:1998ig,Arkani-Hamed:1998nn} and the Randall-Sundrum (RS)  \cite{Randall:1999ee,Randall:1999vf} models assume the existence of 3-branes. A 3-brane is a surface with three spatial dimensions where matter fields are restricted. In the ADD model, the gravitons can propagate in the extra dimensions which are compactified at small scales (bulk).

The idea of EDs also find support from ST \cite{Polchinski:1998rq,Polchinski:1998rr,Zwiebach:2004tj}, the most complete proposal for a theory of quantum gravity as of now. According to ST all particles are different vibrational modes of a closed or open string living in 10 or 11 dimensions. Open strings have endpoints on the 3-brane, whereas closed strings describing force carriers or other multiplets are free to move into the bulk. 

\subsection{Kaluza-Klein Reduction\label{kkr}}

The effect of the presence of EDs can be illustrated with a simple example: the KK reduction on a circle of radius $R$ \cite{Harris:2004mf, Dienes:2002hg}.
\begin{figure*}[htbp]
\centerline{\null\hfill
    \includegraphics*[width=0.7\textwidth]{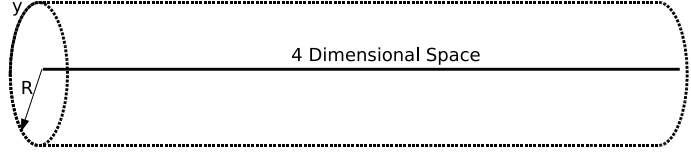}
    \hfill}
\caption{Compactified extra dimension of radius R (dotted line). The solid line denotes 4D spacetime. Adapted from \cite{Dienes:2002hg}.} 
\label{kkcompact}
\end{figure*}
The Klein-Gordon equation for a scalar field $\psi$ of mass $m_{\psi}$ on a 5-dimensional spacetime with topology $M_4\times S_1$, where $M_4$ is the four dimensional Minkowski space and $S_1$ is a circle (see Fig.~\ref{kkcompact}), is
\begin{equation}
\left(\frac{\partial^2}{\partial t^2}-\nabla_5^2+m_{\psi}^2\right)\psi=0,
\label{kgeq}
\end{equation}
where $\nabla_5^2$ is the Laplacian in 5 dimensions:
\begin{equation}
\nabla_5^2=\frac{\partial^2}{\partial x_1^2}+\frac{\partial^2}{\partial x_2^2}+\frac{\partial^2}{\partial x_3^2}+\frac{\partial^2}{\partial y^2}.
\end{equation}
Here $x_i$ ($i$=1,2,3) are the usual three spatial coordinates and $y$ is the coordinate of $S_1$. $\psi(x_i,y,t)$ is periodic under the transformation $y\rightarrow y+2\pi R$. The solution of Eq.~(\ref{kgeq}) is of the form
\begin{equation}
\psi=\Sigma_{l=-\infty}^{\infty}\psi_n(x_i,t)e^{\i p_{5}y},
\label{soln}
\end{equation}
where $p_5=l/R$ (l=$\pm1,\pm2\dots$) is the momentum along the extra dimension. The exponential part of the solution follows from the periodicity condition on $y$. Substituting Eq.~(\ref{soln}) into Eq.~(\ref{kgeq}) it follows
\begin{equation}
\sum_{l=-\infty}^{\infty}\left(\frac{\partial^2}{\partial t^2}-\nabla_4^2+m_{\psi}^2+\frac{l^2}{R^2}\right)\psi=0.
\label{kg}
\end{equation}
Therefore, the scalar field $\psi$ in 4 dimensions is equivalent to an infinite number of KK
states states with masses
\begin{equation}
m^2=m_{\psi}^2+\frac{l^2}{R^2}.
\label{mass}
\end{equation}
The above equation shows that a particle in $n$ dimensions can be described by a collection of 4-dimensional modes. The lightest field $\psi_0$ has mass $m_{\psi}$ (this is the KK zero mode or ground state) and the excited modes are doubly degenerate $m(l)=m(-l)$. The mass of the KK excitation for $R^{-1}\ll m_{\psi}$ is, in natural units,
\begin{equation}
m\sim m_{\psi}+\frac{l}{R}.
\label{mass1}
\end{equation}
Processes with energy $\sim R^{-1}$ should be able to probe excited KK modes. Thus, $R^{-1}$ is the threshold energy for detecting EDs.

Models of EDs can be classified on the basis of  KK excitations. In the universal extra dimension model, all particles experience KK excitations.  In the flat compactification model (e.g. ADD), or in the warped compactification model (e.g. RS), there is a restriction on which particles possess KK excitations. The work in this thesis is concerned with the two latter models. These will be discussed in some detail in Sects.~\ref{add} and ~\ref{rs}.

\subsection{ADD\label{add}}
\begin{figure*}[htbp]
\centerline{\null\hfill
    \includegraphics*[width=0.6\textwidth]{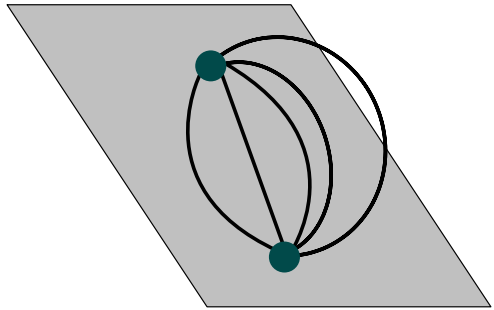}
    \hfill}
\caption{Gravitational lines of force between the two test masses leaks into the EDs. This leads to gravity being weaker on the brane. Adapted from \cite{Dienes:2002hg} .} 
\label{grav2}
\end{figure*}

In the ADD model, all gauge fields except gravity are confined to a 3-dimensional brane. Their quanta do not show KK excitations because they do not feel the effect of the EDs; only gravitons are allowed to propagate in the bulk. According to the ADD model, gravity is a strong force in the higher-dimensional spacetime, but appears weak to a four-dimensional observer due to its``leakage" in the EDs (See Fig.~\ref{grav2}).
In this model, $n$ EDs are compactified on a circle of radius R. The relation between the gravitational constants in $4$ and $n+4$ dimensions, $G_4$ and $G_{n+4}$, can be obtained by considering the $4+n$ dimensional Gauss' Law \cite{Arkani-Hamed:1998nn}

\begin{equation}
\begin{split}
\left(\hbox{Net gravitational flux over a closed surface C} \right)\\=
S_{(3+n)} G_{4+n} \times \hbox{Mass contained within C},
\label{gravgausslaw}
\end{split}
\end{equation}
where $S_d=2\pi^{d/2}/\Gamma(d/2)$ is the surface of an unit $d$-dimensional sphere. The force between two test masses $m_1$ and $m_2$, separated by a distance $r$ is
\begin{equation}
F_{4+n}=G_{4+n}\frac{m_1 m_2}{r^{n+2}} \hbox{~for r$\ll$R}.
\label{addeq1}
\end{equation}
For $r\gg R$, the test masses do not experience the effect of the EDs: the gravitational lines of force are constrained within the volume of the EDs. Therefore, the familiar $1/r^2$ force law is obtained:
\begin{equation}
F_{4}=G_{4}\frac{m_1 m_2}{R^n r^2}.
\label{addeq2}
\end{equation}
The relation between the four-dimensional and the higher dimensional Newton's constants, using Eqs.~(\ref{addeq1}) and~(\ref{addeq2}), is
\begin{equation}
G_{4+n}=R^{D-4}G_{4}\equiv R^{D-4} M_{PL}^{-2},
\label{geq}
\end{equation}
where $D=n+4$ is the total number of spacetime dimensions. Since $R$ in Eq.~(\ref{geq}) has the dimensions of inverse mass, $G_{4+n}$ has dimensions of $({\rm mass})^{2-D}$. The $D$ dimensional Planck's constant $M_{\star}$ is defined as
\begin{equation}
M_{\star}=G_{D}^{\frac{1}{2-D}}.
\end{equation}

The first equality of Eq.~(\ref{geq}) can also be obtained through dimensional analysis \cite{Zwiebach:2004tj}. The four-dimensional gravitational potential $\Phi_{g(4)}$ satisfies
\begin{equation}
\nabla^2\Phi_{g(4)}=4\pi G_{4}\rho_m,
\label{gravpot}
\end{equation}
where $\rho_m$ is the mass density. In $D$ dimensions, Eq.~(\ref{gravpot}) can be rewritten as
\begin{equation}
\nabla^2\phi_{g(n+4)}=4\pi G_{n+4}\rho_m.
\label{gravpotd}
\end{equation}
The left hand sides of Eq.~(\ref{gravpot}) and Eq.~(\ref{gravpotd}) have identical dimensions. Therefore, it follows
\begin{equation}
G_{4+n}\frac{M}{R^{n+3}}=G_{4}\frac{M}{R^3}\Rightarrow G_{4+n}=G_{4}R^{D-4}.
\label{relation}
\end{equation}
Using the above equation and Eq.~(\ref{geq}), the relation between the fundamental scales in $4$ and $n+4$ dimensions is 
\begin{equation}
M_{\rm PL}^2\sim V_n M_{\star}^{n+2}.
\label{reln}
\end{equation}
If the volume of the EDs is large in $M_{\star}^{-1}$ units, $M_{\star}$ can be of the order of one TeV.

Setting $M_{\star} \sim$ 1 TeV in Eq.~(\ref{reln}) implies that the radius of the EDs is $R\sim 10^{32/n - 19}$ meters.  $n$=1 implies $R$=$10^{10}$ km, i.e. a large extra dimension with a size of the order of the solar system. The case $n=2$ is ruled out from astrophysical considerations \cite{Eboli:1999aq}. Gravity has not been tested to scales of $\sim 10^{-6}$ mm or $\sim 10^{11}$ TeV (for $n=3$), and hence one might expect to see deviations from Newton's laws at this scale\footnote{Tests of the Gravitational Inverse-Square Law give bounds on the size of an extra dimension:  currently it is $\le$  44 $\mu m$ \cite{Kapner:2006si}}. 

As outlined in Sect.~\ref{kkr}, compactification of EDs causes the appearance of towers of KK modes. From Eq.~(\ref{mass}), compactification of $n$ extra spatial dimensions of radius $R$ would result in 
\begin{equation}
m_{n}(\textbf{l})=\frac{1}{R} \sqrt{l_1^2+l_2^2+...+l_n^2}\,,
\end{equation}
where $m_{n}(\textbf{l})$ denotes the mass of the $n-th$ KK mode $\textbf{l}$. Depending on the number of EDs,
the graviton modes could be very light. For example, the mass of the lowest graviton mode
is $\sim$ 25 MeV for 6 EDs . Experimental signatures of the ADD model would be production of KK excitations of gravitons and virtual gravitons. Since the mass splitting of the KK modes, $\Delta m=1/R\sim10^{-32/n}$ TeV is  extremely small, a large number of KK modes would be produced in high energy collisions.

In the ADD model, gravitons are expected to decay producing jets and missing energy (\missET\ ), i.e., a jet or photon recoiling against ``nothing". Detection of such events would establish the existence of
EDs. Searches for graviton production have been conducted or are already underway: data from LEP and Tevatron have been analyzed to search processes like  $e^{+}e^{-}\rightarrow \gamma+$\missET\  and $e^{+}e^{-} \rightarrow Z+$\missET\ , $p\bar{p} \rightarrow jet+ $\missET\ and $p\bar{p} \rightarrow \gamma+$\missET\ \cite{PDG}. The combined LEP 95\% CL lower bounds on the fundamental Planck mass and the upper bounds on the size of EDs are listed in Table~\ref{ledtable} \cite{Ask:2004dv}.
%$e^{+}e^{−} \rightarrow Z+$\missET\ channel is not of primary importance because the $Z$ decay into leptons is rare and it is in this channel that the $Z$ is identified.%
%
\begin{table}[htbp]
\caption{LEP lower bounds on $M_{\star}$ and upper bounds on the size of the EDs as a function of the number of EDs for the ADD model.}
\begin{center}
\begin{tabular*}{0.41\textwidth}{@{\extracolsep{\fill}}c|ccc}
\hline\hline
$n$ &  $M_{\star}$ (TeV) & R (mm) &\\
\hline
3 	& $> 1.20$ & $< 2.6 \times 10^{-6}$ & \\
\hline
4 	& $> 0.94$ &  $< 1.1 \times 10^{-8}$ & \\
\hline
5	& $> 0.77$ & $< 4.1 \times 10^{-10}$  & \\
\hline 
6	& $> 0.66$ & $< 4.6  \times 10^{-11}$ &\\
\hline
\end{tabular*}
\end{center}
\label{ledtable}
\end{table}
\subsection{RS\label{rs}}
The simplest RS model assumes the existence of one extra dimension and two 3-branes. SM fields are restricted to one of the branes while gravitons are allowed to exist also in the bulk. In the RS scenario, the hierarchy problem is solved by assuming a single warped extra dimension. 
\begin{figure*}[t]
\centerline{\null\hfill
    \includegraphics*[width=0.75\textwidth]{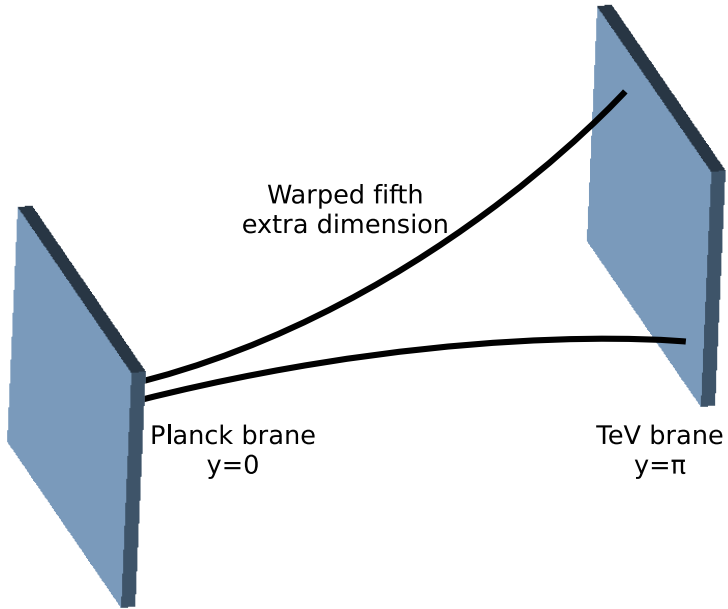}
    \hfill}
\caption{RS model of EDs. The fifth extra dimension is warped leading to the weakness of gravity on our brane. SM fields are restricted to the TeV brane. The graviton can propagate in the extra dimension.} 
\label{rspic}
\end{figure*}
The perceived weakness of gravity is not due to the presence of a large flat extra-dimensional bulk as in the ADD model, but follows from the curvature of the extra dimension. RS compactification is on an orbifold which is obtained by applying a circular compactification on the extra dimension $y$, i.e.  $y\rightarrow y+2\pi r$, where $r$ is the compactification radius of the extra dimension $y$, and identifying $y=-y$. The extra-dimensional coordinate takes values between $0$ and $\pi r$. Thus, the resulting manifold can be seen as a line segment of length $\pi r$. 

\begin{figure*}[t]
\centerline{\null\hfill
    \includegraphics*[width=0.75\textwidth]{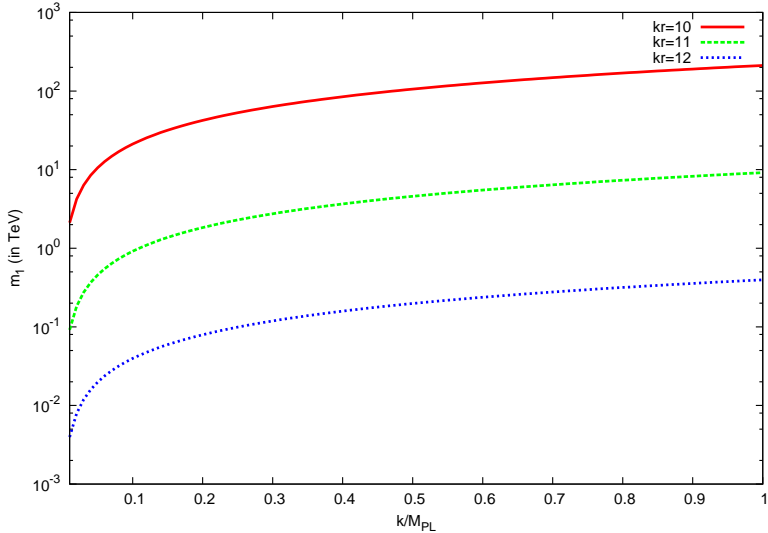}
    \hfill}
\caption{Mass of the first graviton excitation $m_1$ versus $c$ for $kr$=10 (solid red line), 11 (dashed green line) and 12 (dotted blue line).} 
\label{ckpl}
\end{figure*}

The branes in the RS model are located at the orbifold fixed points $y=0$ (Planck  brane) and $y=\pi r$ (TeV brane). SM particles are localized on the TeV brane and gravity can propagate in the bulk. The Planck scale is related to the fundamental scale of gravity in $5$ dimensions by
\begin{equation}
M_{PL}^2=8 \pi\frac{M_\star^3}{k}(1-e^{-2\pi kr})\,,
\label{rseq}
\end{equation}
where $k$ is of the order of the Planck scale. 
The hierarchy between the fundamental scales is removed by the presence of the warp factor $e^{-2\pi kr}$. If $k r$ is in the range 10-12, TeV scale masses can be generated from the fundamental Planck scale $\sim 10^{16}$ TeV. A field on the TeV brane with a mass $k x_n$ has a physical mass $m_n$ of
\begin{equation}
m_{n}=k x_n e^{-k r \pi}\,,
\end{equation}
where $x_n$ are the roots of the Bessel function of first order $J_1(x_n)$. The mass of the lowest graviton mode in the RS scenario is $O$(TeV). The strength of the  graviton-matter coupling is determined by the scale $\Lambda=(1/\sqrt{8 \pi})M_{PL}e^{-k\pi r}$.
%Here, $\Lambda_{\pi} \sim$ 1 TeV. 
In the following sections the ratio~$c=k/M_{PL}$ is chosen to be approximately $10^{-2}$ \cite{Davoudiasl:1999jd}. This leads to a graviton mass of the order of $\sim$ 1 TeV, which could be probed at the LHC (see Fig~\ref{ckpl}). The mass splitting between the first and second excited states of the graviton is $\Delta m=m_2-m_1\sim O(1)$ TeV.

Real gravitons in hadronic collisions would be produced via $q\bar{q}\rightarrow gG, qg \rightarrow qG ~\hbox{and} ~gg \rightarrow gG$. The
experimental signatures would be multijet or jets+photons along with missing $E_T$. Relevant processes for virtual gravitons are $f\bar{f}\rightarrow G^{\star}
~\hbox{and}~gg \rightarrow G^{\star}$. 
%The decay of virtual gravitons would enhance certain SM processes. 
Gravitons in the RS scenario behave differently from ADD gravitons. Being heavy, they quickly decay into SM
particles  as in $G^\star \rightarrow f\bar{f}, gg, \gamma\gamma, Z^0Z^0 ~\hbox{and}~W^+W^-$. 

\section[Black hole production]{Black hole production \label{bhc}}
%\section{Introduction\label{BHLHC}}
%
% from int journal of mod phys-marco and hossenfelder, kanti, gingrich
In scenarios with LEDs, $pp$ collisions at the LHC could produce TeV-mass BHs\footnote{In this thesis we consider only BHs. For reviews on string balls and branes see \cite{PhysRevD.66.036007,Dimopoulos:2001qe,Ahn:2002mj,Jain:2002kf,Anchordoqui2002302} and references therein.}. A non-rotating BH is described by the Schwarzschild solution in ($n$+4) dimensions:
\begin{equation}
ds^2=-R(r)dt^2+R(r)^{-1}dr^2+r^2d\Omega_{n+2}^2,
\label{schmetric}
\end{equation}
where 
\begin{equation}
R(r)=1-\left(\frac{R_{BH}}{r}\right)^{n+1}.
\end{equation}
The radius of the BH as a function of the BH mass $M$ is
\begin{equation}
R_{BH}(M)=\frac{1}{\sqrt{\pi}~M_{\star}}\left\{\left[\frac{8\Gamma(\frac{n+3}{2})}{(2+n)}\right]\frac{M}{M_{\star}}\right\}^\frac{1}{n+1}.
\label{rbh}
\end{equation}
Thus, a fundamental scale-mass BH has a radius of the order of $M_{\star}^{-1}$ in natural units. BH radii for different values of $n$ are shown in Table~\ref{tablebhradius} for M=4 TeV and $M_{\star}$=1 TeV. 
%However, note, that the higher dimensional BH radius is $\sim 10^{-4}$ fm which could be probed by the LHC.
%while the LHC could probe distances of the order of $10^{-5}$ fm.
%
\begin{table}[htbp]
\caption{BH radius as a function of $n$ for a BH mass of M=4 TeV.}
\begin{center}
\begin{tabular*}{0.85\textwidth}{@{\extracolsep{\fill}}c|ccccccc}
\hline\hline
 n & 1 & 2 & 3 & 4 & 5 & 6 & \\
\hline
$R_{BH}\times 10^{-4}$ fm 	& 3.64 & 2.45 & 2.10 & 1.98 & 1.93 & 1.93 & \\
\hline
\end{tabular*}
\end{center}
\label{tablebhradius}
\end{table}

\begin{figure*}[htbp]
\centerline{\null\hfill
    \includegraphics*[width=0.6\textwidth]{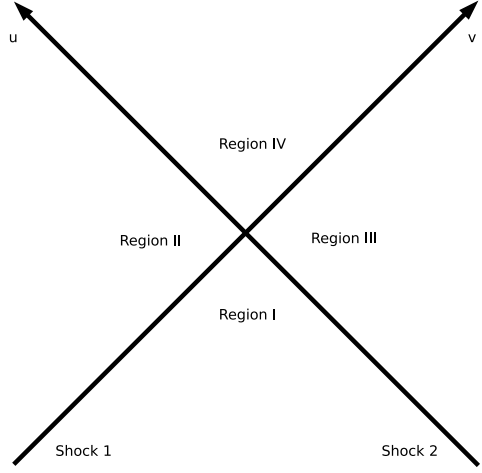}
    \hfill}
\caption{The $u$ and $v$ lines represent two shock waves moving in opposite directions, colliding at $u$=$v$=0. Region $I$ is before the collision. Regions $I,II \hbox{and}~III$ are linear. Region $IV$ is highly non-linear and curved, a closed trapped surface is formed. The BH is formed in this region.  }
\label{shock}
\end{figure*}

BH production in particle collisions would occur as follows. According to Thorne's hoop conjecture \cite{hoop}, a BH of mass $M$ is formed when an object is compacted in all directions such that
\begin{equation}
C < 2\pi R_{BH}(M),
\label{thorne}
\end{equation}
where C is the circumference of the region where the object is compacted into and $R_{BH}$ is the Schwarzschild radius
for a BH of mass $M$. Assuming no gravitational radiation emission the  black disk (BD) cross section for head on collision is
\begin{equation}
\hat{\sigma}(\hat{s};n)=\pi R_{BH}^2,
\label{bd}
\end{equation}
where $\sqrt{\hat{s}}$ is the CM energy of the colliding partons. A more realistic model assumes some CM energy being lost as gravitational radiation (see Ref.\ \cite{Cardoso:2005jq} for a more detailed discussion) and non-zero impact parameter. Conservation of angular momentum implies that BHs formed with a non-zero impact parameter are spinning. They are described by the Kerr solution. A spinning BH has a smaller radius than a non-rotating BH of equal mass
\begin{equation}
R_{BH}(M,J)=\frac{1}{\sqrt{\pi}~M_{\star}}\left\{\left[\frac{8\Gamma(\frac{n+3}{2})}{(2+n)}\right]\frac{M}{M_{\star}}\right\}^\frac{1}{n+1}\left[1+\frac{(n+2)^2 J^2}{4 R_{BH}^2 M}\right]^{\frac{-1}{n+1}},
\end{equation}
where $J$ is the angular momentum of the BH. To estimate the gravitational energy loss, the colliding particles are described as two Aichelburg-Sexl shock waves (see Fig.~\ref{shock}) \cite{Aichelburg:1970dh}; the overlap of the shock waves (region $IV$) forms a trapped-surface (TS) which sets a lower limit to the mass of the BH \cite{Yoshino:2002tx,Yoshino:2005hi}\footnote{For an alternative estimate of the collisional gravitational loss, see Ref.\ \cite{Berti:2003si}.}. Depending on the model, BH masses range from 60\% (TS model) to 100\% (BD model) of the total center-of-mass energy.

The cross section at the parton level for the Schwarzschild BH formation process is obtained from Eq.~(\ref{bd}) as
\begin{equation}
\hat{\sigma}_{ij \to BH}(\hat{s},n) =\frac{1}{M_{\star}^2}\left[\frac{8\Gamma(\frac{n+3}{2})}{(2+n)}\right]^\frac{2}{n+1}\left\{\frac{\hat{s}}{M_{\star}^2}\right\}^\frac{1}{n+1},
\label{pcross}
\end{equation}
where $i,j$ are the incident partons with CM energy $\hat{s}$. The parton cross section is independent of the type of incoming partons and depends only on $M, M_{\star}$ and $n$.  If the energy loss due to graviton emission is neglected, the total cross section is obtained by summing the contributions from all possible parton pairs:
\begin{equation}
\begin{split}
\sigma_{pp \to BH}(s,n)&\\ 
&=\sum_{ij}\int_{\frac{\hat{s}}{s}}^1dx\,\int_x^1 \frac{dx'}{x'}\, f_i(x',Q)f_j(x/x',Q)\sigma_{ij\rightarrow BH}(\hat{s};n)\,,
\end{split}
\label{totcross}
\end{equation}
where $s$ is the squared CM energy of the $pp$ collision, $Q$ is four-momentum transfer, $f_i(x',Q)$ are the parton distribution functions (PDFs), $x'$ are the parton momentum fractions and $x$=$x'_i x'_j=\frac{\hat{s}}{s}$. 
%TeV BHs may carry color, electric charge and angular momentum.
% It is interesting to note that increasing the center-of-mass energy at the LHC does not necessarily result in the formation of a BH:  this is 
% because even though parton cross section increases with increasing energy the  parton distribution functions decrease with increasing energy.

H.~Yoshino \& Y.~Nambu and H.~Yoshino \& V.~S.~Rychkov \cite{Yoshino:2002tx,Yoshino:2005hi} have shown that the energy loss due to gravitational radiation is significant for large impact parameters and large number of EDs. This reduces the BH mass from $M= \sqrt{\hat{s}}~\hbox{to}~M= y(z)\sqrt{\hat{s}}$, where $z=b/b_{max}$, $b_{max}$ is the maximum impact parameter and $y(z)$ ranges between $0\dots1$ \cite{Anchordoqui:2003jr}. This condition puts a lower cutoff on the parton momentum fractions. The lower bound on the parton momentum fraction is obtained by the condition $x_{min}=M_{min}^2/[s~y^2]$ where $M_{min}$ is the minimum mass for BH formation \cite{Anchordoqui:2003jr}. The inelastic BH cross section is 
\begin{equation}
\begin{split}
\sigma_{pp \to BH}(s,n) &\\
&=\sum_{ij}\int_0^1 2zdz \int_{x_{min}}^1dx\,\int_x^1 \frac{dx'}{x'}\, f_i(x',Q)f_j(x/x',Q)\sigma_{ij\rightarrow BH}\,.
\end{split}
\label{totcross1}
\end{equation}

\begin{figure*}[htb]
\centerline{\null\hfill
    \includegraphics*[width=0.5\textwidth]{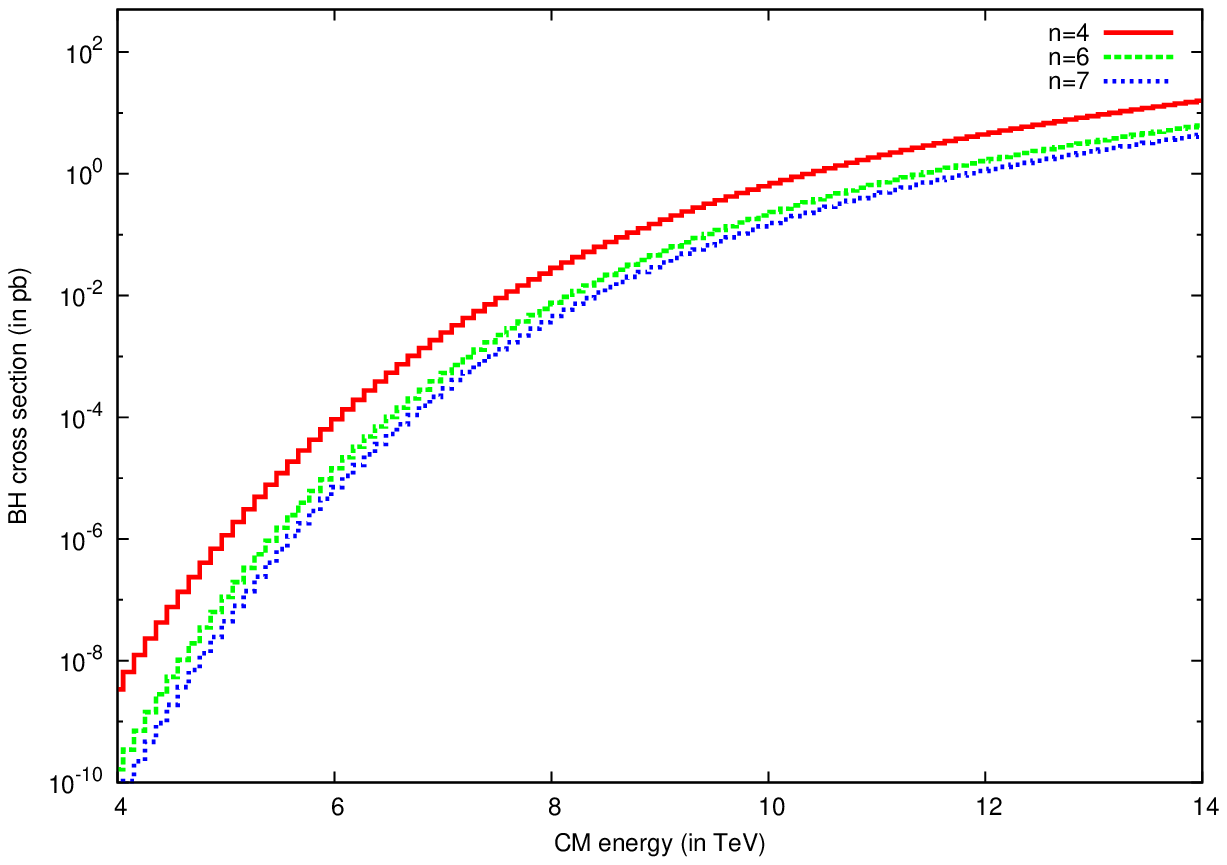}
    \null\hfill
    \includegraphics*[width=0.5\textwidth]{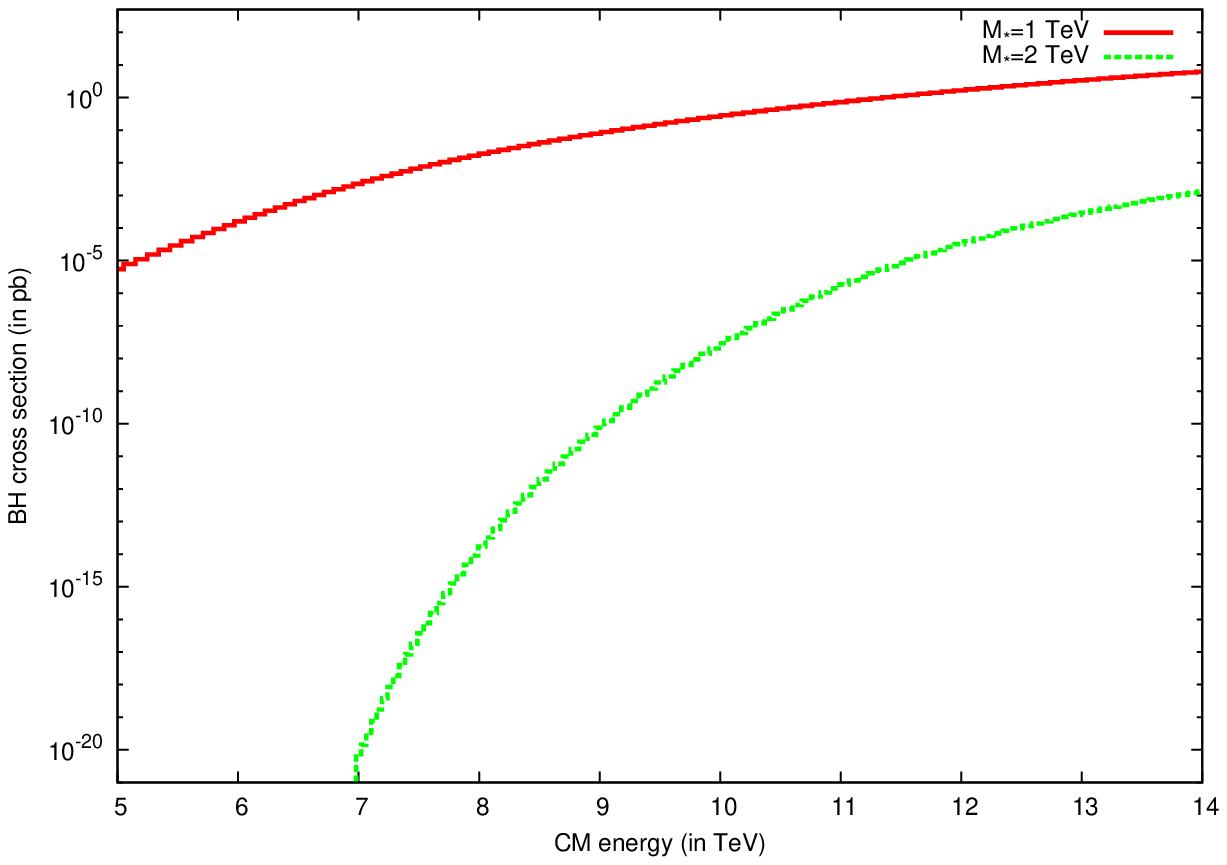}
    \null\hfill}
\caption{Left Panel: BH cross section for different numbers of extra spatial dimensions as a function of the CM energy. The fundamental scale $M_{\star}$ is assumed to be 1 TeV. Right Panel: BH cross section for $M_{\star}$=1 TeV (dashed) and  $M_{\star}$=2 TeV (dotted).}
\label{revbh1}
\end{figure*}

The total BH cross section~(\ref{totcross1}) is shown in Fig.~\ref{revbh1}. The cross section increases with energy. For $6$ extra spatial dimensions and fundamental scale $M_{\star}$= 1 TeV, $\sigma_{pp\rightarrow BH}\sim$ 6 pb at 14 TeV. This would turn the LHC into a BH factory with $\sim 2\times10^5$ BHs per year at a luminosity of $10^{33} cm^{-2} s^{-1}$!\footnote{Lepton colliders are expected to produce an even higher number of BHs than hadron colliders at equal CM energy because the collision energy is not distributed among the hadron constituents.} The BH cross section decreases with the increase in the number of spacetime dimensions and for increasing $M_{\star}$. 

\begin{figure*}[htb]
\centerline{\null\hfill
    \includegraphics*[width=0.5\textwidth]{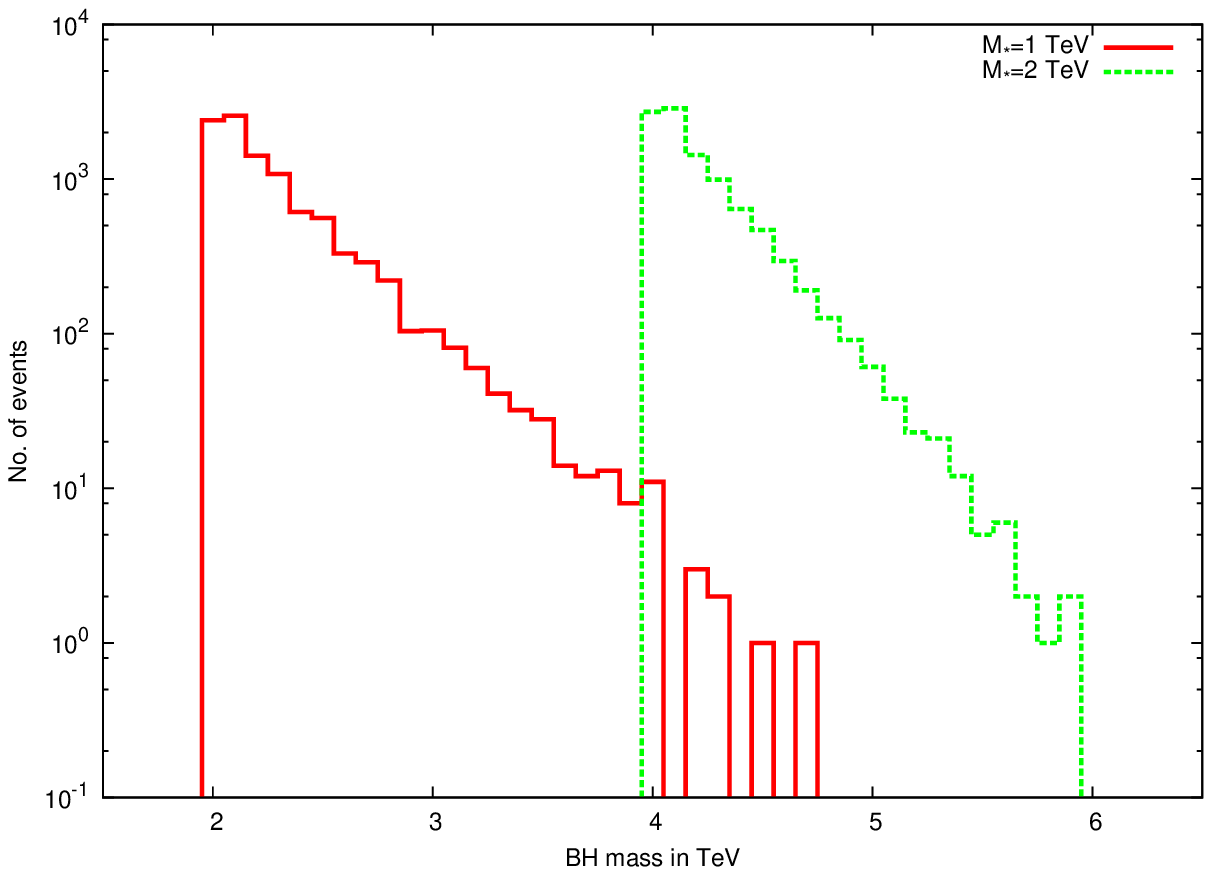}
    \null\hfill
    \includegraphics*[width=0.5\textwidth]{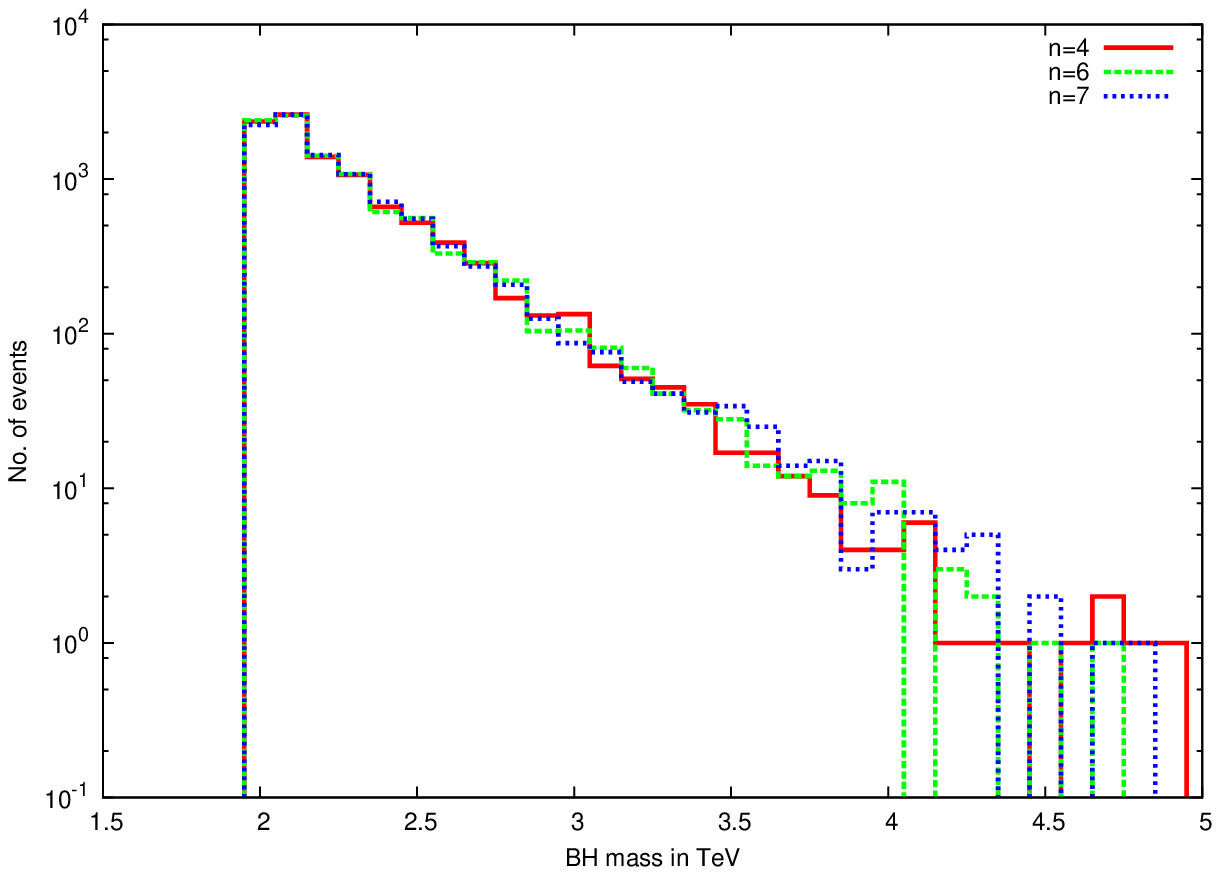}
    \null\hfill}
\caption{Left Panel: BH mass for different values of $M_{\star}$. Right Panel: BH mass for different values of extra spatial dimensions, $n$. Both plots are for 10000 events.} 
\label{bhmass}
\end{figure*}

The distribution of BH masses is shown in Fig.~\ref{bhmass} for different values of $M_{\star}$ and $n$. Higher values of $M_{\star}$ lead to more massive BHs.
%:  a large value of the fundamental scale implies a larger BH mass so that our fundamental assumption $M_{BH}> M_{\star}$ remains valid. This is necessary 
% because as long as the above assumption holds, the BH be treated as  a classical object.
%

TeV BHs decay immediately after formation. The decay starts via loss of excess multipole moments (balding phase). In this phase, the BH also loses mass by emitting gravitational radiation. In the spindown phase the BH loses angular momentum. This is followed by the evaporation phase where the BH loses the bulk of its mass by emitting Hawking radiation \cite{Hawking:1974sw}. 

The Hawking temperature of a $(n+4)$ dimensional BH is 
\begin{equation}
T_H=\frac{n+1}{4\pi R_{BH}},
\label{bhtemp}
\end{equation}
and its lifetime is approximately
\begin{equation}
\tau\sim\frac{1}{M_{\star}}\left(\frac{M}{M_{\star}}\right)^\frac{n+3}{n+1}.
\label{bhtime}
\end{equation}
\begin{table}[htbp]
\caption{BH temperature and lifetime as a function of $n$ for a BH with mass M=4$M_{\star}$=4 TeV.}
\begin{center}
\begin{tabular*}{0.85\textwidth}{@{\extracolsep{\fill}}c|ccccccc}
\hline\hline
 n & 1 & 2 & 3 & 4 & 5 & 6 & \\
\hline
$T_H$ (GeV)					& 86 & 192 & 298 & 397 & 487 & 570 & \\
\hline
$\tau$ ($\times 10^{-26}$ s)	& 1.05 & 0.66 & 0.53 & 0.46 & 0.42 & 0.39 & \\
\hline
\end{tabular*}
\end{center}
\label{tablebhtemptime}
\end{table}
The BH temperature and lifetime for a BH with mass $M$=4 TeV and $M_{\star}$=1 TeV are shown in Table~\ref{tablebhtemptime}. Higher-dimensional BHs are hotter and have shorter lifetimes than their four-dimensional counterparts.
%The temperature and decay time of astrophysical BHs\footnote{Assuming a mass of 4$\times10^6$ solar masses for the BH at the center of the Milky Way.} is $\sim 10^{-19}$ eV and $\sim 10^{83}$ years (larger than the age of the universe!), respectfully. Therefore, astrophysical BHs are not likely to be discovered by Hawking %emission as their temperature is considerably less than the cosmic microwave background temperature of  $\sim$ 3K. On the contrary, as discussed below, Hawking emission forms the basis of studying extra-dimensional BHs.  
%
\begin{figure*}[htbp]
\centerline{\null\hfill
    \includegraphics*[width=0.6\textwidth]{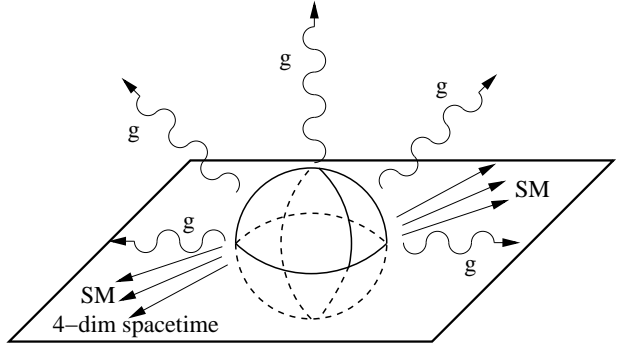}
    \hfill}
\caption{Illustration of the Schwarzschild evaporation of a BH, showing brane and bulk emission \cite{Cavaglia:2002si}. SM particles are emitted on the brane. The graviton escapes both in the brane and bulk.} 
\label{evaporation}
\end{figure*}

The Hawking process ends when the BH reaches a mass comparable to the fundamental scale. At this stage a final $n$-body decay or remnant production (Planck phase) may occur. SM particles produced during the BH decay phase are emitted on the brane and can be experimentally detected \cite{Emparan:2000rs,Cavaglia:2003hg} (see Fig.~\ref{evaporation}). 
%For a more detailed discussion of the evaporation and Planck phases of TeV BHs see Refs.~\cite{Cavaglia:2002si}. 
As SM particles have overall more degree of freedoms (dofs) than bulk particles, brane
emission is greater than bulk emission. This result becomes less and less significant in higher dimensions. For example, the power loss due to gravitons is negligible in four dimensions whereas it is  $\sim$ 25\% in 11 dimensions \cite{Cardoso:2005mh,Cardoso:2005vb}.

\section[String Resonances]{String Resonances\label{SA}}
%\section{Introduction\label{SA}}
%
ST  is  the leading candidate for the unification of the fundamental forces. The string scale $M_s$ is defined as \cite{Zwiebach:2004tj} 
\begin{equation}
l_s=\sqrt{\alpha^{\prime}}=\frac{1}{M_s},
\label{eq1}
\end{equation}
where $\alpha^{\prime}$ is the slope parameter with units of inverse energy squared. The slope parameter is related to the four dimensional Newton's constant $G_{4}$ \cite{Dienes:2002hg} by
\begin{equation}
\alpha^{\prime}=\frac{G_{4}}{g_s^2}.
\label{aplhap}
\end{equation}
The strength of string interactions is controlled by the string coupling $g_s$. Combining Eq.~(\ref{eq1}) and Eq.~(\ref{aplhap}), one obtains 
\begin{equation}
M_s=g_s M_{\star}.
\label{eq2}
\end{equation}
Since string effects are expected to appear just before quantum gravity effects set in, the string coupling is
generally assumed to be of order one. A small coupling also justifies the use of perturbative analysis \cite{Dienes:1996du}. We have seen that in theories with EDs the fundamental quantum gravity scale may be as low as a TeV.  Thus string effects might appear at the TeV scale. If this is the case, SRs would be observed at the LHC before the onset of non-perturbative quantum gravity effects such as BH production. 
String excitation modifications to SM amplitudes may even exceed amplitude modifications due to KK gravitons.  Detection of string events
through corrections to SM amplitudes  \cite{Burikham:2004su, Anchordoqui:2008ac, Anchordoqui:2008di} at the LHC would be the most direct evidence of ST and EDs. 

In Chapter~\ref{simanaly} we will investigate the $pp\rightarrow \hbox{string resonance} \rightarrow \gamma+$ \hbox{jet process} \cite{Anchordoqui:2008ac}\footnote{For a discussion of SR in the dileptonic and diphotonic channels see for e.g. \cite{Cullen:2000ef,Cornet:2001gy,Friess:2002cc,Burikham:2003ha,Burikham:2004su}.}. The relevant process for $pp\rightarrow\gamma+jet$ events is gluon-gluon scattering: $gg\rightarrow g\gamma$. The string amplitude for  this process is \cite{Anchordoqui:2008ac}
\begin{equation}
\begin{split}
|M(gg\rightarrow g\gamma)|^2 = g_s^4 Q^2 C(N)\left\{\left[\frac{s\mu(s,t,u)}{u}+\frac{s\mu(s,u,t)}{t}\right]^2\right\}\\
+ g_s^4 Q^2 C(N)\left\{(s\longleftrightarrow t)+(s\longleftrightarrow u)\right\},
\end{split}
\label{string_amp}
\end{equation}
where $s$, $t$ and $u$ are the Mandelstam variables and
\begin{equation}
\mu(s,t,u)=\Gamma(1-u)\left(\frac{\Gamma(1-s)}{\Gamma(1+t)}-\frac{\Gamma(1-t)}{\Gamma(1+s)}\right).
\end{equation}
Here $N$=3 is the number of $D$-branes needed to generate the eight gluons of the SM,
$C(N)=\frac{2(N^2-4)}{N(N^2-1)}$ is a constant parameter, and $Q^2=\frac{1}{6}\kappa^2 \cos^2\theta_W\sim 2.55
\times 10^{-3}$, where $\kappa^2$=0.02 and $\theta_W$ are the mixing parameter and the Weinberg angle, respectively.
The values of these parameters are chosen as in Ref.~\cite{Anchordoqui:2008ac}.

In the limit $s\rightarrow nM_s^2$, the string amplitude possesses poles at $n$=$s/M_s^2$, where $n$ is an integer. The amplitude has the form of Veneziano amplitudes \cite{Polchinski:1998rq, Polchinski:1998rr,Cullen:2000ef}, a feature common to all string models.
The limiting value of the amplitude at the $n=$odd poles is
\begin{equation}
|M(gg\rightarrow g\gamma)|^2=g_s^4 Q^2
C(N)\frac{4}{(n!)^2}\frac{s^4+u^4+t^4}{M_s^4[s-nM_s^2]}\left\{\frac{\Gamma(t/M_s^2+n)}{\Gamma(t/M_s^2+1)}\right\}^2.
\label{string_amp1}
\end{equation}
The behavior of the amplitude at the even poles is obtained from Eq.~(\ref{string_amp1}) with the substitutions
$s\rightarrow t$ and $n\rightarrow m=t/M_s^2$ in the square bracket term. 
The singularities of the amplitude are smeared with a fixed width $\Gamma=0.1$ for all $n>1$ and as
\begin{equation}
\begin{split}
|M(gg\rightarrow g\gamma)|^2 \sim \frac{4g^4 Q^2
C(N)}{M_s^4}\left\{\frac{M_s^8}{(s-M_s^2)^2+(\Gamma^{J=0} M_s)^2}\right\}\\
+\frac{4g^4 Q^2
C(N)}{M_s^4}\left\{\frac{t^4+u^4}{(s-M_s^2)^2+(\Gamma^{J=2} M_s)^2}\right\},
\end{split}
\label{eq4}
\end{equation}
for $n$=1~\cite{Anchordoqui:2008ac}. Equation~(\ref{eq4}) includes a correction for spin dependent widths: $\Gamma^{J=0}=0.75 \alpha_s M_s$ and $\Gamma^{J=2}=0.45 \alpha_s M_s$, where $\alpha_s=g_s^2/4\pi$ is the strong coupling constant. The presence of the poles indicates the formation of SR.  The total cross section for the $pp \rightarrow \gamma+jet$ event is obtained by integrating the parton cross section over the parton distribution functions of the protons \cite{Pumplin:2002vw}
\begin{equation}
\sigma_{pp \rightarrow string \rightarrow \gamma+jet}=\int_{\frac{\hat{s}}{s}}^{1} dx \int_{x}^{1} dx' \int dt~
f_1(x',Q) f_2(x/x',Q) \frac{d\sigma}{dt},
\label{crosssec1}
\end{equation}
where 
\begin{equation}
\frac{d\sigma}{dt}=\frac{|M(gg\rightarrow g\gamma)|^2}{16 \pi s^2}.
\end{equation}
%
%The limits on $x_1$ and $x_2$ are $0,1$. 
The limits on $t$ are fixed by the conditions $s+u+t=0$ and $|t||u|/s \ge P_{Tmin}^2$ where $P_{Tmin}$ is the minimum transverse momenta of the two outgoing particles of the $2\times2$ scattering.
\begin{figure}[htbp]
\centerline{\null\hfill
    \includegraphics*[width=0.7\textwidth]{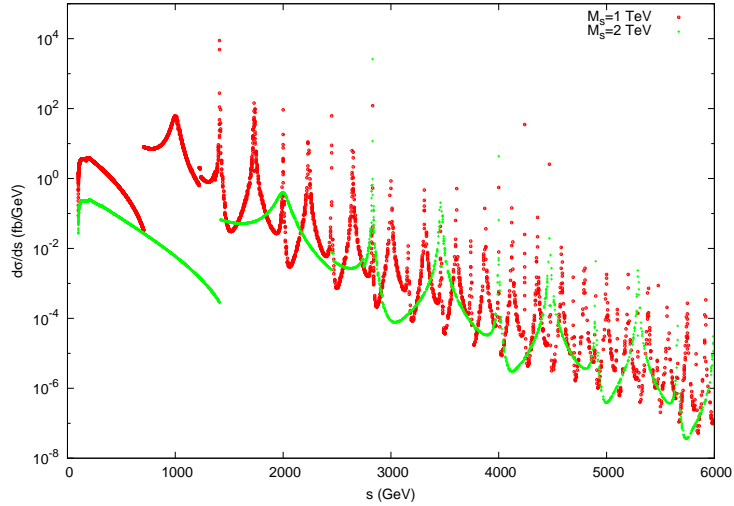}
    \null\hfill}
\caption{Differential cross section of string events for $M_s$= 1 TeV (red dots) and $M_s$= 2 TeV (green
crosses) with $P_{Tmin}$=50 GeV. SRs are clearly seen when $\hat{s}=nM_s^2$.}
\label{dsigds}
\end{figure}
Figure~\ref{dsigds} shows the differential cross section of the $pp\rightarrow \gamma+jet$ process
\begin{equation}
%\frac{d\sigma}{ds}=\int \int dx_2~dt \frac{2 \sqrt{s}}{x_2 E_{CM}^2} f_1(x_1,Q)f_2(x_2,Q) \frac{d\sigma}{dt}.
\frac{d\sigma}{ds}=\int \int dx'~dt \frac{2 \sqrt{s}}{x' E_{CM}^2} f_1(x/x',Q)f_2(x',Q) \frac{d\sigma}{dt},
\end{equation}
where we have used the CTEQ6D PDFs.
\begin{figure}[htbp]
\centerline{\null\hfill
    \includegraphics*[width=0.5\textwidth]{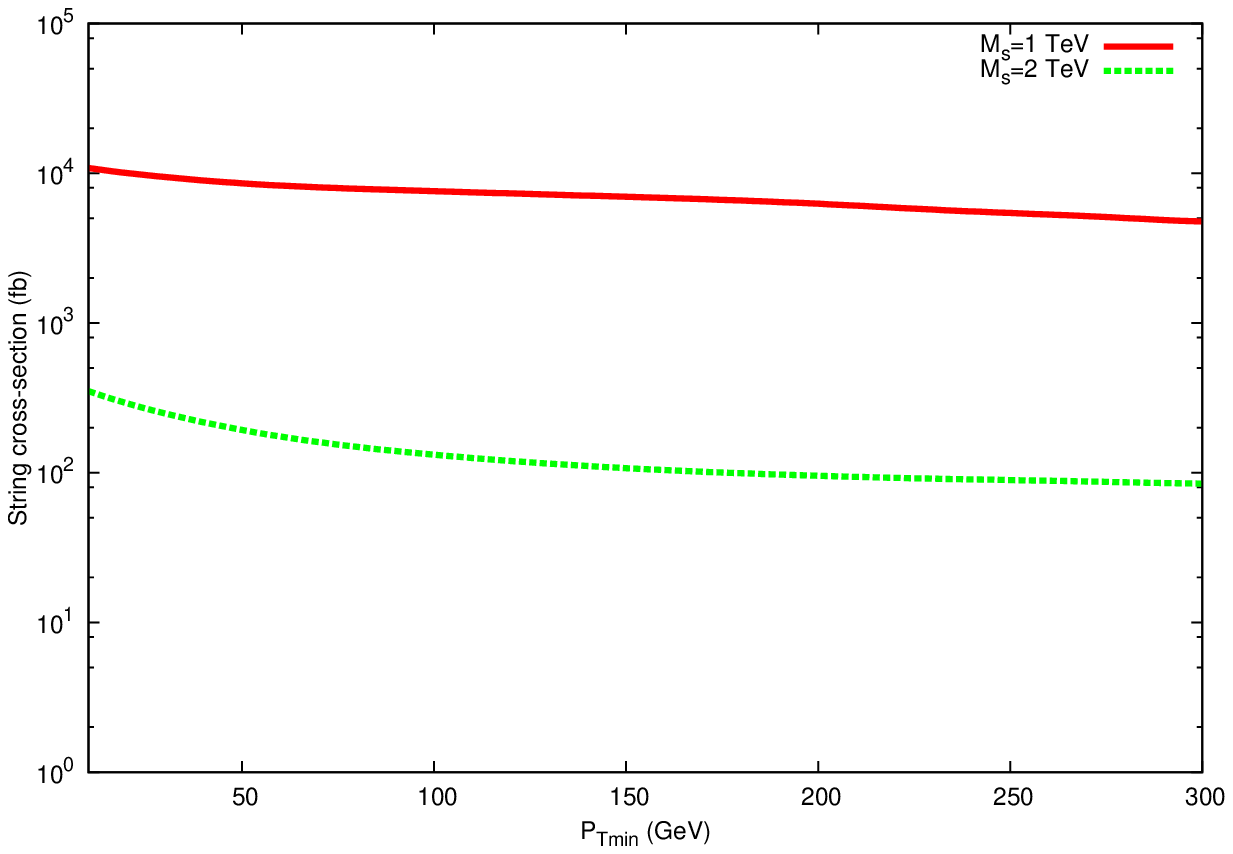}
    \null\hfill
    \includegraphics*[width=0.5\textwidth]{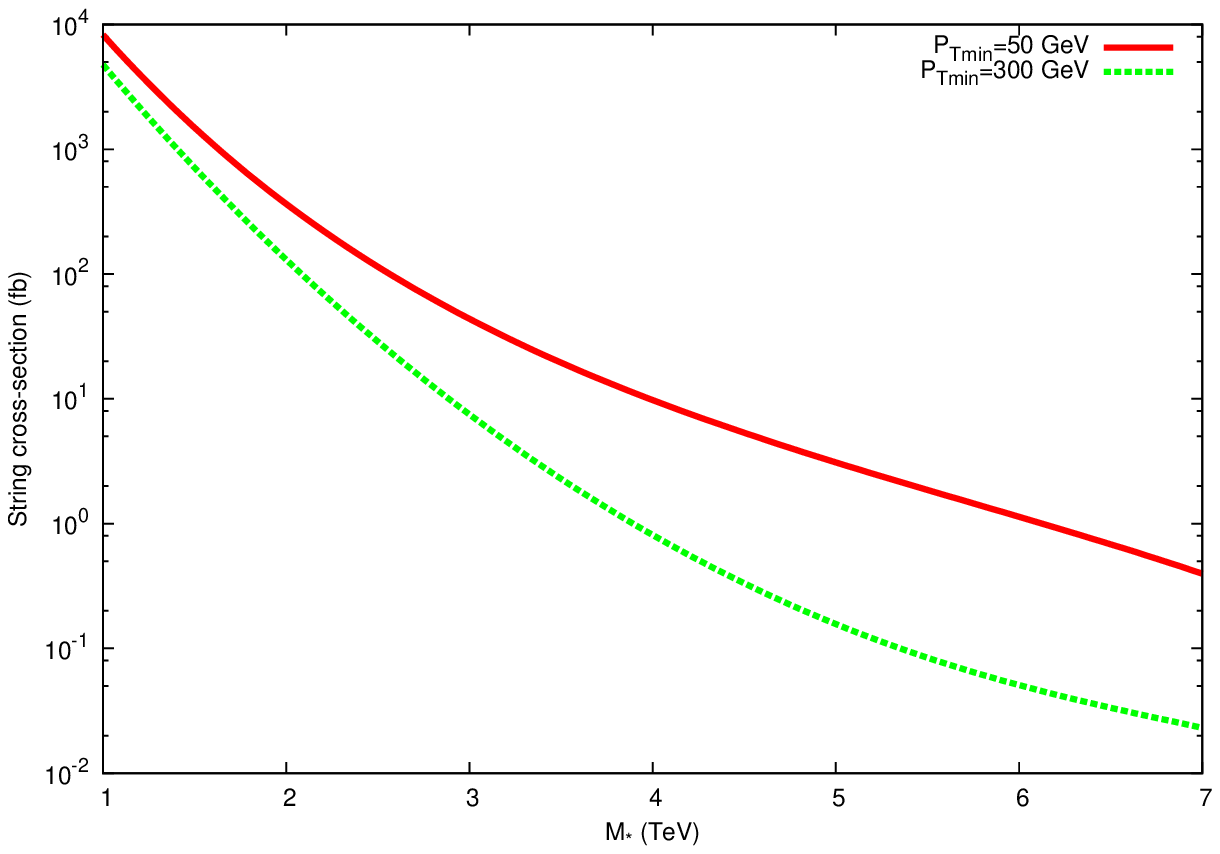}
    \null\hfill}
\caption{Left panel: String cross section for $M_s$= 1 TeV (red dots) and $M_s$= 2 TeV (green crosses). The cross section for $M_s$= 1 TeV is $\sim$ 44 times larger than the cross section for $M_s$= 2 TeV. Right panel: String cross section for the two extreme values of $P_{Tmin}$, 50 GeV (solid red line) and 300 GeV (dashed green line) as a function of $M_{\star}$.}
\label{cross}
\end{figure}
The left panel of Fig.~\ref{cross} shows the total cross section as a function of $P_{Tmin}$.  The string cross section for $M_s$=1 TeV (solid red line) and
the cross section for $M_s$=2 TeV (dashed green line) are $\sim$ 5$\times 10^4$ and $10^3$ times less than SM,
respectively. 

The string cross-sections for $P_{Tmin}$ = 50 GeV (solid red line) and 300 GeV (dashed
green line) are shown in the right panel of Fig.~\ref{cross}, respectively. 
%Fig.~\ref{cross_50_300} shows the string cross section for $P_{Tmin}$ =50 and 300 GeV respectively. 
For $P_{Tmin}$ = 300 GeV, the string cross section is $\sim$ 100 times less than  $P_{Tmin}$ = 50 GeV when $M_{\star}$=7 TeV. The choice $P_{Tmin}$=50 GeV leads to a  signal-to-background ratio of $\sim$ 73.  The string cross section is highly suppressed $w.r.t.$ the SM cross section for lower values of $P_{Tmin}$; for example $\frac{\sigma_{string}}{\sigma_{SM}}\sim10^{-5}$ for $P_{Tmin}$~=~10 GeV. Thus, the discrimination of string events from the SM background is more difficult for events with lower $P_{Tmin}$. Both the SM background and the signal are substantially reduced for higher values of $P_{Tmin}$. For example, at 300 GeV they are reduced by a factor of $\sim$ 98\% and $\sim$ 42\% $w.r.t.$ values at $P_{Tmin}$~=~50 GeV, respectively. The optimal signal-to-background ratio is obtained for $P_{Tmin}\lesssim$ 100 GeV.

\chapter[Physics of Particle Collisions]{Physics of Particle Collisions\label{partcoll}}
In this chapter we briefly discuss the main features of CERN's LHC and introduce kinematical and dynamical quantities which will be used in our analysis.
\section{The Large Hadron Collider\label{LHC}}
\begin{figure*}[ht]
\centerline{\null\hfill
    \includegraphics*[width=0.7\textwidth]{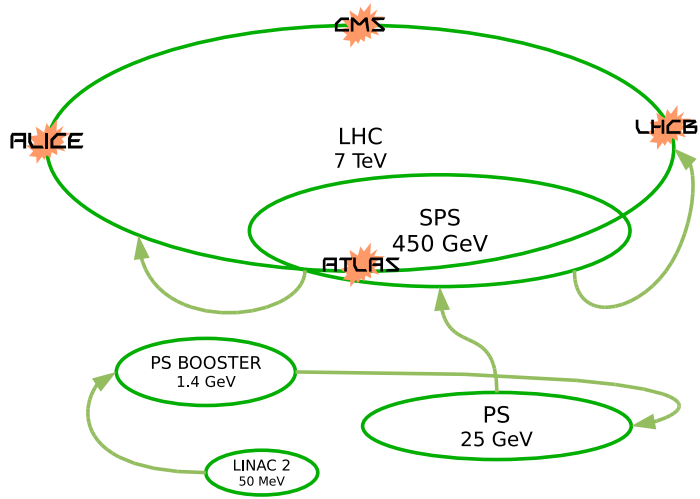}
    \hfill}
\caption{Illustration of the LHC ring structure showing four of the six experiments. The energy of the protons is increased in succession by the use of various smaller accelerators. Protons, first injected into LINAC, are emitted with an energy of 50 GeV. Next, the PS booster accelerates the protons to 1.4 GeV before it enters the Proton Synchrotron (PS). 25 GeV protons from PS is fed to the SPS (Super Proton Synchrotron), accelerating them more than 10 times their present energy to 450 GeV. In the final phase, the protons enter the LHC where they reach their final energy of 7 TeV per beam.  Adapted from \cite{lhcguide:2008}.} 
\label{lhcring}
\end{figure*}
\begin{figure*}
\begin{center}$
\begin{array}{cc}
    \includegraphics*[width=1\textwidth]{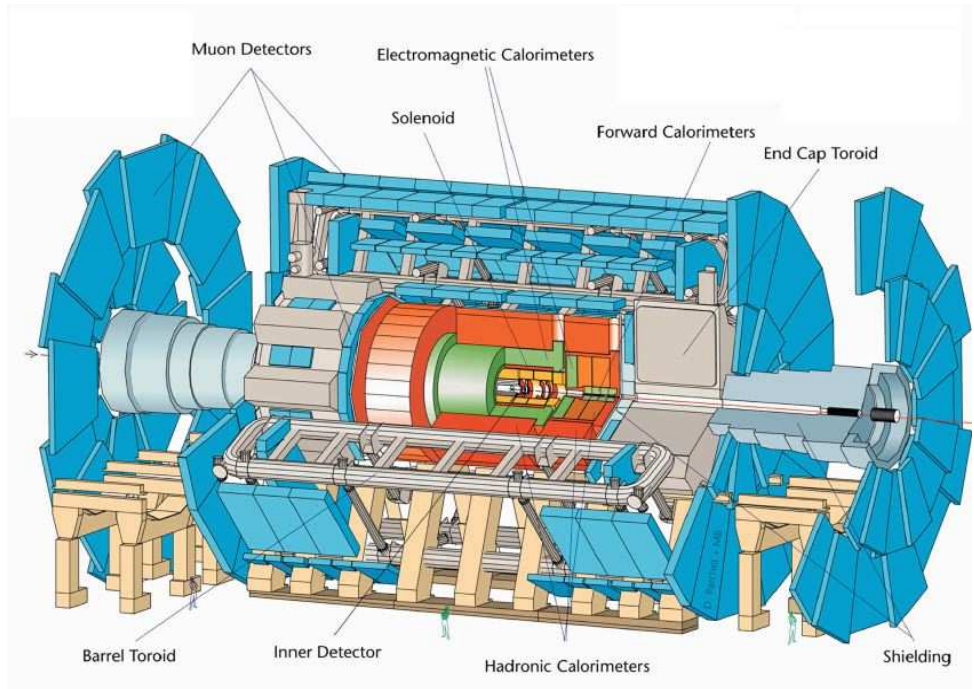}\\
    \includegraphics*[width=1\textwidth]{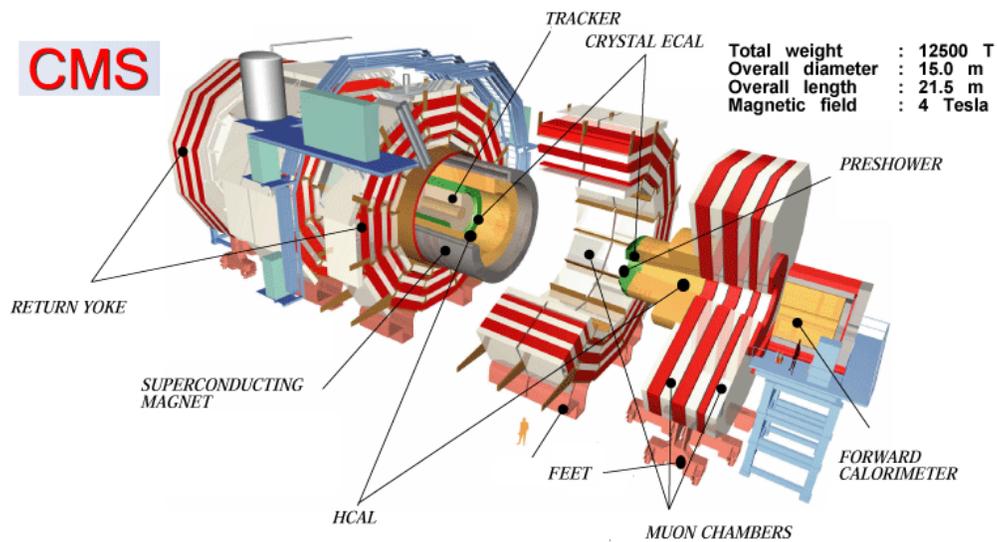}
    \end{array}$
    \end{center}
\caption{Cross section of the ATLAS and CMS  \cite{CERN:web} detectors at CERN.} 
\label{atlas_cms}
\end{figure*}

The LHC is a circular proton-proton collider with a circumference of 27 km. The LHC is host to various  experiments: A Large Ion Collider Experiment (ALICE),  A Toroidal LHC ApparatuS (ATLAS) \cite{Armstrong1:1994}, Compact Muon Solenoid (CMS)  \cite{Armstrong2:1994}, the Large Hadron Collider beauty (LHCb) experiment, the Large Hadron Collider forward (LHCf) experiment and the TOTal Elastic and diffractive cross section Measurement (TOTEM) experiment. The ATLAS and CMS (see Fig.~\ref{atlas_cms}) are general-purpose high-luminosity detectors designed to cover a wide range of physics from the search for the Higgs \cite{Higgs:1964ia,Englert:1964et,Guralnik:1964eu,Higgs:1966ev,Djouadi:2005gi,ATLAS:1999,Drozdetsky:2007zza,Kcira:2007ty} and  SUSY \cite{Wess:1973kz,Weinberg:2000cr,Baer:1995nq,Baer:1995va,Paige:1997xb} to EDs \cite{Arkani-Hamed:1998rs,Antoniadis:1998ig,Arkani-Hamed:1998nn,Randall:1999ee,Randall:1999vf, Appelquist:2000nn,Cembranos:2006gt}. The main feature of the ATLAS detector is its magnet system consisting of superconducting magnet coils. The CMS is built with the same goals as ATLAS but it offers an excellent muon tracking system. The LHC is based on the ‘barrel plus endcaps’ design; a cylindrical detector covers the central region and two flat circular ‘endcaps’ cover the angles close to the beam: ALICE and LHCb detectors are exceptions as they have asymmetric shapes. ATLAS  and CMS detectors are entrusted with the task of studying events with large transverse momentum $P_T$, a signature common to SUSY, EDs and string-mediated interactions.

\begin{figure*}[ht]
\centerline{\null\hfill
    \includegraphics*[width=0.9\textwidth]{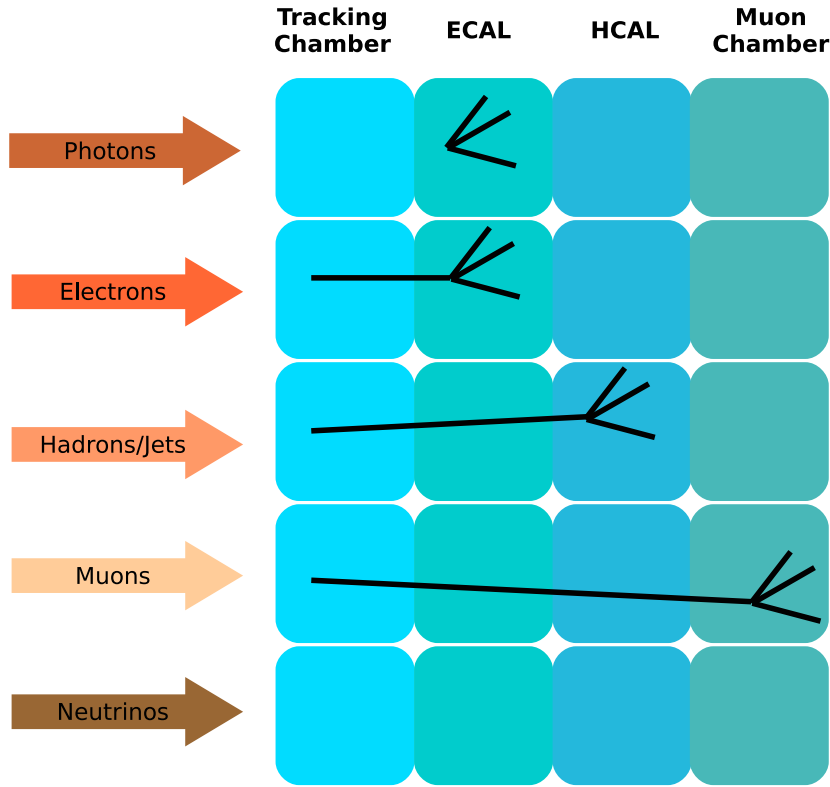}
    \hfill}
\caption{Passage of charged particles through matter in particle colliders. Charged particles leave a track in the tracking chamber, neutral particles shower without leaving a track. Neutrinos are undetected. Adapted from~\cite{lhcguide:2008}.} 
\label{passage}
\end{figure*}

The primary purpose of the LHC detectors is to measure particle positions, charges, velocities, masses and energies. To achieve this goal, the detectors are composed of various sub-detectors. The calorimeter is the most important component as it measures the energy of the particles. It encloses the tracking system that electronically records the path of charged particles as they pass through matter. Calorimeters work on the following principle: they measure the ionization energy deposited in the shower when a particle interacts with a dense medium (e.g. lead). The electromagnetic calorimeter (ECAL) absorbs electrons and photons, the hadronic calorimeter (HCAL) records the energy of hadrons (e.g. pions and jets). Photons and neutrons are detected only by the calorimeter. Muons are identified by the long tracks in the calorimeter and the bulk of their energy is deposited in the muon chamber outside the calorimeter (See Fig.~\ref{passage}.). At the LHC, the total $P_T$ of the colliding particles is assumed to be zero. Neutrinos are weakly interacting particles and cannot be directly detected, but their presence can be inferred due to missing energy.  

The LHC follows a right-handed coordinate system where the beam direction is along the $z$-axis and the vertical axis is $y$. The coordinates can also be expressed in terms of ($r,\eta,\phi$), where $\phi=\tan^{-1} (y/x)$ and the pseudorapidity is $\eta=-\ln[\tan(\theta/2)]$. The true rapidity of a particle defined as atanh$(p_z/E)$, where $p_z$ is the momentum along the $z$ direction and $E$ is the energy of the particle. The rapidity for massless particles is the same as their pseudorapidity.
\section{Luminosity\label{lum}}
The performance of the LHC is described by its instantaneous luminosity $L$, which is a function of the  number of particles in each bunch($n_b$),  the bunch crossing frequency $f$ and the bunch area overlap $A$ \cite{PDG}:
\begin{equation}
L=f\frac{n_{b1} n_{b2}}{A}.
\end{equation}
The proton beam at the LHC consists of $\sim$ 3000 bunches, each containing about $10^{11}$ protons. Particle bunches are squeezed at the interaction points to increase the chances of collision to a cross-sectional area of about 10 sq. mm \cite{lhcguide:2008}. The luminosity can be enhanced by increasing the number of particles in each bunch and by increasing the frequency of collision. The latter can be achieved by properly focusing the beam at the interaction points. 

The instantaneous luminosity has units of $cm^{-2} s^{-1}$. The number of events $N$ for a given process in the time $\Delta t$ is
\begin{equation}
N=\sigma\int_0^{\Delta t} L(z)dz,
\label{lum}
\end{equation}
where $\sigma$ is the total cross section. The integral in Eq.~(\ref{lum}) is known as the integrated luminosity. The design luminosity of the LHC is $10^{34 }cm^{-2} s^{-1}$, to be achieved within three years of running. The LHC initial or startup luminosity is $10^{29 }cm^{-2} s^{-1}$.
\section{Event Shape variables\label{evtshape}}
Event shape variables are commonly used in particle physics to describe the spatial distribution of the outgoing particles. They include sphericity, $2^{\textrm{nd}}$ Fox-Wolfram moment and thrust \cite{Sjostrand:2006za,PYTHIA:Web}. 
%In \texttt{PYTHIA} they are implemented using the routines \texttt{PYSPHE}, \texttt{PYFOWO}, \texttt{PYTHRU} and \texttt{PYJMAS} respectively.
\begin{itemize}
\item Sphericity:
\begin{itemize}
\item[]
\begin{equation}
S=\frac{3}{2}(\lambda_2+\lambda_3),
\end{equation}
where $\lambda_2$ and $\lambda_3$ are the two eigenvalues of the sphericity tensor $S^{\alpha\beta}$
\begin{equation}
S^{\alpha\beta}=\frac{\Sigma_i~\textbf{p}_i^\alpha\cdot\textbf{p}_i^\beta}{\Sigma_i~|\textbf{p}_i^2|}.
\end{equation}
Here $\alpha,\beta=1,2,3$ and $\textbf{p}_i$'s are the final state momenta of particles. An isotropic event is characterized by
values of $S\sim$ 1 whereas $S\sim$ 0 for ``jetty" events.
\end{itemize}
\item $2^{\textrm{nd}}$ Fox-Wolfram moment:
\begin{itemize}
\item[]
\begin{equation}
H_2=\Sigma_{ij}\frac{|\textbf{p}_i||\textbf{p}_j|}{E_{vis}^2}~P_2(\cos\theta_{ij}),
\end{equation}
where $\cos\theta_{ij}$ is the angle between hadrons $i$ and $j$, $E_{vis}$ is the total visible
energy of the event and $P_2(x)$ is the Legendre polynomial of order 2. Values of $H_2$ closer to zero indicate spherical
events whereas $H_2\sim$ 1 indicates that the event is dominated by jets.
\end{itemize}
\item Thrust:
\begin{itemize}
\item[]
\begin{equation}
T=max_{|n|=1}\frac{\Sigma_i|\textbf{n}\cdot\textbf{p}_i|}{\Sigma_i|p_i|},
\end{equation}
where $n$ is the unit vector along the thrust axis. The thrust axis is defined as the axis which maximizes projected momenta. Isotropic events have $T\sim \frac{1}{2}$ and dijets have $T\sim$ 1.
\end{itemize}
\end{itemize}
\section{Invariant Mass and Jet Masses}
Another two useful quantities in high energy physics are the invariant mass  of outgoing particles and jet mass. They are defined as
\begin{itemize}
\item Invariant Mass:
\begin{itemize}
\item[]
For a system of $m$ particles, the invariant mass is defined as
\begin{equation}
M_m\equiv(\sum_m E_m)^{2}-(\sum_m \textbf{p}_m)^{2}\,.
\label{M121}
\end{equation}
where $E (\textbf{p})$ is the energy (momenta) of each particle. In a $2\times 2$ collision the invariant mass of the outgoing particles is
\begin{eqnarray}
M_{12}&=&(E_1+E_2)^{2}-(\textbf{p}_1+\textbf{p}_2)^{2},\\
&=&M_1^2+M_2^2+2(E_1~E_2-\textbf{p}_1\cdot\textbf{p}_2).
\label{invmasseq}
\end{eqnarray}
in natural units. Neglecting the masses of the outgoing particles, as in often the case in high energy collisions, the invariant mass of the system is
\begin{equation}
M_{12}=\sqrt{(E_1+E_2)^2-(\textbf{p}_1 +\textbf{p}_2)^2}=
\sqrt{2p_1p_2(1-\cos\theta)}\,,
\label{M12}
\end{equation}
where $\theta$ is the angle between the two outgoing particles. The invariant mass is an important quantity in high energy physics because it predicts the mass of the particle formed in resonance.
\end{itemize}
\item Jet Masses:
\begin{itemize}
\item[]
The particles of an event are randomly assigned into two groups and their squared invariant mass, $M_1$ and $M_2$, is calculated. The heavy and light jet masses are defined as the larger and smaller of $M_1$ and $M_2$ that minimize $M_1^2+M_2^2$, respectively.
\end{itemize}
\end{itemize}

\chapter[Simulation Techniques]{Simulation Techniques \label{simtech}}
%\section{Event Simulations\label{evt_sim}}
%
This chapter briefly introduces the simulation programs which will be used for our analysis: \texttt{ISAJET}~(ver.\ 7.75) \cite{Paige:2003mg}, \texttt{PYTHIA}~(ver.\ 6.406) \cite{Sjostrand:2006za,PYTHIA:Web}, \texttt{CATFISH} \cite{Cavaglia:2006uk,Cavaglia:2007ir} and the string event generator. SUSY simulations will be carried out using a combination of \texttt{ISAJET} and \texttt{PYTHIA}. Simulations of graviton processes will use \texttt{PYTHIA}. BH simulations will be carried out using the \texttt{CATFISH} Monte Carlo (MC) generator. SR and SM simulations will be done with \texttt{PYTHIA}. The setup for each simulation is:
\begin{itemize}
\item SUSY:
\begin{itemize}
\item The MSSM mass spectrum is generated with \texttt{ISAJET};
\item The mass spectrum in SLHA format is fed into \texttt{PYTHIA};
\item SUSY processes except SM Higgs production are simulated;
\item Unstable SM particles and sparticles are hadronized or decayed with \texttt{PYTHIA}.
\end{itemize}
\item Graviton decay:
\begin{itemize}
\item $f\bar{f}\rightarrow G$ and $gg\rightarrow G$ processes are simulated with \texttt{PYTHIA} as internal processes.
\end{itemize}
\item BHs:
\begin{itemize}
\item The cross section of the BH event is calculated in the CM frame;
\item The initial BH mass is sampled from the differential cross section;
\item The BH is decayed through Hawking mechanism and final $n$-body event (or remnant);
\item Unstable quanta are hadronized with \texttt{PYTHIA}.
\end{itemize} 
\item String resonances:
\begin{itemize}
\item The differential cross section for the $pp\rightarrow\gamma+jet$  process~\cite{Anchordoqui:2008ac} is calculated;
\item The total cross section is obtained from Eq.~(\ref{crosssec1});
\item The resonance decay is simulated with \texttt{PYTHIA}.
\end{itemize}
\item SM background:
\begin{itemize}
\item $t\bar{t}, Z^0, W, WZ^0, WW~\hbox{and}~Z^0Z^0$ SM processes are simulated with \texttt{PYTHIA}.
\end{itemize}
\end{itemize} 
\subsubsection{PYTHIA}
\begin{figure*}[t]
\centerline{\null\hfill
    \includegraphics*[width=1\textwidth]{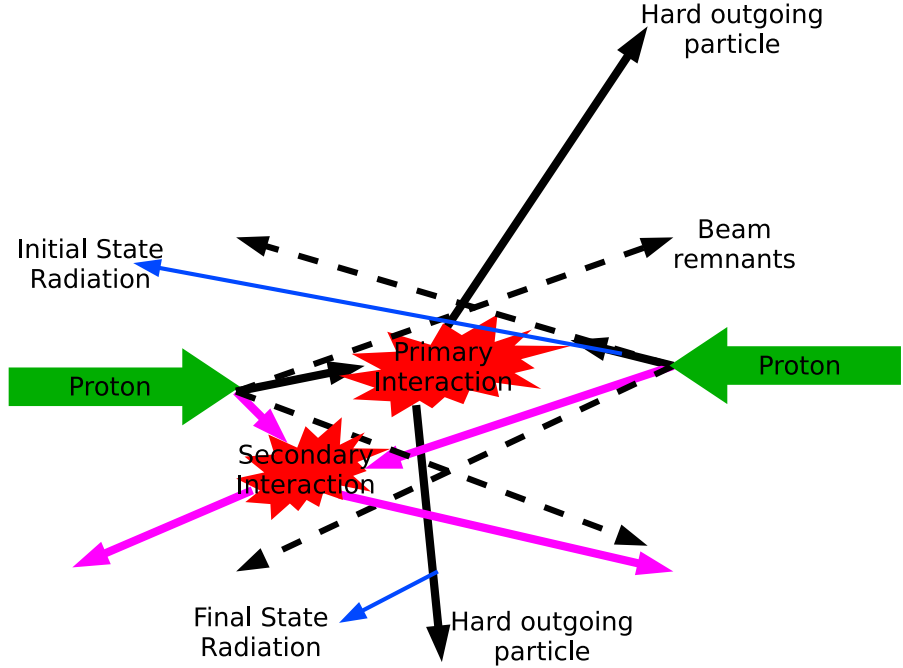}
    \hfill}
\caption{$2\times 2$ interaction showing initial and final state radiation, multiple interactions and beam remnants. Adapted from \cite{PhysRevD.65.092002}.} 
\label{2by2int}
\end{figure*}

\texttt{PYTHIA} is a \texttt{FORTRAN}\footnote{Starting with version 8, \texttt{PYTHIA} is written in \texttt{C++}.} MC code used for high energy particle collisions. Typically, a high energy collision is a complex process producing a plethora of particles. The main purpose of \texttt{PYTHIA} is to generate a detailed event history, for example the $2\times2$ hard-process, initial- and final-state radiation, beam remnants, fragmentation, decays etc., for a generic detector. Events are generated using the following three main steps (see also Fig.~\ref{2by2int}):
\begin{itemize}
\item {\it Hard interaction}. The primary hard interaction is generated for process under study through the appropriate cross section. The hadrons are described by a set of parton distribution functions $f(x,Q^2)$ which give the probability of finding a parton with a certain momentum fraction $x$ of the total energy of the proton at the energy scale $Q^2$.
%denotes the energy scale where the function is evaluated. It therefore follows that
%
%\begin{equation}
%\Sigma_i\int_0^1 xf_i(x,Q^2)dx=1,
%\end{equation}
%
%where $i$ denotes a particular parton.
%
 \texttt{PYTHIA} currently contains about 300 hard processes which are classified according to the number of final particles ($2\rightarrow 1$ and $2\rightarrow 2$ hard interactions) or according to the physical processes. When classifying hard processes based on  their physical properties, \texttt{PYTHIA} currently supports about 17 different scenarios. Some of the processes relevant for this thesis include hard QCD processes, $W/Z$ production, new gauge boson production, SUSY and extra dimensions. 
\item {\it Initial and final state radiation}. Incoming- and outgoing- state radiate before and after the hard interaction through soft or semi-hard processes. For example an initial quark  produces a photon as in $q\rightarrow q\gamma$. 
\item {\it Quark-gluon confinement}. Quarks and gluons produced in the interaction are hadronized to produce stable particles.
\end{itemize}

Each event in \texttt{PYTHIA} is stored as an event record in a common block called \texttt{PYJETS}.  Origin, present status, momentum, energy, mass and production vertices of the particle are recorded.  The \texttt{PYLIST} command provides a summary of the information stored in \texttt{PYJETS} as shown in Table~\ref{pythiaoutput}. Lines 1$\dots$9, show the event history, denoted by the exclamation marks around the particle names. The first two lines of the output show the initial 7 TeV protons along the $z$ direction. Lines 3,4 (5,6) show the initial (final) state radiation particles. Line 7 contains the graviton produced from $gg$ fusion which decays into a $q\bar{q}$ pairs (lines 8 and 9). The particles produced in the final state are shown from line 10 onwards. Particle names in brackets signify that these particles further decay. The last line of the output shows the total energy and momentum of the interaction.
\begin{table}[t]
\caption{Abridged \texttt{PYTHIA} output showing a $pp$ collision to produce a graviton (G*). Orig refers to where the current particle originated from. Other symbols have the usual meaning. All quantities are in GeV.}
\begin{center}
\begin{tabular*}{1\textwidth}%
{@{\extracolsep{\fill}}ccccccccc}                
\hline\hline
\multicolumn{8}{c}{Event listing (summary)}&\\
\hline
 I & particle/jet & orig & $p_x$ & $p_y$  &  $p_z$   &   E   & m&\\
\newline\newline
    1 &!p+! &  0 & 0.000 & 0.000 & 7000.000 & 7000.000  &  0.938&\\
    2 &!p+! &  0  &  0.000  &  0.000 & -7000.000 & 7000.000 &   0.938&\\
\hline
    3 & !g!   &       		    1  &  1.363 &   0.703& 1868.599 &1868.600  &  0.000&\\
    4 &!g!      &    		    2   &-1.025 &   0.267&-2704.946& 2704.946   & 0.000&\\
    5 &!g!     &     		    3  &  2.289 &  -1.080& 1097.417 &1097.419 &   0.000&\\
    6 &!g!     &     		    4  & 16.514 &  13.034& -240.005 & 240.925 &   0.000&\\
    7 &!G*!  &	    0  & 18.80  & 11.95  &857.41& 1338.34 & 1027.38&\\
    8 &!u!          	&	    7 &-443.276 & -232.577&  358.721 & 615.846 &   0.330&\\
    9 &!$\bar{u}$!      & 		    7 & 462.079 & 244.531  &498.691  &722.499 &   0.330&\\
\hline
   10 & (G*) & 	    7   &18.80 &  11.95 & 857.41& 1338.34 &1027.38&\\
   11 &($\bar{d}$)    &		    4    &5.352  & -4.349  &  3.821   & 7.891    &0.330&\\
   12& (g)      	&	            4    &1.490  & -3.330  & -0.438   & 3.675    &0.000&\\
   13 &(g)      &		            4    &0.253  &  0.121   &-0.255    &0.379    &0.000&\\
   14 &(g)     &  		    4    &1.849  & -6.459   &-7.163    &9.820    &0.000&\\    
   \vdots&\multicolumn{8}{c}{}\\
   185 &pi+   &         137 &  -2.183  &  0.833  &  1.493  &  2.776  &  0.140&\\
    \vdots&\multicolumn{8}{c}{}\\
\hline
    \multicolumn{2}{r}{sum:}   &    &    0.00 &    0.00 &    0.00 &14000 &14000&\\
\hline\hline    
\end{tabular*}
\end{center}
\label{pythiaoutput}
\end{table}    
    
\subsubsection{ISAJET}
\texttt{ISAJET} is another \texttt{FORTRAN} MC for simulating high energy $pp$ collisions. It is similar to \texttt{PYTHIA}, but it is primarily used for parton level studies. Events are generated as follows in four steps:
\begin{itemize}
\item Computation of the cross section of the hard scattering process and simulation of the hard scattering.
\item Simulation of initial state and final state radiation.
\item Hadronization of partons.
\item Jet simulation to describe the ``soft" partonic interactions, i.e. multiple interactions, beam remnants etc.
\end{itemize}
\texttt{ISAJET} stores particle information, momentum, origin and decay products, in a common block.  Information about the production vertex in not stored by \texttt{ISAJET}. Contrary to \texttt{PYTHIA}, particles are produced/decayed at the origin. Therefore, it lacks detector response simulation. In this thesis \texttt{ISASUSY}, included with \texttt{ISAJET}, will be used only to generate SUSY particle masses. 
\subsubsection{CATFISH}
\texttt{CATFISH}  \cite{Cavaglia:2006uk,Cavaglia:2007ir} is  \texttt{FORTRAN} MC generator for simulating BH events at the LHC. \texttt{CATFISH} interfaces to the \texttt{PYTHIA} MC fragmentation code. Simulations using \texttt{CATFISH} are controlled by a set of external parameters and switches:
\begin{itemize}
\item {\it Fundamental Planck scale}, $1\leq M_{\star}\leq 14$.
\item {\it Number of large EDs}, $3\leq n\leq 7$. As discussed in Sect.\ref{add}, models with one or two large EDs are excluded. BH production in warped scenarios such as the RS models (see Sect.\ref{rs}) are not considered in \texttt{CATFISH} since most theoretical studies concerning black holes at colliders have been derived for a flat extra-dimensional scenario.
\item {\it Gravitational loss model}. Options include the no graviton loss BD model and the YN and YR trapped surface models.
\item {\it Minimum BH mass at formation}, $M_{min}$ in units of $M_{\star}$.
\item {\it Quantum BH mass threshold at evaporation}, $Q_{min} \sim 1$ TeV. $Q_{min}$ is defined as the mass of the BH at the end of the evaporation phase.
\item {\it Number of quanta at the end of BH decay}, $2 \leq n_p\leq 18$.
\item {\it Momentum transfer model in parton collision}. Two choices are allowed: M or inverse Schwarzschild radius.
\item {\it Dilepton Invariant Mass Parameters}. (See Sect.~\ref{dilep}.)
\end{itemize}

\begin{table}[b!]
\begin{center}
\begin{tabular*}{0.7\textwidth}{@{\extracolsep{\fill}}c|cccc|cc}
\hline\hline
Particle& Color & Flavor & Charge & Spin & dof &\\
\hline
Quarks & 3 & 6 & 2 & 2 & 72 &\\
\hline
Gluons & 8 &   & 1 & 2 & 16    &\\
\hline
Leptons &  & 3 & 2 & 2 & 12    &\\
\hline 
Neutrinos &  & 3 & 2 & 1 & 6    &\\
\hline
$W,Z$ &   &   & 3 & 3 & 9   &\\
\hline
$\gamma$ &   &   & 1 & 2 & 2    &\\
\hline
Higgs &   &   & 1 &   & 1    &\\
\hline
Graviton &   &   & 1 &   & 1    &\\
\hline
\end{tabular*}
\end{center}
\caption{Dof's of the known SM particles, Higgs and the graviton.}
\label{doftable}
\end{table}
\texttt{CATFISH} simulations neglect the energy loss in the balding phase since it is poorly understood. Since emissivities of rotating BHs are not known for all fields, the description of the evaporation phase is also approximated. For simulation purposes, BH generators rely on the Schwarzschild phase. The total decay multiplicity in \texttt{CATFISH} is
\begin{equation}
N=\frac{(n+1)S}{4\pi}\,\frac{\sum_i c_i{\cal R}_i\Gamma_{{\cal R}_i}}
{\sum_j c_j{\cal P}_j\Gamma_{{\cal P}_j}}\,.
\label{multin}
\end{equation}
The decay multiplicities per species $N_i$ are
\begin{equation}
N_i=N\,\frac{c_i{\cal R}_i\Gamma_{{\cal R}_i}}
{\sum_j c_j{\cal R}_j\Gamma_{{\cal R}_j}}\,,
\label{multinspecies}
\end{equation}
where $\Gamma_{{\cal P}_i}$ and $\Gamma_{{\cal R}_i}$ are the relative emissivities of Ref.~\cite{Cardoso:2005mh,Cardoso:2005vb}, $S$ is the initial entropy of the BH, ${\cal P}_s$ and ${\cal R}_s$ are spin-dependent power and emissivity normalization factors, respectively and $c_i$ is the number of dof  of the particle species $i$.

Table~\ref{doftable} shows the values of  $c_i$  for SM particles. The overall dof of a species is the product of the particle's color, flavor, charge and spin \cite{Gingrich:2007fk}. For example, quarks come in 3 color states, 6 flavor states, 2 charge states and two spin states. Therefore, the degree of freedom is $3\times6\times2\times2=72$. The ratio of the number of dofs for quarks:bosons:leptons is 72:29:18 with a $\sim$ 5:1 ratio between hadrons and leptons.

\subsubsection{STRING EVENT GENERATOR}
SR simulations are implemented as an external process in \texttt{PYTHIA} using the \texttt{UPINIT} and \texttt{UPEVNT} subroutines. Simulation parameters include $M_s$, $P_{Tmin}$ and cuts on the emitted particles. 

%\section[SUSY and BH Event Analysis]{SUSY and BH Event Analysis\label{evt_analysis_susbh}}
\chapter[Phenomenology of New Physics at the LHC]{Phenomenology of New Physics at the LHC \label{simanaly}}
In this section we present our analysis of BH, SUSY, graviton and SR events at the LHC. Part of the original work in this chapter is based on Refs.~\cite{Roy:2007fx,Roy:2008we,Roy:2009pc}.
\section{SUSY and BH event analysis \label{susybhevt}}
\begin{figure*}[t!]
\begin{center}$
\begin{array}{cc}
    \includegraphics*[width=0.5\textwidth]{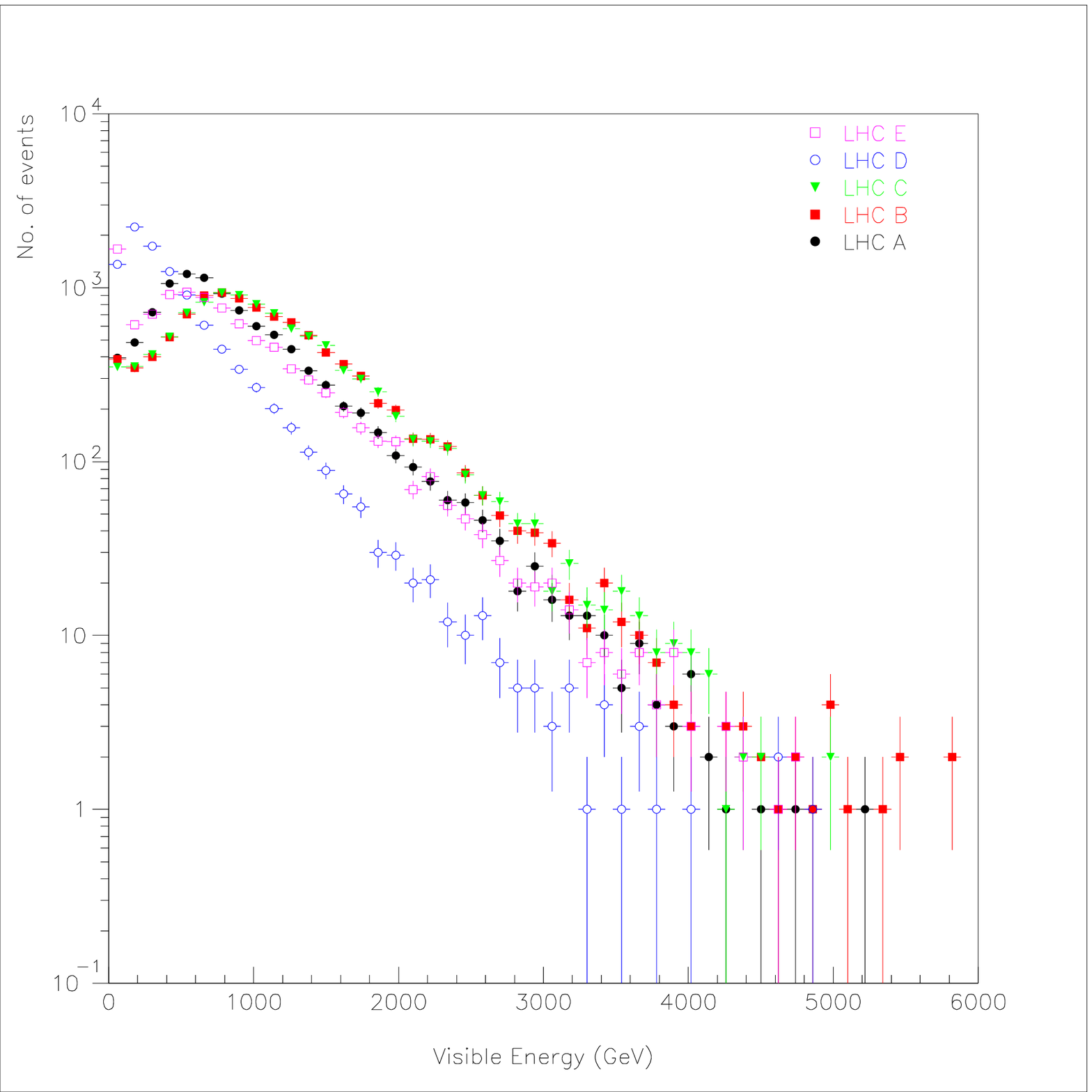}&
    \includegraphics*[width=0.5\textwidth]{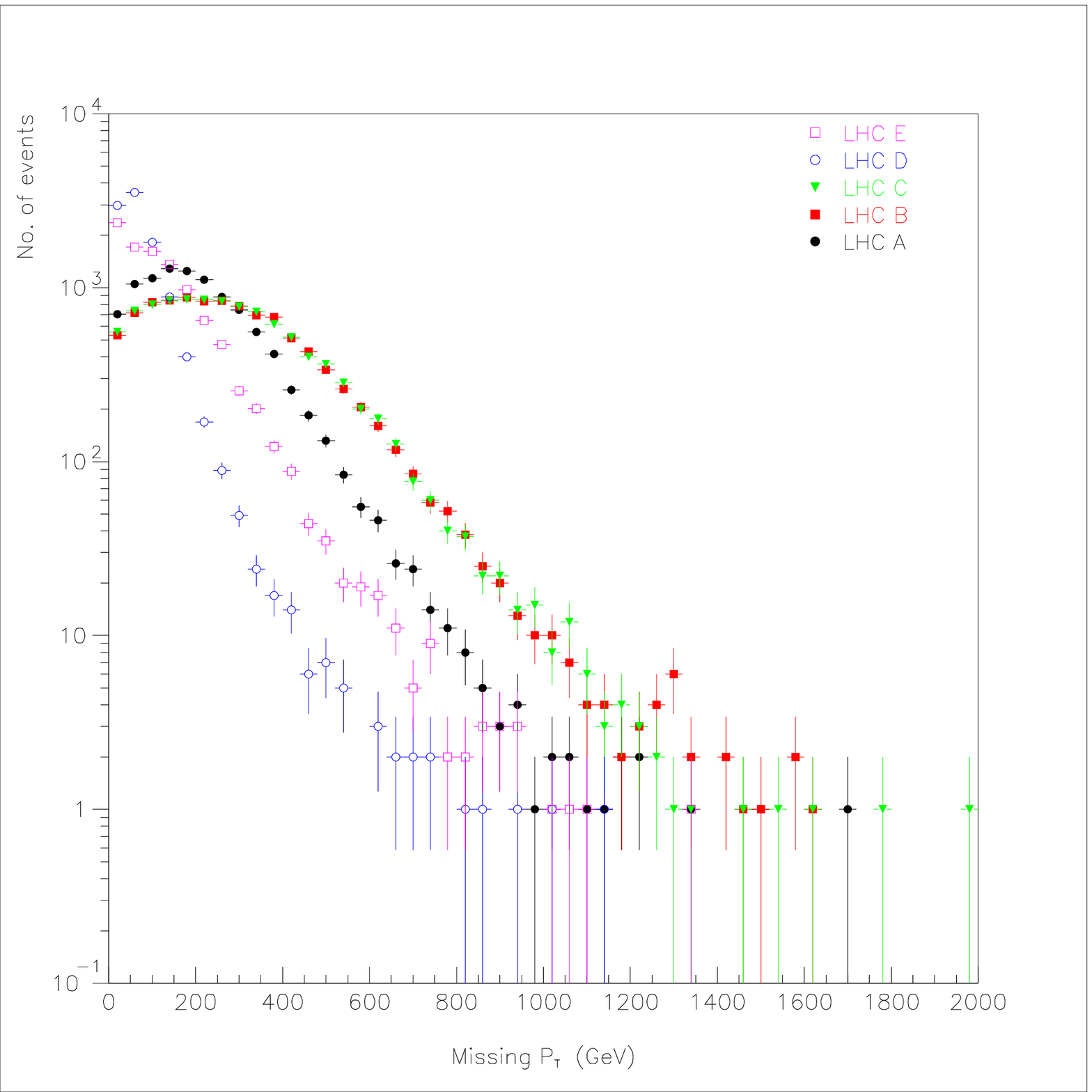}\\
    \multicolumn{2}{c}{\includegraphics*[width=0.5\textwidth]{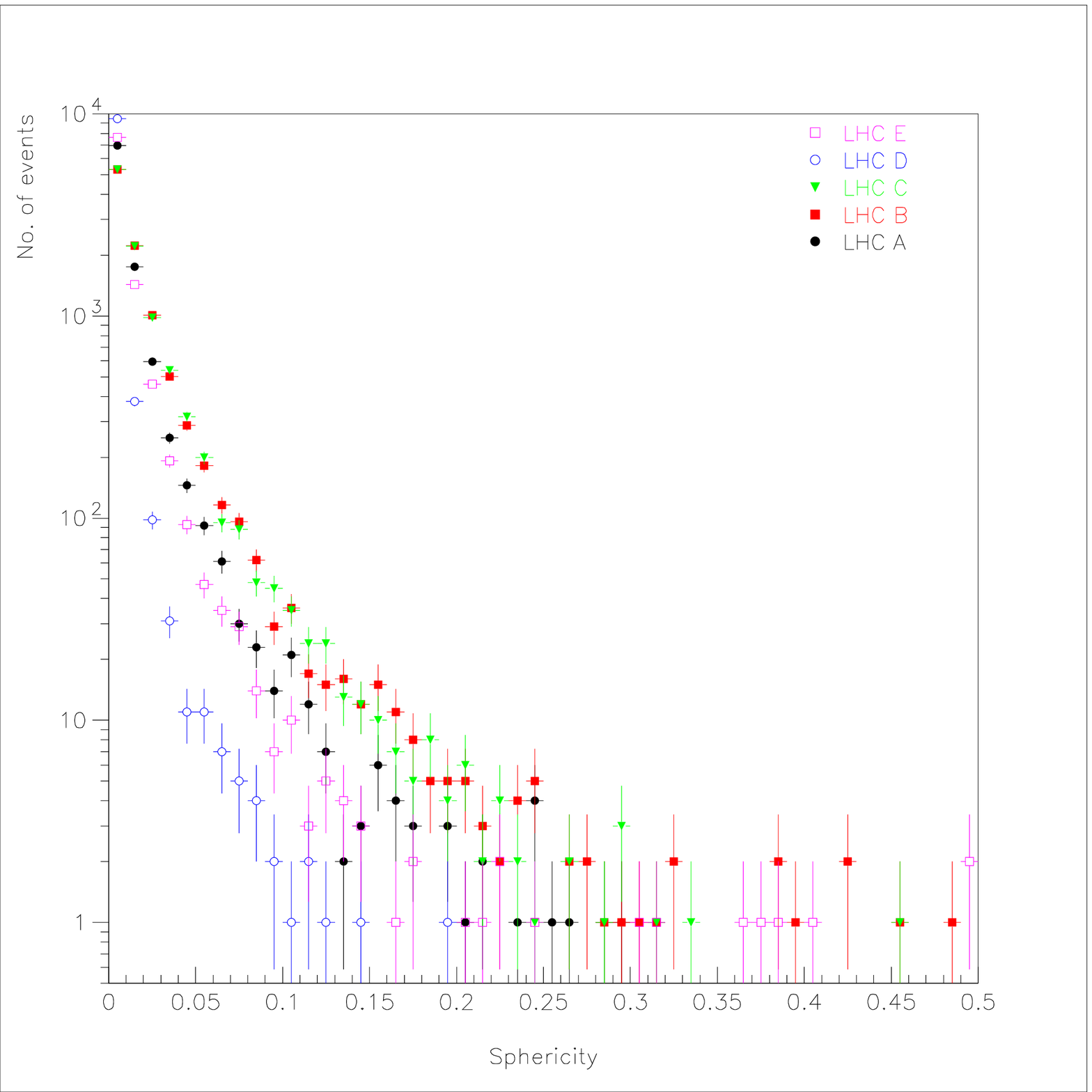}}
    \end{array}$
    \end{center}
\caption{Comparison of visible energy (top left), \missPT\ (top right) and sphericity (bottom) for
10000 events for the five ATLAS points of Table \ref{table1} (A: filled black circles, B: filled red squares, C: filled
green triangles, D: open blue circles and E: open pink squares).} 
\label{fig1_1}
\end{figure*}

Our analysis of SUSY and BH events is based on visible energy, missing momentum of leptons and hadrons, event shape
variables and high-$P_T$ leptons. Fig~\ref{fig1_1} shows visible energy, \missPT\ and sphericity distribution for the five ATLAS points of Table~\ref{table1}, where all SUSY processes
except SM Higgs production have been implemented. For the purposes of the first part of our analysis, the difference between the five points
is not significant and any of them can be chosen as SUSY benchmark. In the following, we will consider point A. This is
justified by the fact that point A allows for SUSY Higgs production \cite{Hinchliffe:1996iu}. Since BHs may evaporate into
Higgs (see Sect.~\ref{bhc} below), a meaningful comparison of SUSY and BH events requires the presence of the Higgs
channel in both models. Moreover, distinguishability of SUSY and BH events must be assessed by minimizing the differences
between the two models. Since BH events are characterized by up to several TeV of \missPT\, SUSY points
with large \missPT\, such as point A, must be considered. 

The parameters for the BH benchmark model are fundamental Planck scale $M_\star=1$ TeV, minimum BH mass $M_{min}=2$ TeV, classical-to-quantum threshold $Q_{min}=1$ TeV, six extra-dimensions ($n=6$) and two-body final decay ($n_p=2$). Particles produced in the initial-radiation phase are removed by imposing $P_T$ cuts of
$5$ GeV and $15$ GeV for leptons and photons+hadrons, respectively \cite{Cavaglia:2006uk,Cavaglia:2007ir}. 

In Sect.~\ref{e_m} we will discuss missing energy and momentum followed by an analysis of event shape variables in Sect.~\ref{shape}.  BH events tend to be more spherical than SUSY events due to the spherical nature of the Hawking
radiation; this is specially evident for high-mass BHs. The formation of a stable BH remnant at the end of the evaporation
phase also helps to discriminate SUSY and BH events because of the large amount of energy which is carried away by the
remnant. 

Analysis of isolated dilepton events is another powerful method to distinguish the two models and will be discussed in Sect.~\ref{dilep}. This is due to the fact that SUSY dileptons typically originate from a single decay chain whereas leptons are rarely emitted by BHs
(the hadron-to-lepton ratio is approximately 5:1) and are uncorrelated; i.e. they can be emitted at any angle w.r.t.\ beam
axis. 
\subsection{Energy and Momentum \label{e_m}}
\begin{figure*}[t]
\centerline{\includegraphics*[width=0.75\textwidth]{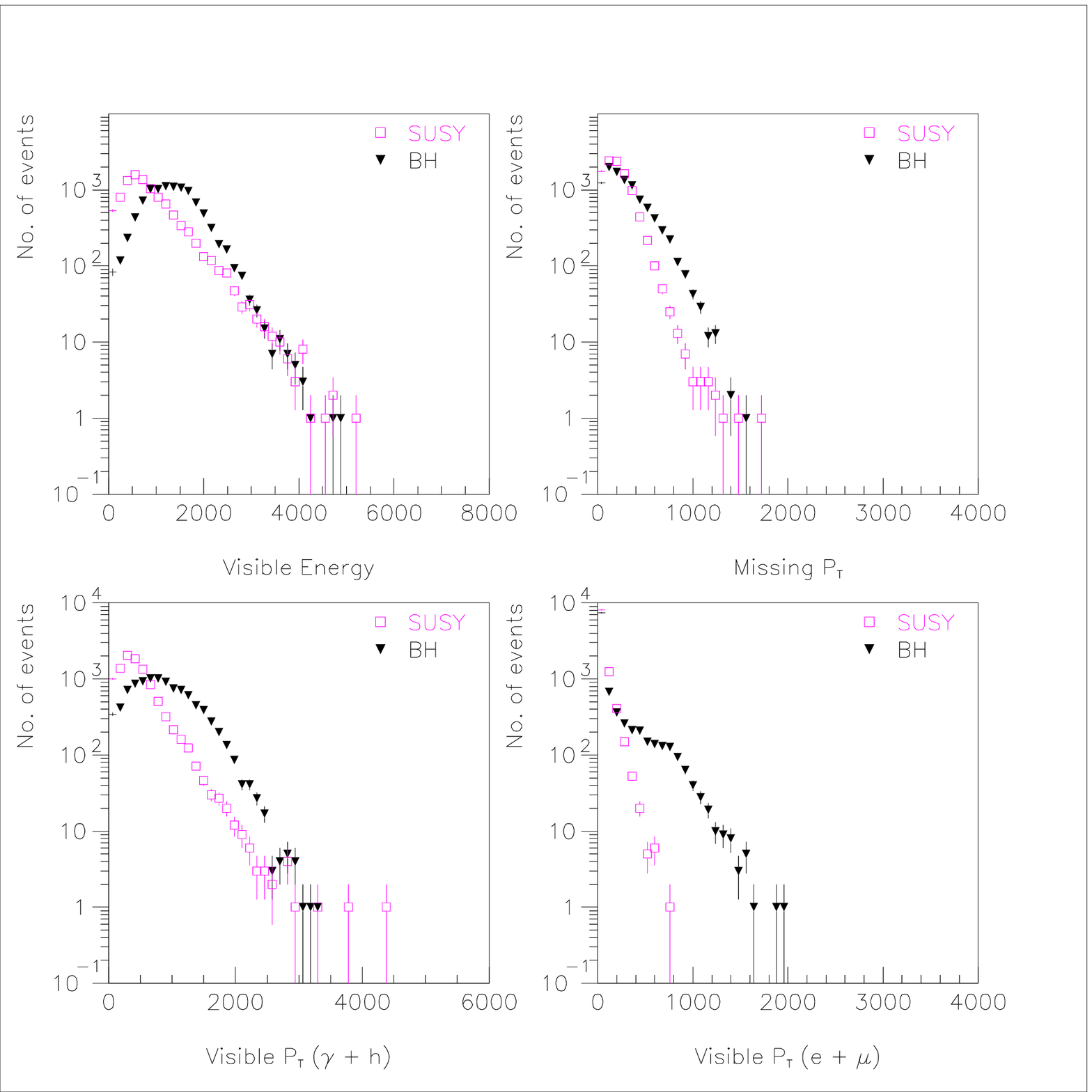}}
\caption{Comparison of 10,000 SUSY and BH benchmark events at the LHC. Visible energy and
\missPT\ (top panels) are comparable due to the presence of
invisible channels in both models. Leptons with large $P_T$ provide instead an
effective discriminator (bottom right panel).}
\label{susybh1}
\end{figure*}

Figure~\ref{susybh1} shows visible energy, \missPT\, and $P_T$ of leptons and
hadrons \& photons for 10,000 SUSY and BH benchmark events. The amount of visible energy and \missPT\ is comparable for
the two scenarios, even in the absence of a BH remnant. This is due to the presence of invisible channels in both
models: neutrinos+gravitons for the BH and LSP for SUSY. The flavor of the decay products is a better discriminator.
SUSY interactions do not produce leptons with energy above the TeV since isolated leptons are produced by the decay of
sparticles with typical energy of less than a few hundred GeV. On the contrary, quanta produced in the BH decay are
characterized by an average energy $E\sim M/N$, where the multiplicity $N$ is less than 10 for typical BHs at the LHC.
Since Hawking evaporation does not distinguish leptons from hadrons, hard leptons with energy up to several TeV are
likely to be produced during the BH decay. This suggests that isolated leptons may provide a powerful means to
discriminate the two models. This is indeed the case, as we shall see in Sect.~\ref{dilep}.
%
%\clearpage

\begin{figure*}[t]
\centerline{\null\hfill
    \includegraphics*[width=0.75\textwidth]{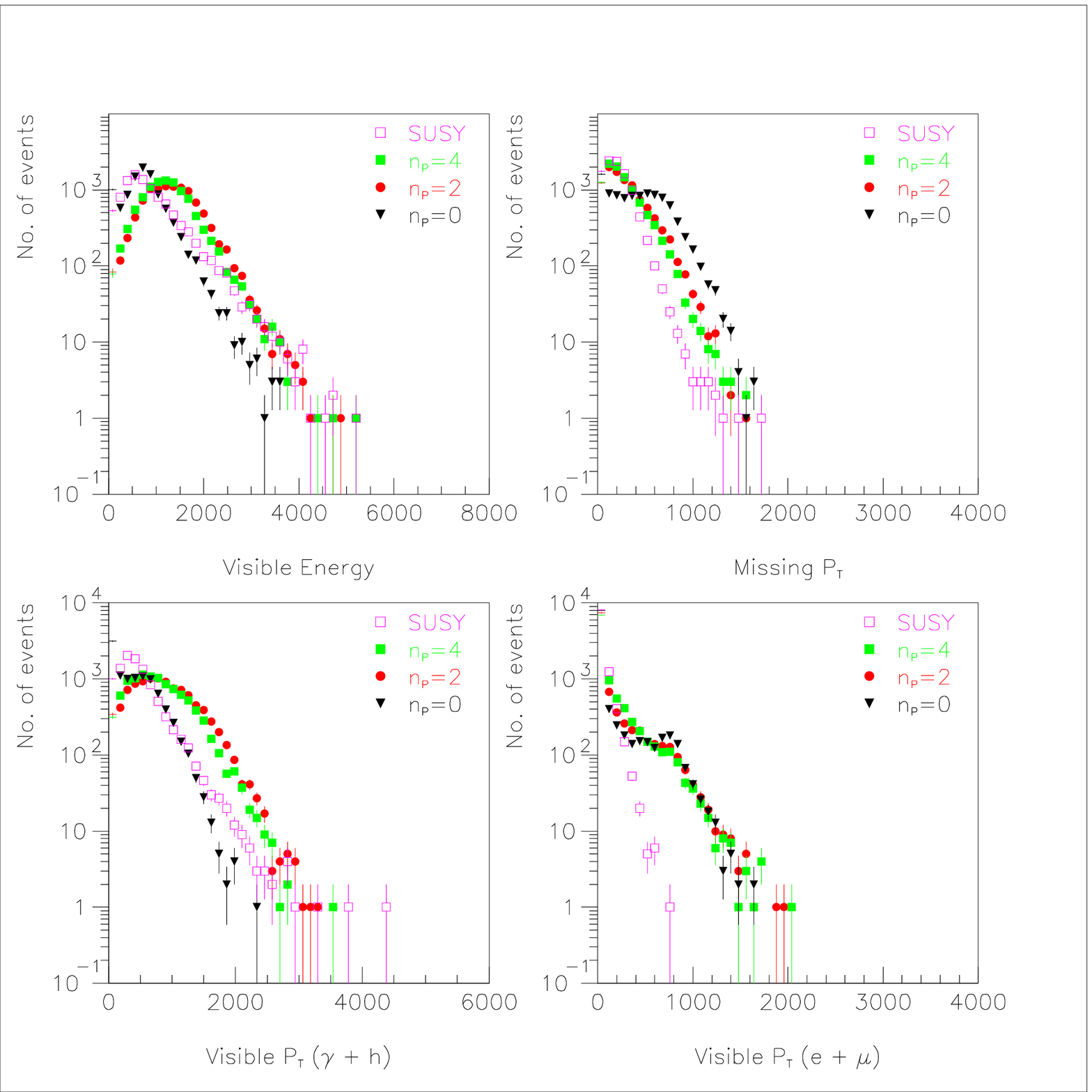}
    \null\hfill}
\caption{Distribution of visible energy, \missPT\ and transverse momenta of leptons and hadrons \& photons. SUSY plots
are shown as pink open squares. The four plots show the effect of different decay modes in the
Planck phase of ten-dimensional BHs: remnant formation ($n_p=0$, black filled triangles), two-body decay ($n_p=2$, red
filled circles) and four-body decay ($n_p=4$, green filled squares). The fundamental Planck scale is $M_\star=1$ TeV.}    
\label{susybh2}
\end{figure*}

Fig.\ \ref{susybh2} shows how variations in the BH Planck phase affect the observables of
Fig.\ \ref{susybh1}. The plots compare zero~(BH remnant)-, two- and four-body decays. The BH is expected to have shed its
electric and color charges by the time the remnant is formed\footnote{For an alternative scenario see, Ref.\
\cite{Koch:2005ks}.}. The BH remnant is thus undetectable and a source of \missPT\ in addition to
neutrinos and gravitons which are emitted during the Hawking evaporation phase. This leads to a larger difference in
\missPT\ between SUSY and BH models. The visible $P_T$ in hadrons \& photons is sensibly reduced in the
presence of a BH remnant. The BH remnant carries away energy which otherwise would have been emitted in visible
channels (mostly hadrons) during the BH decay phase. The leptonic channel is essentially unaffected by the presence of
a BH remnant since leptons are rarer than hadrons in the BH decay phase. Variations in the energy distribution of the
leptonic channel are thus suppressed compared to the hadronic channel. Changes in the number of final Planckian hard
quanta do not produce significant differences in the distributions; more quanta of lower energy behave statistically
like less quanta with higher energy. Provided that the BH decays at the end of the Hawking phase, it is thus safe to
set the number of Planckian quanta to $n_p=2$ (or $n_p=4$), although BHs may decay in different numbers of particles on
a event-to-event basis. Variations in the classical-to-quantum threshold $Q_{min}$ are also not expected to cause
significant differences in the energy/momentum distributions. A higher threshold increases the emission in the Planck
phase while decreasing Hawking radiation. Since these phases differs only in relative greybody factors, the effect is
too small to be detected. 
%\clearpage
%
%\newline

\begin{figure*}[ht]
\centerline{\null\hfill
    \includegraphics*[width=0.75\textwidth]{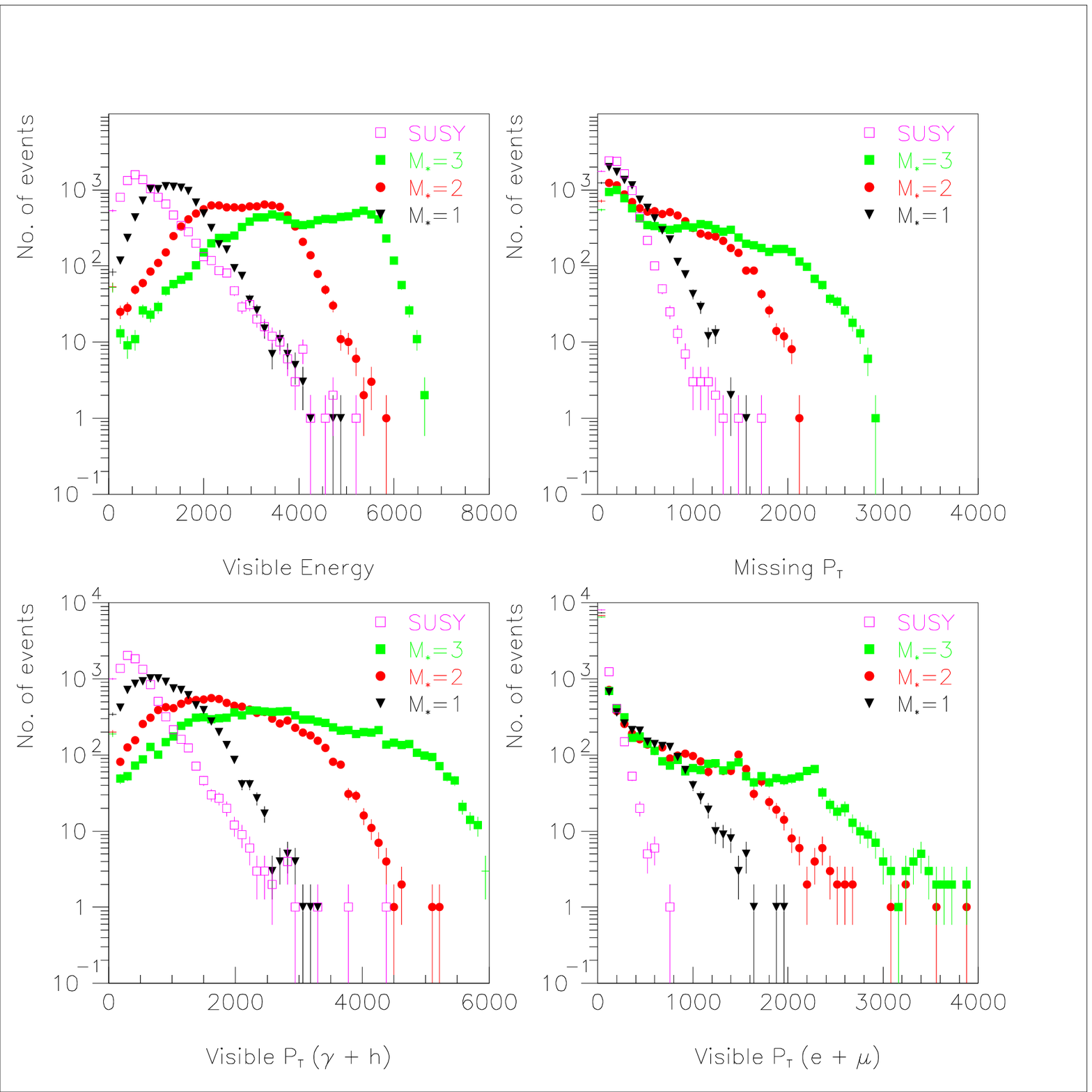}
    \null\hfill}
\caption{Distribution of visible energy, \missPT\ and transverse momenta of leptons and hadrons \& photons. SUSY plots are
shown as pink open squares.  The  plots show the effect of varying the fundamental Planck scale: $M_\star=1$ TeV (black
filled triangles), $M_\star=2$ TeV (red filled circles) and $M_\star=3$ TeV (green filled squares). The ten-dimensional
BHs decay in two hard quanta at the end of the evaporation phase.}    
\label{susybh3}
\end{figure*}

Fig.\ \ref{susybh3} shows visible energy, \missPT, and visible transverse momenta of
leptons and hadrons \& photons for different values of the fundamental Planck scale. Higher values of $M_\star$ lead to
more massive BHs, higher multiplicity and more energetic quanta. This causes a significant increase in
missing and visible momenta. If the value of the fundamental Planck scale happens to be large, BHs are likely to be
detected and easily distinguished from SUSY through detection of highly-energetic isolated leptons and hadronic jets.
\missPT\ of several TeV would also be observed. However, the LHC is expected to produce very light
BHs (see Fig.~\ref{bhmass}) therefore such an analysis may not provide the most effective discriminators \cite{Cavaglia:2006uk,Cavaglia:2007ir, Meade:2007sz,Kanti:2002nr,Kanti:2002ge,Harris:2003eg,Ida:2002ez,Cornell:2005ux}. The study of leptonic final states alleviates this problem. See Sect.~\ref{dilep}.
%
%
%\newpage
\subsection{Event shape variables\label{shape}}
\begin{figure*}[t]
\begin{center}$
\begin{array}{cc}
    \includegraphics*[width=0.46\textwidth,totalheight=0.15\textheight]{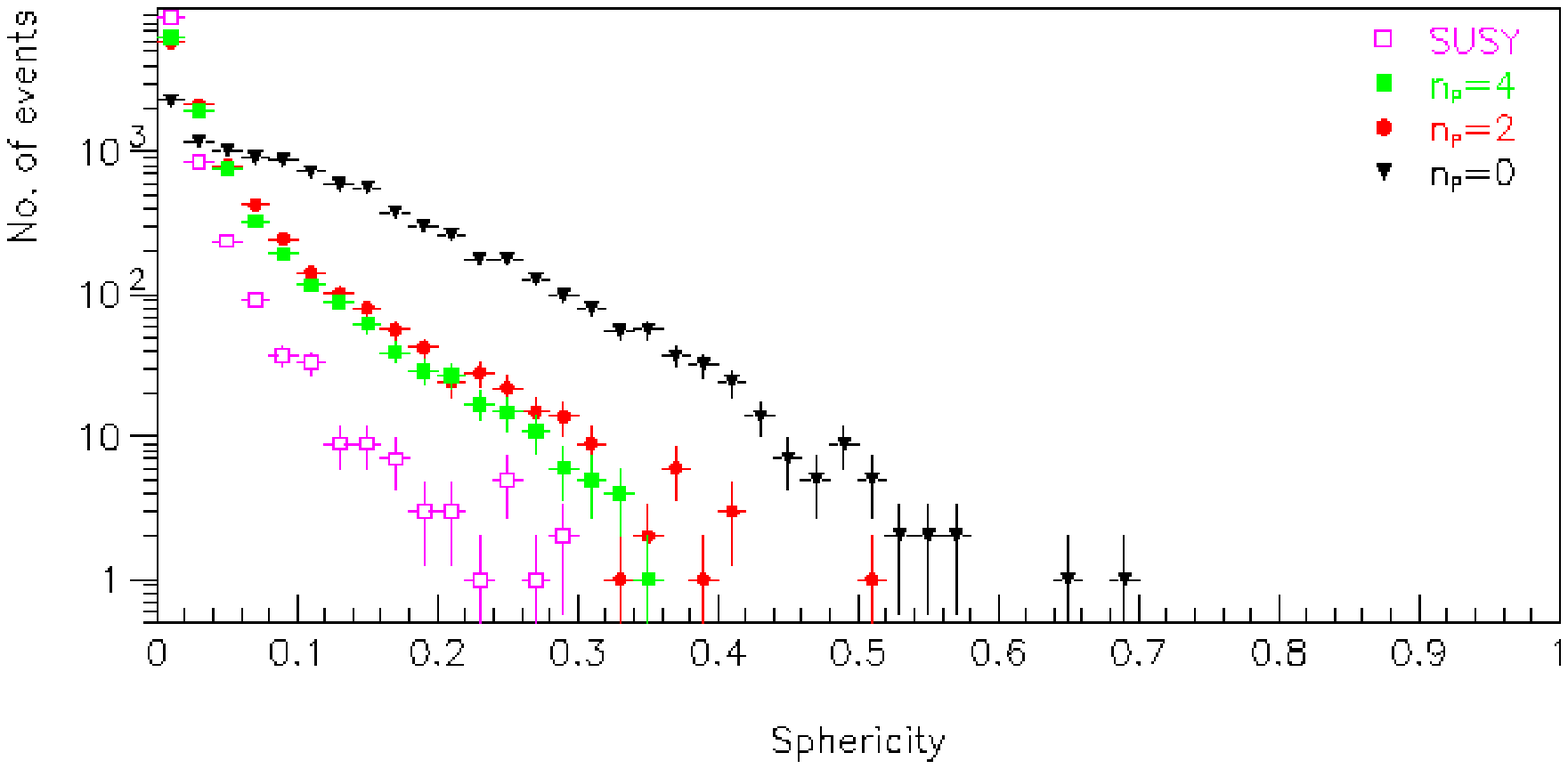}&
    \includegraphics*[width=0.46\textwidth,totalheight=0.15\textheight]{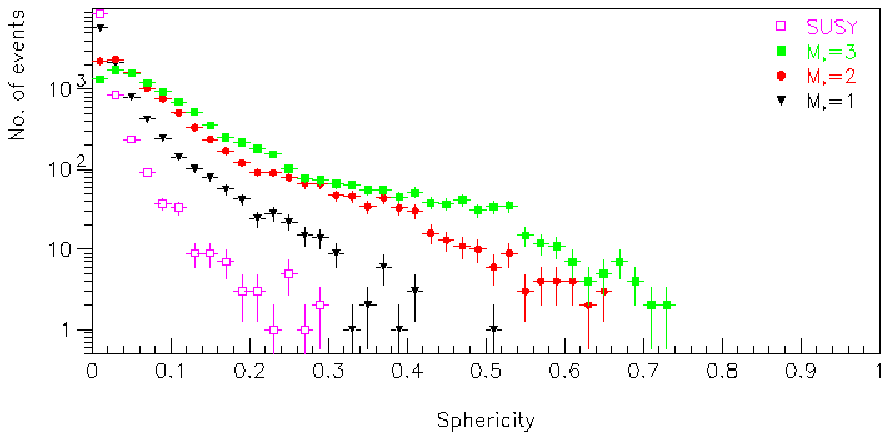}\\
    \includegraphics*[width=0.46\textwidth,totalheight=0.15\textheight]{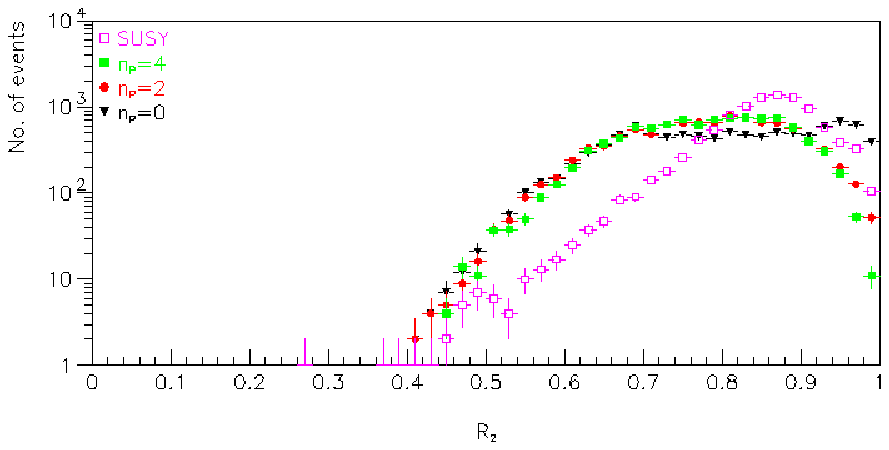}&
    \includegraphics*[width=0.46\textwidth,totalheight=0.15\textheight]{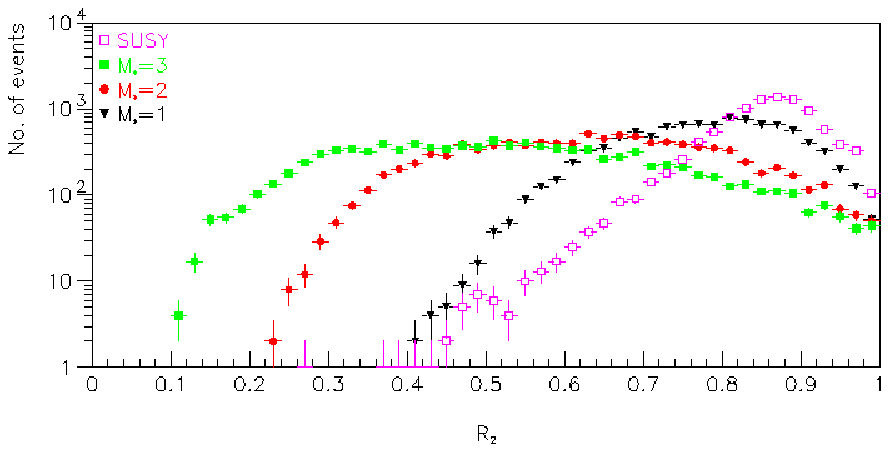}\\
    \includegraphics*[width=0.46\textwidth,totalheight=0.15\textheight]{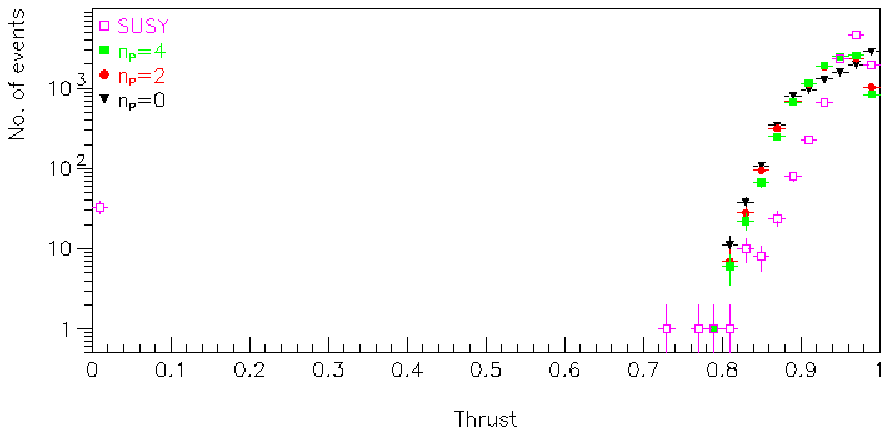}&
    \includegraphics*[width=0.46\textwidth,totalheight=0.15\textheight]{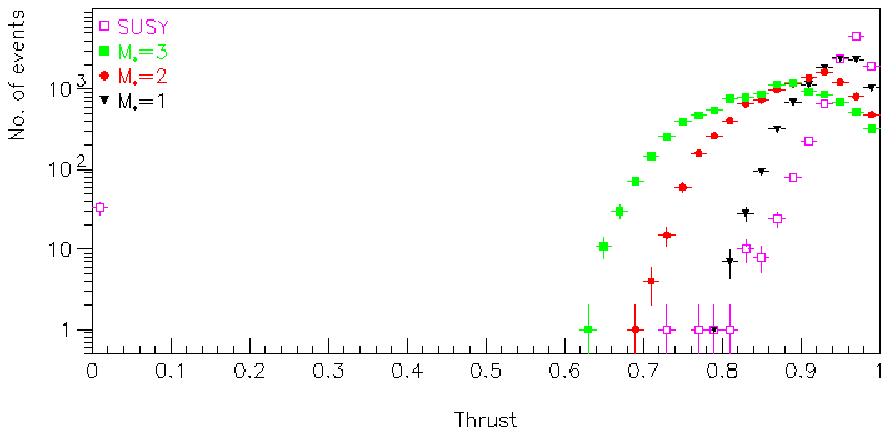}
    \end{array}$
    \end{center}
\caption{Sphericity (top panels), $2^{\rm nd}$ Fox-Wolfram moment (middle panels) and thrust (bottom panels) for 10,000 BH and SUSY events. The left panels show the effect of different Planckian decay modes: BH remnant (black filled triangles), two-body decay (red filled circles) and four-body decay (green filled squares). The fundamental scale is $M_\star=1$ TeV and the number of EDs n=6. The right panels show the effect of different fundamental scales: $M_\star=1$ TeV (black filled triangles), 2 TeV (red filled
circles) and 3 TeV (green filled squares). The ten-dimensional BHs decay in two quanta at the end of the Hawking phase.} 
\label{susybh4}
\end{figure*}

Event shape variables, such as the sphericity, $2^{\rm nd}$ Fox-Wolfram moment and thrust, and jet masses defined in Chapter~\ref{partcoll} can be used to complement the above analysis. Our analysis shows that BH events are more spherical because of the nature of Hawking radiation and the production of larger number of jets for SUSY decays. Formation of a BH remnant and high values of the fundamental scale lead to a significant higher sphericity than SUSY events (top panels of Fig.~\ref{susybh4}). The $2^{\rm nd}$ Fox-Wolfram moment (middle panels of Fig.~\ref{susybh4}) is stable versus changes in the BH Planck phase and provides a good SUSY/BH discriminator. BH models with higher $M_\star$ can be differentiated more easily from SUSY events.

\begin{figure*}[t]
\begin{center}$
\begin{array}{cc}
    \includegraphics*[width=0.46\textwidth,totalheight=0.15\textheight]{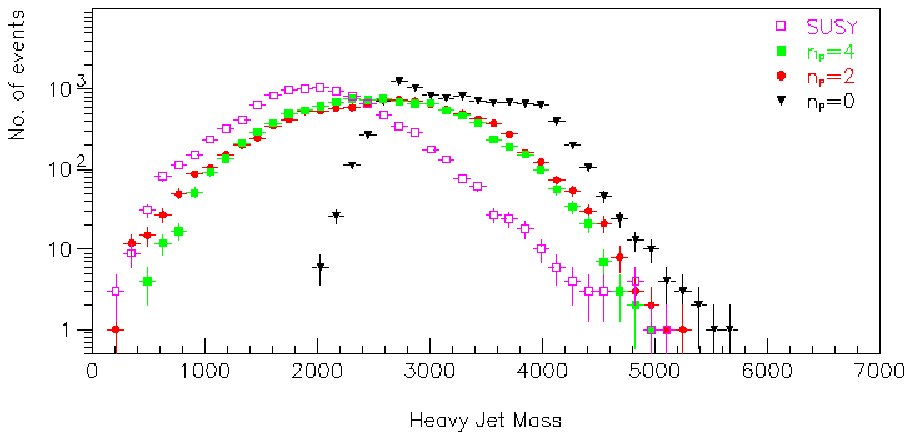}&
    \includegraphics*[width=0.46\textwidth,totalheight=0.15\textheight]{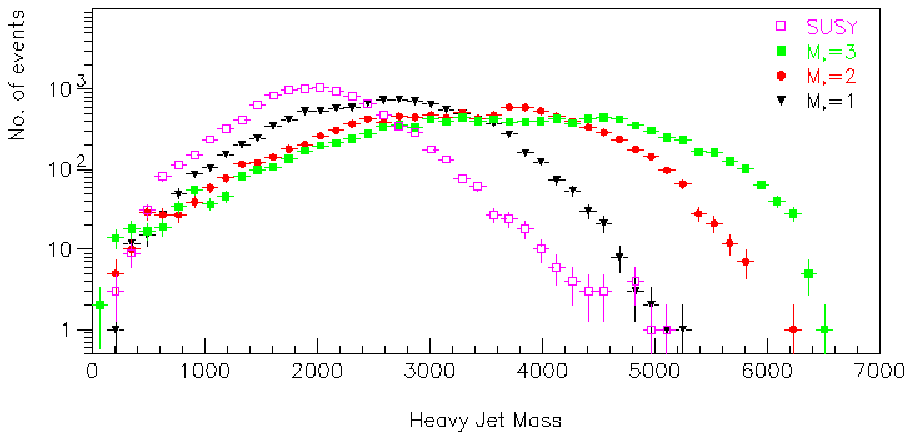}\\
    \includegraphics*[width=0.46\textwidth,totalheight=0.15\textheight]{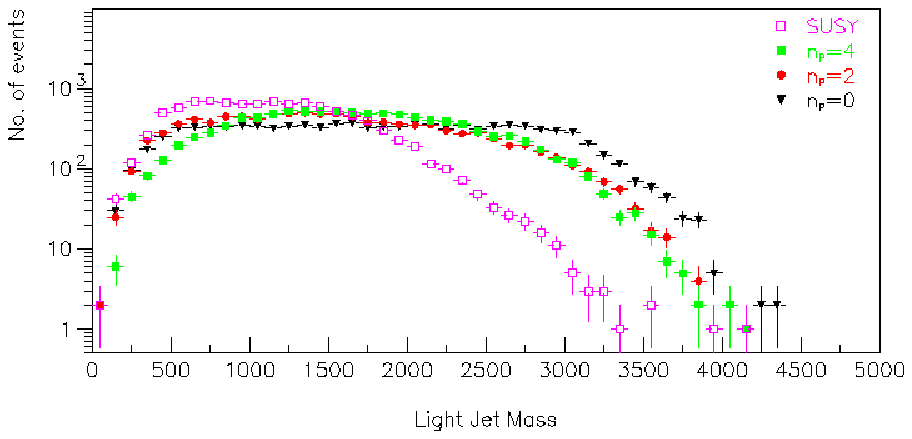}&
    \includegraphics*[width=0.46\textwidth,totalheight=0.15\textheight]{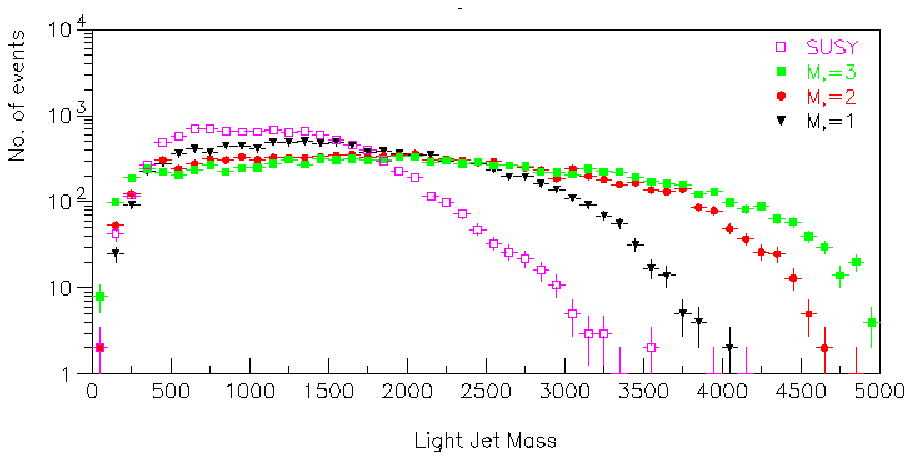}\\
    \includegraphics*[width=0.46\textwidth,totalheight=0.15\textheight]{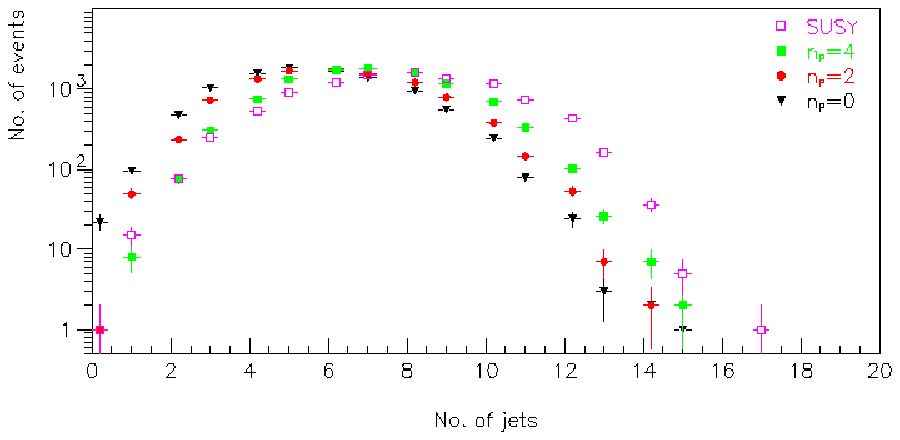}&
    \includegraphics*[width=0.46\textwidth,totalheight=0.15\textheight]{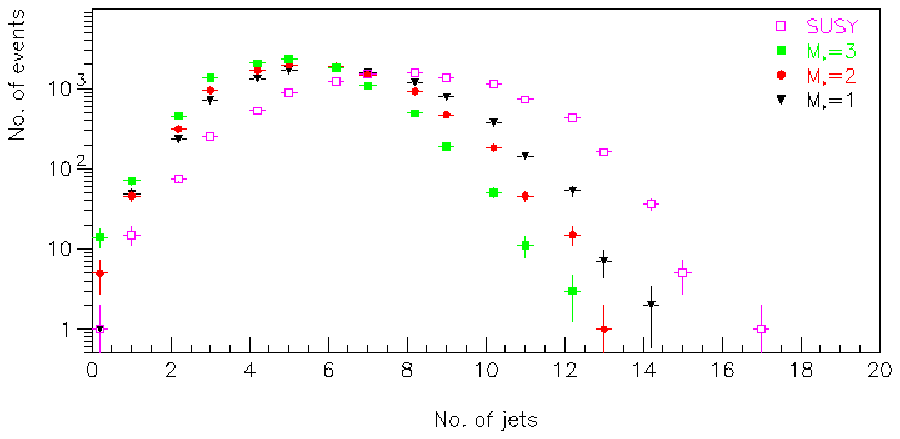}
    \end{array}$
    \end{center}
\caption{Heavy and light jet masses (top and middle panels) and number of jets (bottom panels) for 10,000 BH and MSSM events. Symbols are like in previous figures.} 
\label{susybh5}
\end{figure*}

Similar conclusions can be reached by looking at jet masses and number of jets. SUSY events generate more and lighter
jets than the BH model due to copious production of quarks (Fig.\ \ref{susybh5}). The difference is again especially
significant for high values of $M_\star$ and in the presence of BH remnants. Absence of sub-$Q_{min}$ hard jets could
provide strong evidence for BH remnant production. (See the suppression of heavy jets below the classical-to-quantum
threshold $Q_{min}=2$ TeV in the top leftmost panel of Fig.~\ref{susybh5}.)
%\clearpage
%
\subsection{High-$P_T$ leptons \label{dilep}}
SUSY events can be categorized by the number of high $P_T$ leptons. Events with no isolated
leptons can occur when a SUSY decay chain ends in the production of jets. Events with a single isolated high $P_T$ lepton
are difficult to identify because of the large SM background from $W$ decays into leptons. Events with two isolated high
$P_T$ leptons are characterized by a lower signal-to-background ratio (due to the low branching ratio into leptons) with respect to zero lepton or one lepton events. Nevertheless, dileptons provide a clear signature because the presence of two isolated leptons can be accurately measured with early LHC data. 

Production of gluinos and squarks dominate SUSY production at the LHC. The bulk of the decay modes of these sparticles are characterized by the production of the lighter chargino and two neutralinos. Thus, the decay of the second lightest neutralino $\tilde{\chi}_2^0$ forms an important part of this analysis. Charginos decays are more difficult to analyse as they involve either a missing neutrino or quark jets. Two body decays of neutralinos have large branching ratios. If two body decays are not kinematically allowed, then the three body leptonic mode dominates \cite{Armstrong2:1994}. A list of dominant SUSY interactions leading to dileptons at the LHC is described below for the LHC points of Table~\ref{table1}. 
\begin{figure*}[t]
\begin{center}$
\begin{array}{cc}
   \includegraphics*[viewport=0 200 575 725,width=0.5\textwidth]{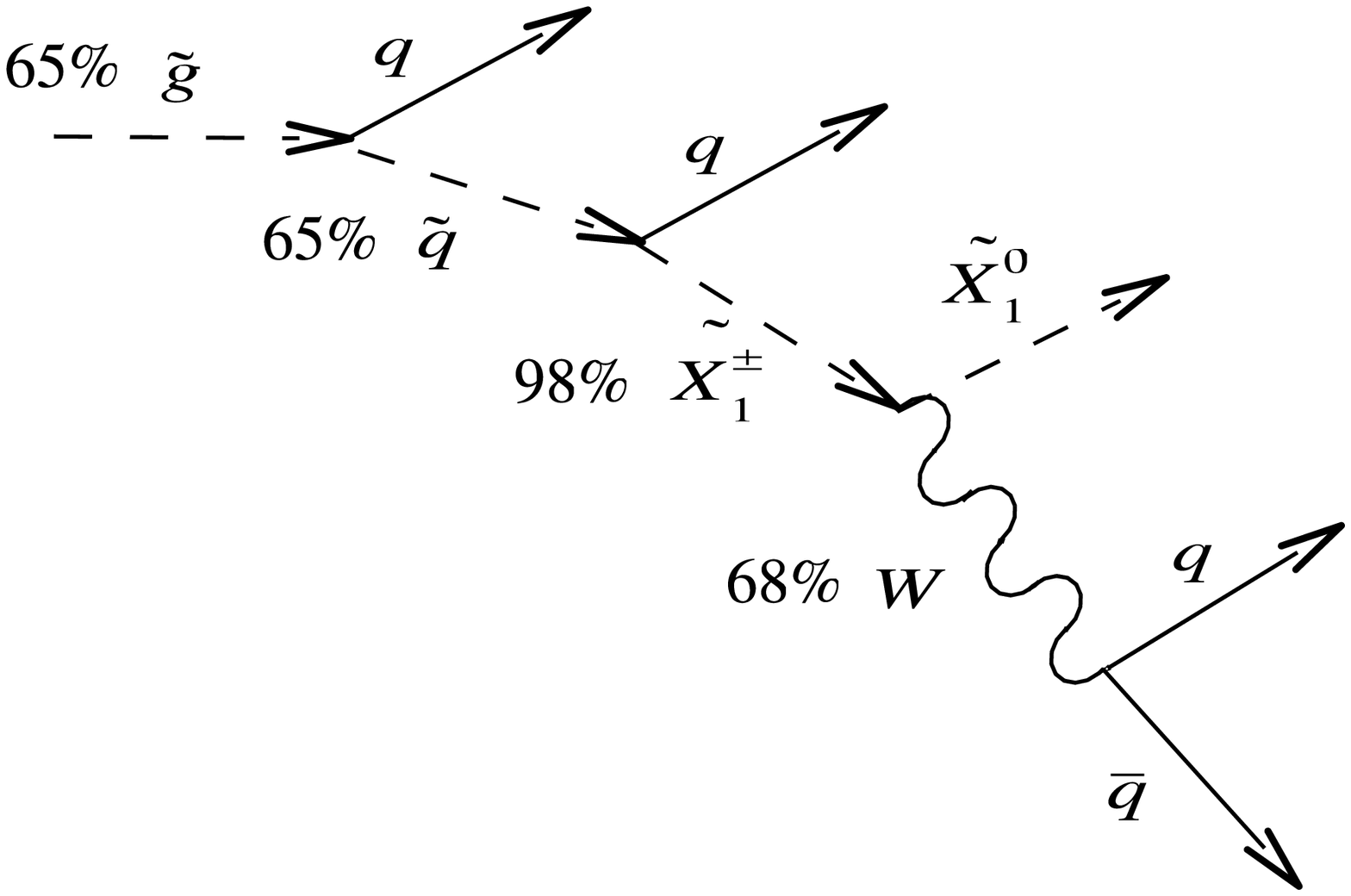}&
    \includegraphics*[viewport=0 200 575 725,width=0.5\textwidth]{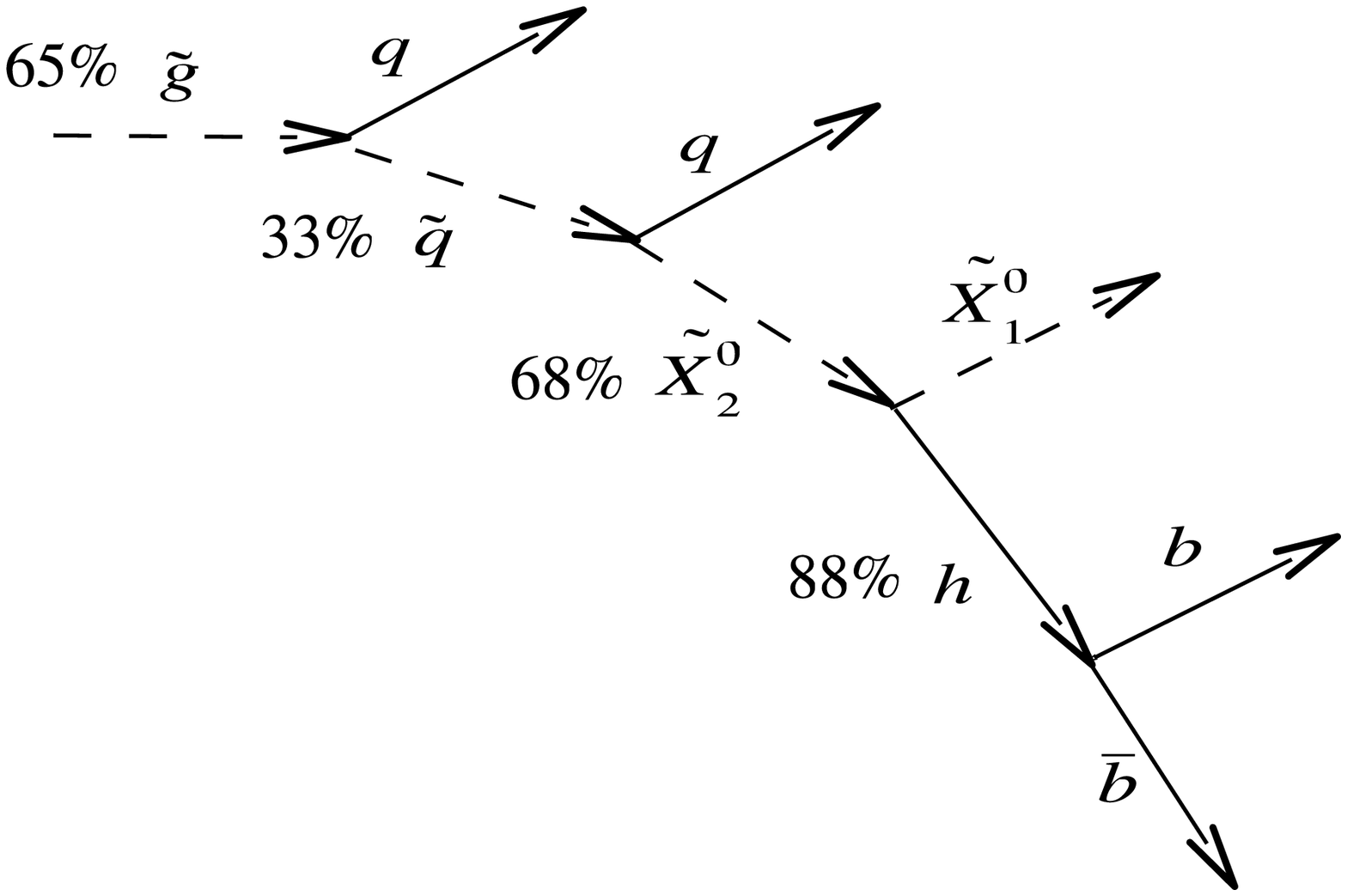}\\
     \multicolumn{2}{c}{\includegraphics*[viewport=0 200 575 725,width=0.5\textwidth]{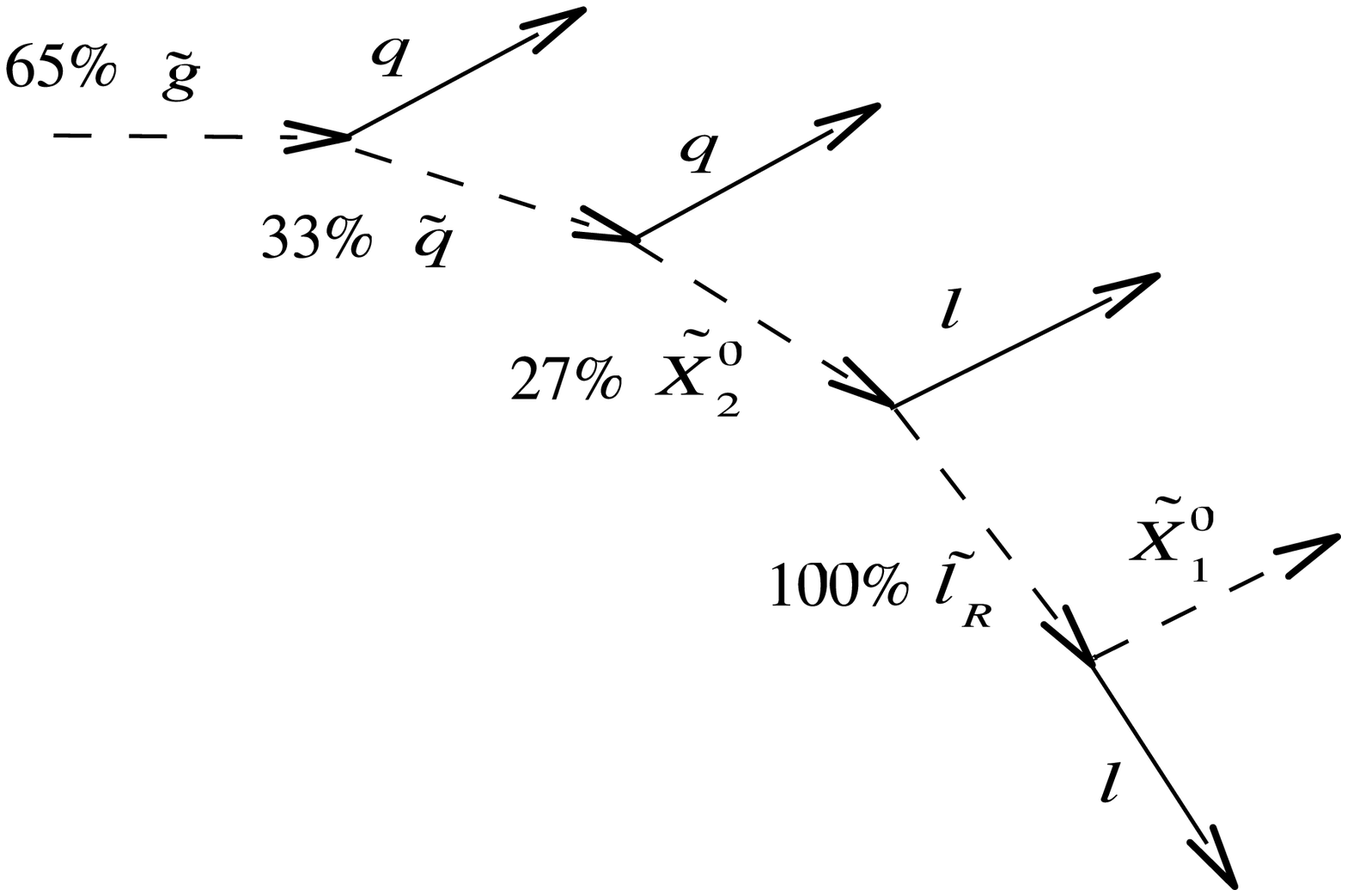}}
    \end{array}$
    \end{center}
\caption{The top three SUSY decay chains at point A and their branching ratios.} 
\label{diagtable}
\end{figure*}

The relevant decay chains for ATLAS point A are shown in Fig.~\ref{diagtable}. The third decay chain allows the separation of  leptons from the hadronic background \cite{atlas111}.  Requiring the leptons to be isolated and have high $P_T$ (in excess of 15 GeV) ensures that they were indeed produced from the above decay chain and not from the initial $2\times2$ hard interaction. In the rest frame of the second lightest neutralino, the dilepton invariant mass is
\begin{equation}
M_{ll}=\left[M_{\tilde{\chi}_2^0}^2+M_{\tilde{\chi}_1^0}^2-2~M_{\tilde{\chi}_2^0}~M_{\tilde{\chi}_1^0}
\left(1+\frac{P_{\tilde{\chi}_1^0}^2}{M_{\tilde{\chi}_1^0}^2}\right)\right]^{1/2}\,.
\label{invmass}
\end{equation}
The invariant mass is a function of the masses of the particles involved in the decay.
Since the momentum of the LSP is not constrained, the invariant mass distribution shows an edge at
$\sim 100$ GeV. This is because $M_{ll}$ reaches a maximum when the LSP is produced at rest. In that case the dilepton invariant mass is the difference between the masses of the LSP and $\tilde{\chi}_2^0$ which is $\sim$ 100 GeV for point A.

Heavy sparticles are produced at points B and C, where $m_0$=400 GeV and $m_{1/2}$=400 GeV. The dominant decay mode at
these points is $\tilde{\chi}_2^0\rightarrow\tilde{\chi}_1^0 h$ because $\tilde{\chi}_2^0$ decay into sleptons is forbidden due
to the high mass of the slepton  \cite{ATLAS:1999,Drozdetsky:2007zza,Kcira:2007ty}. Dileptons are produced from the decay of heavy charginos. Thus, the dilepton invariant mass plot does not have an edge as observed at point A. The large value of $\tan\beta$ at point C
increases production of $Z^0$ bosons and thus the number of dileptons.

Point D ($m_0$=200 GeV, $m_{1/2}$=100 GeV) is characterized by light SUSY particles, thus sparticle production rates
are large \cite{Hinchliffe:1996iu}. The dominant SUSY processes, where the percentages in parenthesis are the corresponding BRs, is $\tilde{g}g$  with
$\tilde{g}\rightarrow\tilde{b}\bar{b}~\hbox{(89\%)}~\rightarrow\tilde{\chi}_2^0b\bar{b}~\hbox{(86\%)}~\rightarrow\tilde{\chi}_1^0b\bar{b}l^{+}l^{-}~\hbox{(33\%)}$
\cite{ATLAS:1999,Drozdetsky:2007zza,Kcira:2007ty}. The above interactions imply that sparticle decays at this point are characterized by $b$ jets and
dileptons.

At point E ($m_0$=800 GeV and $m_{1/2}$=200 GeV) gluino production is dominant because $M_{\tilde{g}} <
M_{\tilde{q}}$. Neutralino decay into sleptons is suppressed because of the high mass of the slepton. There is significant $Z^0$
production a feature common to models with large $\tan\beta$. Dominant processes and branching ratios at this point are
$\tilde{\chi}_{3,4}^0\rightarrow~Z^0\tilde{\chi}_{1,2}^0$ (39\%) and $\tilde{\chi}_{2}\rightarrow~Z^0\tilde{\chi}_{1}$ (32\%)
\cite{ATLAS:1999,Drozdetsky:2007zza,Kcira:2007ty}. Dileptons are produced from $\tilde{\chi}_2^0\rightarrow\tilde{\chi}_1^0l^{+}l^{-}$ (6\%)~\cite{ATLAS:1999,Drozdetsky:2007zza,Kcira:2007ty} and from the decay of $Z^0$ into leptons.

Point LM1 \cite{Allanach:2002nj,Ball:2007zza}
($m_0$=60 GeV and $m_{1/2}$=250 GeV) is characterized by a  large value of $\tan\beta$, which leads to a lower mass of $\tilde{\tau}$ and
$\tilde{b}$. This results in the decay of the second lightest neutralino into $\tilde{\tau}$, producing dileptons
via the process  $\tilde{\chi}_2^0\rightarrow\tilde{\tau_1^{\pm}}\tau^{\mp}\rightarrow\tau^{+}\tau^{-}\tilde{\chi}_1^0$ and 
$\tilde{\chi}_2^0\rightarrow\tau^{+}\tau^{-}\tilde{\chi}_1^0$, where the $\tau$ decays in the lepton channel
~\cite{Baer:1998sz,Denegri:1999pe,Andreev:2007fy}. Moreover, $Z^0$ production is enhanced due to higher sbottom
production. This is due to the decay chain
$\tilde{b_1}\rightarrow\tilde{\chi}_{3,4}^0~b,\tilde{\chi}_{3,4}^0\rightarrow~Z^0\tilde{\chi}_{1,2}^0$
\cite{Baer:1998sz, Denegri:1999pe}. Since the mass of the gluino is larger than the squark mass, the dominant process is $\tilde{g}\rightarrow\tilde{q}q$ as at point  A. The branching ratios of the dominant decays are
 $\tilde{\chi}_2^0\rightarrow\tilde{\tau}_1^{\pm}\tau^{\mp}$ (46\%), $\tilde{\chi}_1^{\pm}\rightarrow\tilde{\nu}_l l$ (36\%) and $\tilde{\chi}_2^0\rightarrow\tilde{l}_R l$ (11.6\%) \cite{Ball:2007zza}.  The invariant
mass distribution of dileptons is expected to show an edge due to the process $\tilde{\chi}_2^0\rightarrow\tilde{l}_R l$.

The SM dilepton background contribution can be removed by applying suitable cuts on the transverse momenta of the
leptons~\cite{Bartl:1996dr}:  $P_{Tl} \ge 15$ GeV, $|\eta_l| < 2.5$, isolation cut $\sum_l P_{Tl} < 7$ GeV in a cone of $R=0.2$, and \missPT $\ge 200$ GeV. Here $P_{Tl}$ is the transverse momentum of the leptons, $R=\sqrt{\Delta\eta^2+\Delta\phi^2}$,  where $\phi$ and $\theta$ are the azimuthal and polar
angles of the lepton w.r.t.\ beam axis. 
The use of isolated dileptons as a SUSY signature has been previously discussed in the literature \cite{Hinchliffe:1996iu,Abdullin:1998nv,Chiorboli:2007zz}.
Although the production of leptons is not as high as the production of colored particles, high energy isolated leptons provide a cleaner
environment by allowing the removal of the SM background. The dominant SUSY interaction for opposite-sign, same-flavor
(OSSF) dileptons at ATLAS point A is $\tilde{\chi}_{2}^{0} \rightarrow  l^{\pm} \tilde{l} \rightarrow l^{\pm}l^{\mp}~
\tilde{\chi}_{1}^{0}$ with a 27\% branching ratio (BR) \cite{atlas111}. The maximum dilepton invariant mass for this
interaction is
\begin{equation}
M_{ll}^{max}=m_{\tilde{\chi}_{2}^0}
\left[\left(1-\frac{m_{\tilde{l}}^2}{m_{\tilde{\chi}_{2}^0}^2}\right)
\left(1-\frac{m_{\tilde{\chi}_{1}^0}^2}{m_{\tilde{l}}^2}\right)\right]^{1/2}
\sim 100~{\rm GeV}\,.
\label{mll}
\end{equation}

Dilepton production in BH events differ greatly from dilepton production in the MSSM. There is no single process of dilepton production for BHs. Dileptons are either produced by the BH directly or by the decay of heavier particles such as the $Z^0$ boson, $t\bar{t}$ pairs of a combination of the two. Therefore, the BH dilepton invariant mass does
not show a sharp cut-off at high energy.  Since the decay of top quarks into leptons is rare  \cite{Abbott:1997fv}, and the BR of $Z^0$ into leptons is small, $\Gamma(l^{+}l^{-})/\Gamma_{\rm tot}\sim 0.034$ \cite{PDG}, production of OSSF dileptons is less
frequent for BHs than SUSY events. Our analysis shows that an OSSF dilepton event occurs approximately every
100 BH and 20 SUSY events, with a $\sim$ 1:5 ratio of BH-to-SUSY dilepton events at fixed luminosity. 

\begin{figure*}[ht]
\centerline{\null\hfill
    \includegraphics*[width=0.65\textwidth]{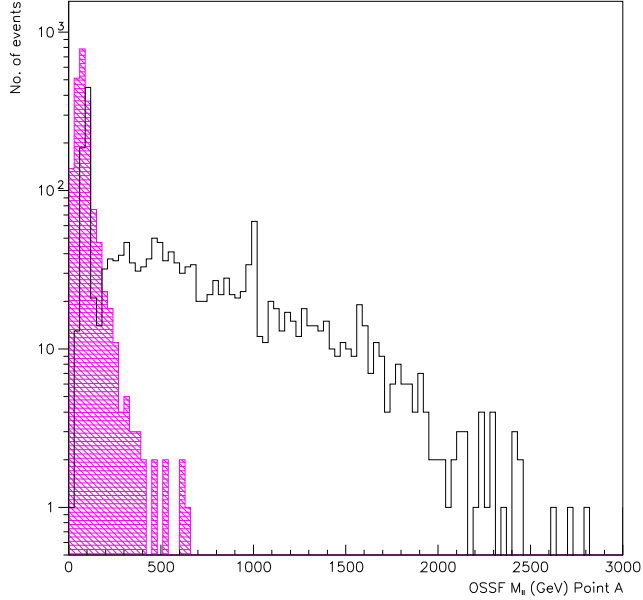}
    \hfill}
\caption{Invariant mass distribution (in GeV) for 2000 SUSY and BH OSSF dilepton events
for point A. The SUSY distribution (shaded pink histogram) shows the typical endpoint due to the
presence of the LSP. The high-$P_T$ tail of the BH distribution is originated by uncorrelated lepton
pairs emitted during the Hawking evaporation phase. The final BH decay is in two-quanta. The SM
background is negligible.} 
\label{fig_susy_pt_a1}
\end{figure*}
Fig.~\ref{fig_susy_pt_a1} shows the invariant mass distribution for 2000 SUSY OSSF dilepton events and 2000 BH OSSF
dilepton events. As was expected, the SUSY invariant mass distribution shows a sharp edge at $\sim$ 100 GeV. The BH
invariant mass distribution is characterized by two peaks, at $\sim 90$ GeV and a smaller peak at 1 TeV, and a tail at
high $P_T$. The first peak is due to dilepton events from single $Z^0$ bosons which are directly emitted by the BH. This
is the dominant channel of OSSF dilepton production in BH events. The peak at 1 TeV is due to dileptons emitted at the
end of the Hawking phase \cite{Cavaglia:2006uk,Cavaglia:2007ir}. The BH mass and the number of final hard quanta at the
end of the Hawking phase have been chosen to be $Q_{min}$=1 TeV and $n_p$=2, respectively. Since the BH at the end of the Hawking phase
is expected to be electrically neutral, isolated OSSF dilepton events can occur. This peak is expected to be smeared out in a more realistic
description of the final BH phase \cite{Roy:2008we}. The high-$M_{ll}$ tail of the distribution is originated by pairs
of uncorrelated leptons from the BH.

\begin{figure*}[htbp]
\centerline{\null\hfill
    \includegraphics*[width=0.65\textwidth]{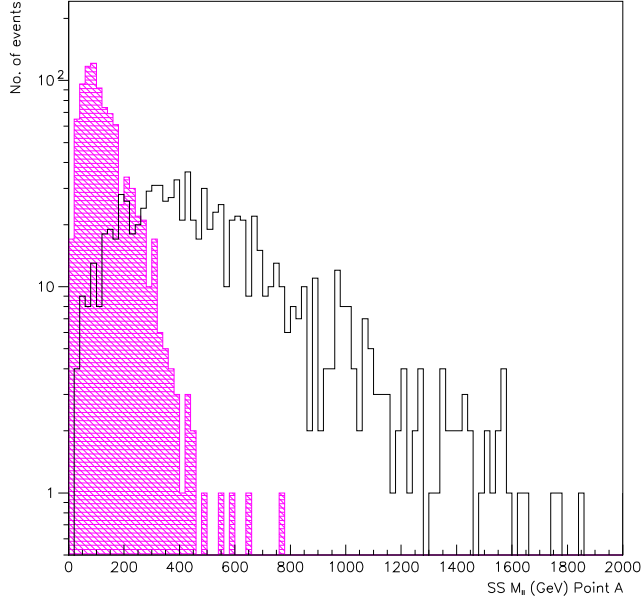}
    \hfill}
\caption{Invariant mass distribution for 900 same-sign dilepton events at point A. The SM background is negligible.} 
\label{fig_susy_pt_a2}
\end{figure*}

Fig.~\ref{fig_susy_pt_a2} shows the invariant mass distribution for 900 same-sign (SS) SUSY and BH events. In SUSY SS
dileptons are produced by $ \tilde{\chi}_{2}^{\pm} \rightarrow  W^{\pm} \tilde{l} l^{\pm}, \tilde{\chi}_{1}^{\pm}
\rightarrow  W^{\pm} \tilde{\chi}_{1}^{0}, \label{chi3}$ and from top quark decay into leptons. SS dileptons from BHs
are originated either from the BH itself or from the decay of heavier quarks and bosons ($t,\bar{t},W,Z^0$). In both
cases SS dilepton events are expected to be  rare because of the low BR of $W$ into leptons ($\sim$ 11\% \cite{PDG}). SS
dileptons are selected using the same cuts of the OSSF channel. The high-$M_{ll}$ tail of the BH invariant mass
distribution can be clearly distinguished from the SS case. Isolation requirements reduce the SM background to virtually
zero \cite{Baer:1989hr,Baer:1991xs}.

\begin{figure*}[htbp]
\centerline{\null\hfill
    \includegraphics*[width=0.65\textwidth]{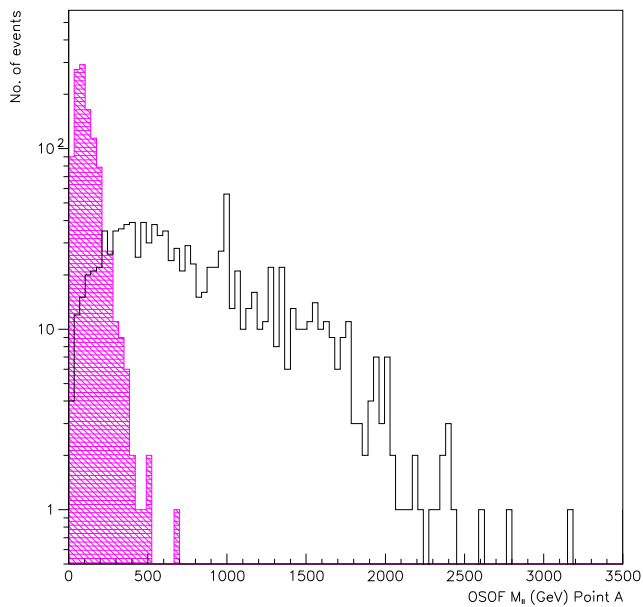}
    \hfill}
\caption{Invariant mass distribution for 1100 OSOF dilepton events at point A. The SM background is negligible.} 
\label{fig_susy_pt_a3}
\end{figure*}

Dilepton events with same sign and/or opposite flavor leptons can also be used as discriminators  (Fig.~\ref{fig_susy_pt_a3}). The ``democratic'' nature of BH decay makes events with same/opposite flavor leptons roughly equally probable, whereas BRs of SUSY events favor same-flavor  dileptons. OS dileptons from the decay of $\tilde{X_2^0}$ are of the same flavor. On the other hand, leptonic decays of charginos or third generation quarks and squarks would produce equal amounts of opposite-sign, opposite flavor (OSOF) dilepton pairs as compared to same flavor lepton pairs \cite{Baer:1995nq,Baer:1995va,Paige:1997xb}.

Table~\ref{table:table1} shows the BRs of same-/opposite-sign,  same-/opposite-flavor isolated dilepton events for SUSY
and BH processes. The dominant channel is the OSSF channel for both models. However, SUSY and BH events can be easily
discriminated by comparing the rate of same-flavor events to the rate of opposite-flavor events. 
\begin{table}[htbp]
\caption{BRs of high-$P_T$ isolated dileptons for SUSY and BH models. 
21,000 and
100,000 events were simulated in the two cases, respectively, yielding approximately 1000
dilepton events. OS(SS) stands for opposite (same) sign and OF(SF) denotes opposite (same)
flavor.}
\begin{center}
\begin{tabular*}{0.80\textwidth}%
{@{\extracolsep{\fill}}c|ccccc}
\hline\hline
~High $P_T$ isolated dileptons~ & $SUSY$ & \% & $BH$ & \% &\\
\hline
OSSF                   & 768 & 73  & 523   & 50  &\\
\hline
SSSF                   & 65  &  6  & 103    & 10  &\\
\hline
OSOF                   & 169  & 16  & 341   & 32  &\\
\hline  
SSOF                   & 52  & 5  & 87   & 8  &\\
\hline
\end{tabular*}
\end{center}
\label{table:table1}
\end{table}

\begin{figure*}[ht]
\centerline{\null\hfill
    \includegraphics*[width=0.65\textwidth]{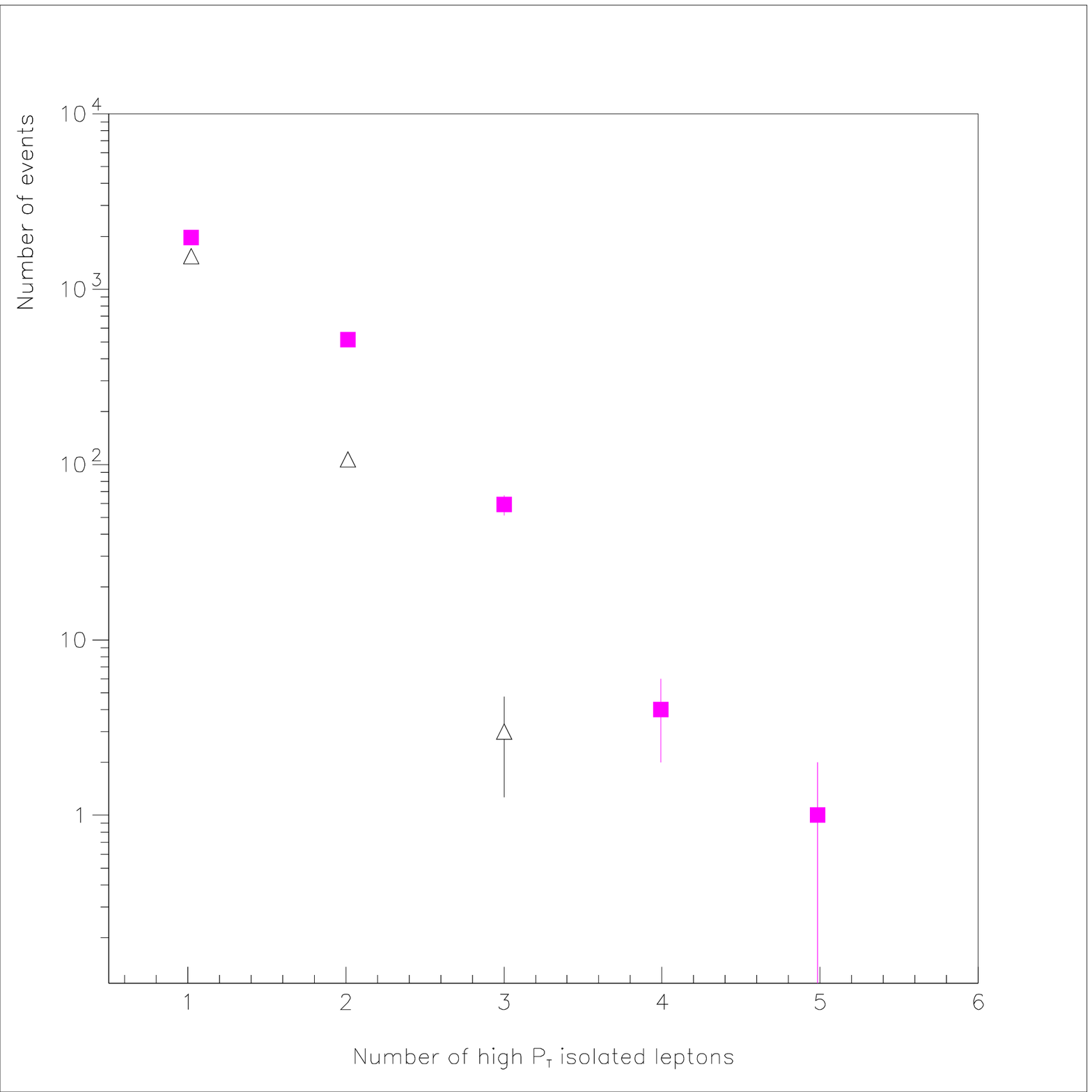}
    \hfill}
\caption{Histogram of the number of events with high-$P_T$ leptons for 10,000 MSSM
(pink filled squares) and BH interactions (black open triangles). The number of BH events with three
isolated leptons is smaller than the number of SUSY events by a factor of $\sim$ 20. The
probability of producing BH events with four or more leptons is virtually zero.} 
\label{FIG7_1}
\end{figure*}

The number of isolated, high-$P_T$ leptons can also be used to complement the dilepton analysis (Fig.~\ref{FIG7_1}). SUSY events are capable of producing up to five isolated leptons from the cascade decay of heavy
sparticles. Events with $\tilde{\chi}_{2}^{0} \tilde{\chi}_{2}^{0}$ or $\tilde{\chi}_{1}^{\pm} \tilde{\chi}_{2}^{0}$ may
produce four or three isolated leptons, respectively \cite{Baer:1995tb}. On the contrary, events with three or more
isolated leptons are very suppressed in BH decays at the LHC energy. Although multilepton events are rare, there is very
little background and they could be effectively used to distinguish the MSSM and the BH model. 

\begin{figure*}[htbp]
\centerline{\null\hfill
    \includegraphics*[width=0.65\textwidth]{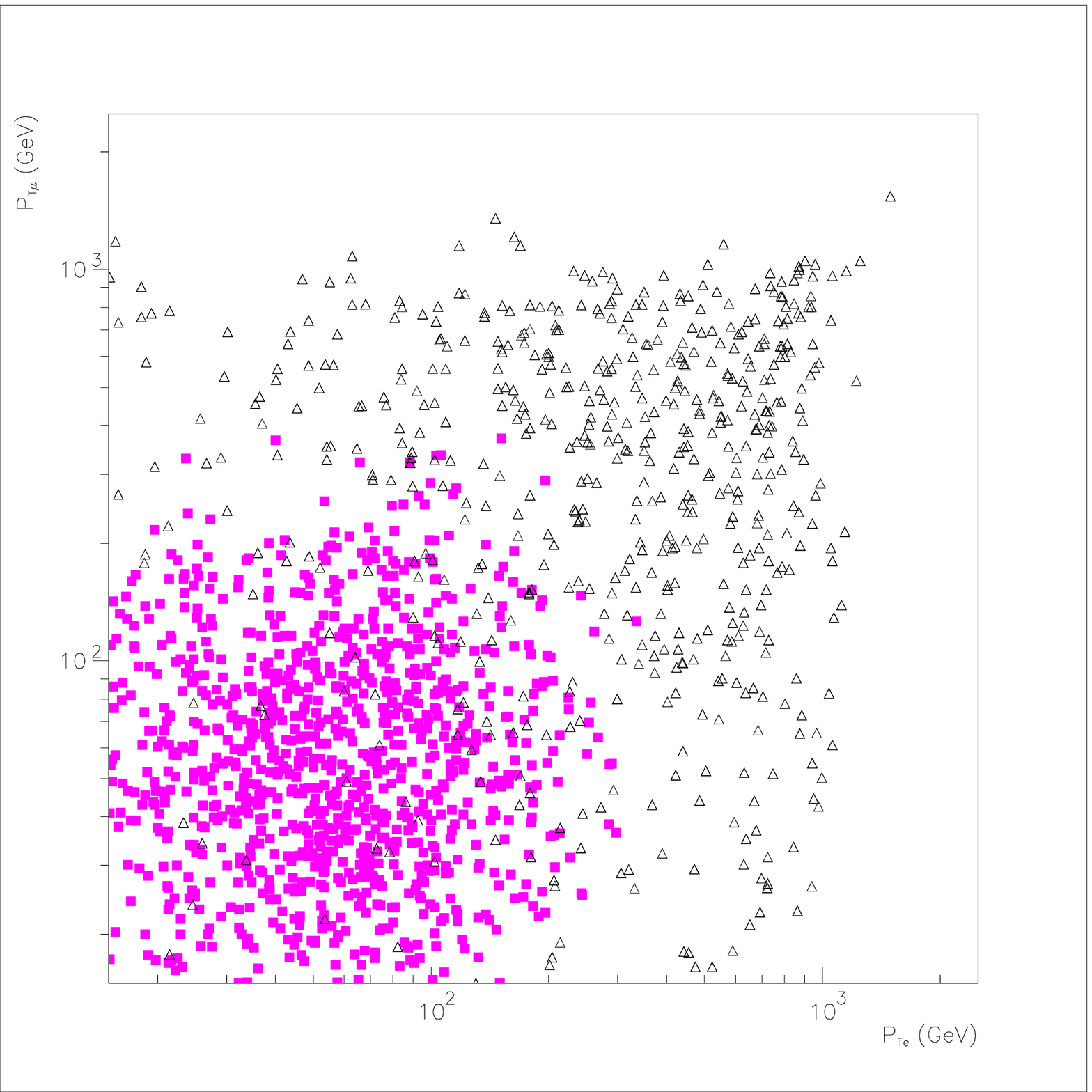}
    \hfill}
\caption{$P_T$ scatter plot for $\sim$ 1000 isolated opposite-flavor dilepton events for SUSY (pink filled squares) and
BHs (black open triangles). BH leptons are harder than SUSY leptons and show a larger spread in $P_T$.} 
\label{FIG7_2}
\end{figure*}

Other effective discriminators can be constructed by looking at dilepton events with same sign and/or opposite-flavor
leptons. The ``democratic'' nature of the BH decay makes all dilepton events roughly equally probable, whereas the MSSM
favors same-flavor dileptons. Presence of hard opposite-flavor leptons is a clear indication of BH decay (Fig.~\ref{FIG7_2}). Our analysis shows that 73\% of SUSY dilepton events are OSSF, compared to only 50\% in the BH model.
Conversely, opposite-flavor events are twice more frequent in the BH model (40\%) compared to the MSSM (21\%). 

\subsection*{Dilepton analysis at other LHC points}
\begin{figure*}[thbp]
\begin{center}$
\begin{array}{cc}
    \includegraphics*[width=0.5\textwidth,totalheight=0.25\textheight]{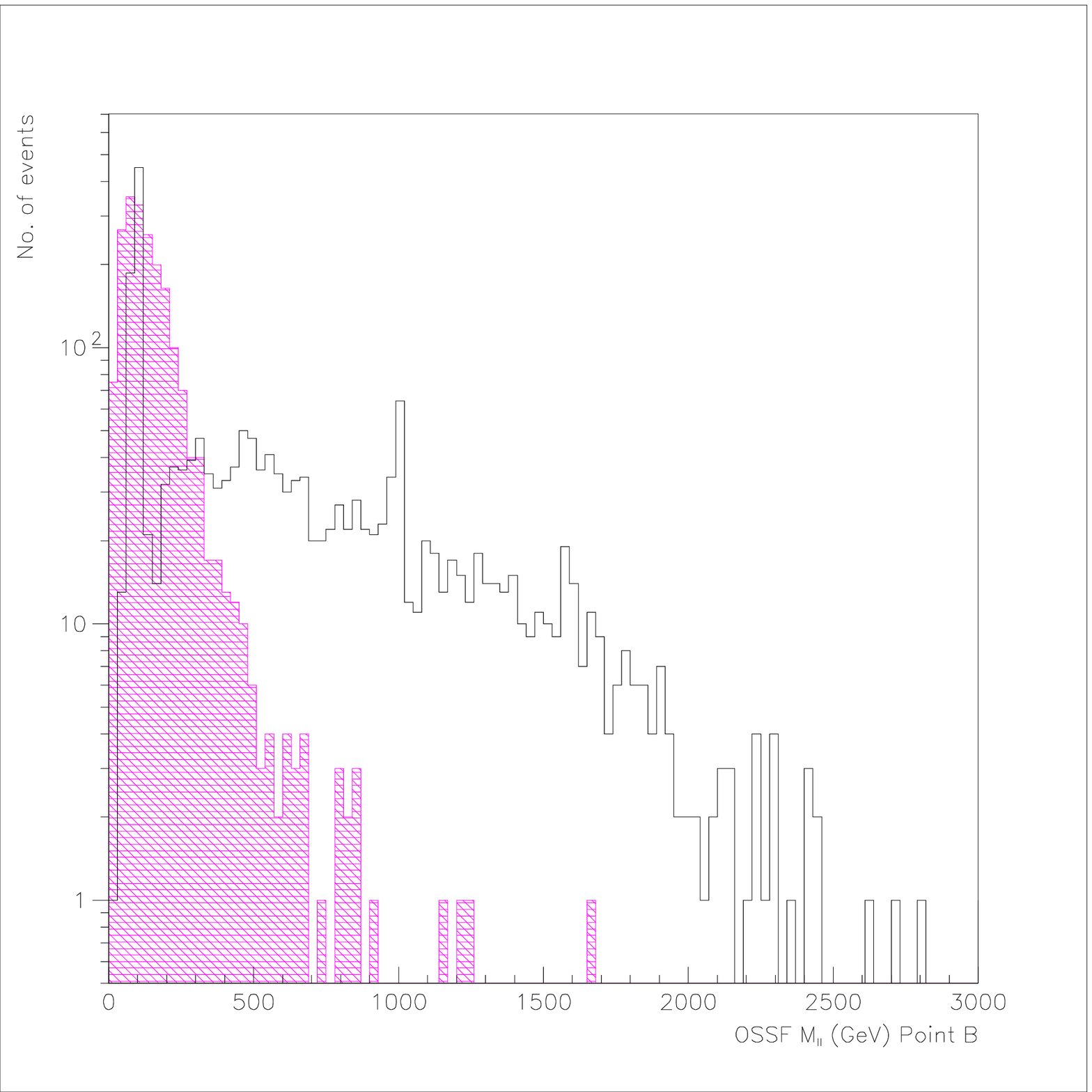}&
    \includegraphics*[width=0.5\textwidth,totalheight=0.25\textheight]{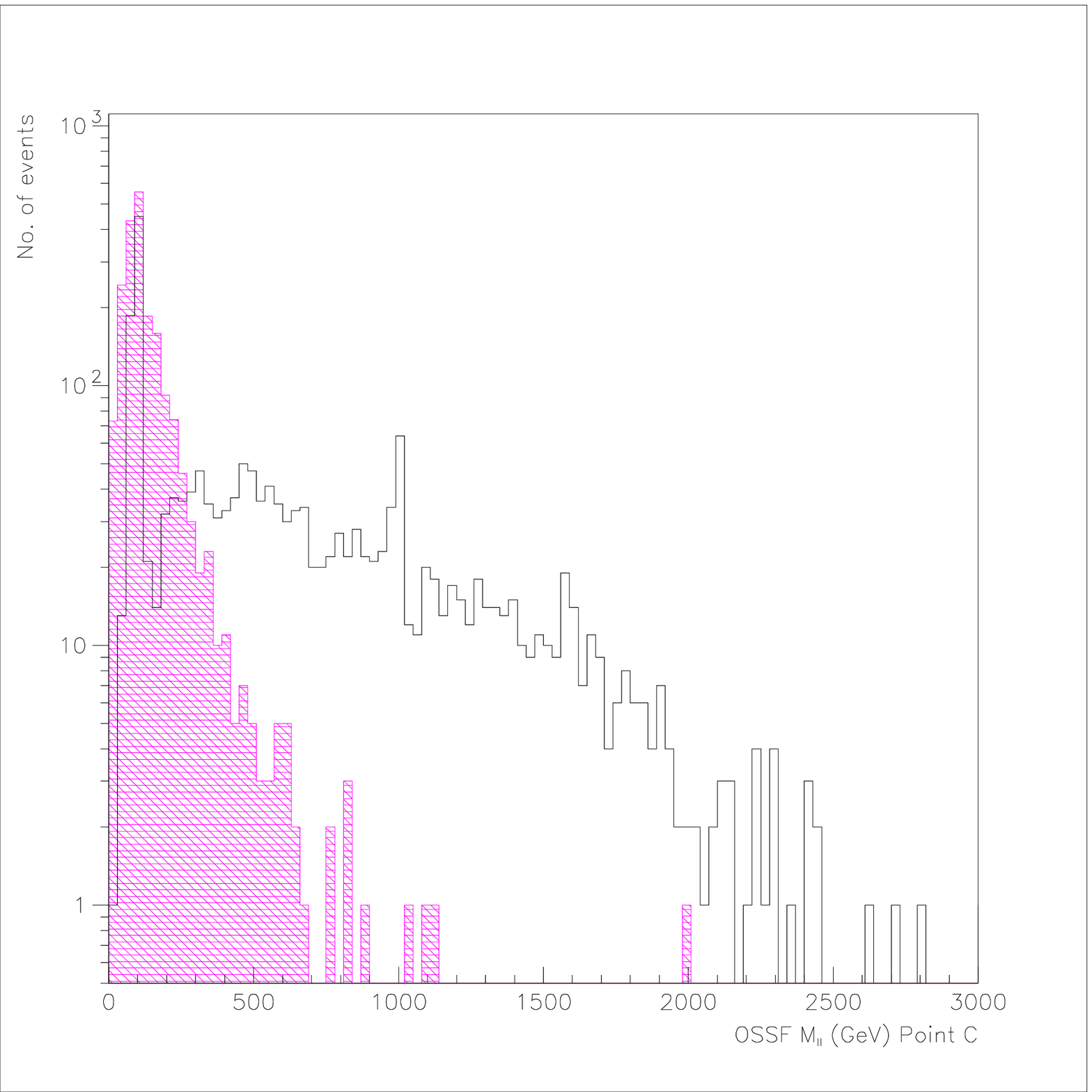}\\
    \includegraphics*[width=0.5\textwidth,totalheight=0.25\textheight]{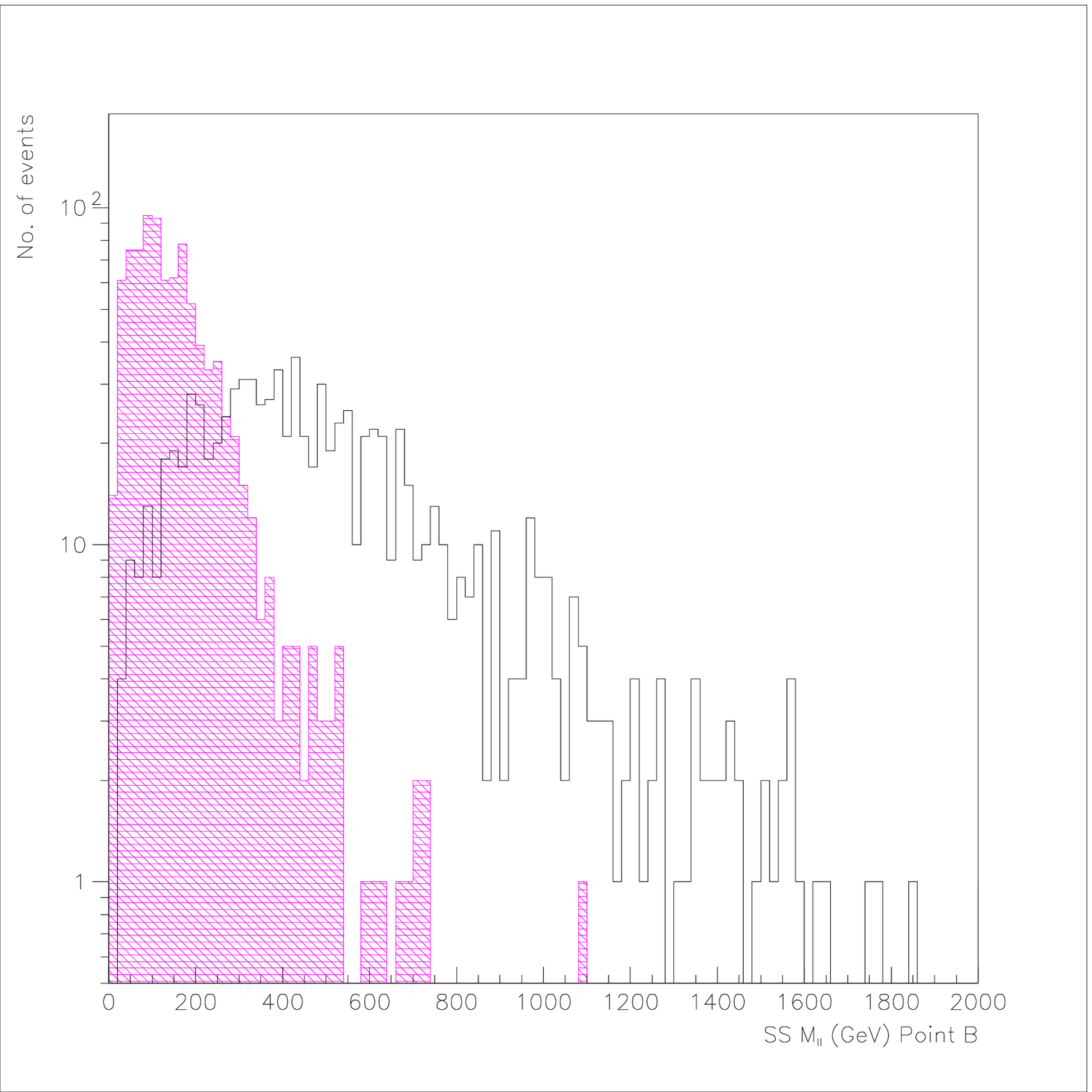}&
    \includegraphics*[width=0.5\textwidth,totalheight=0.25\textheight]{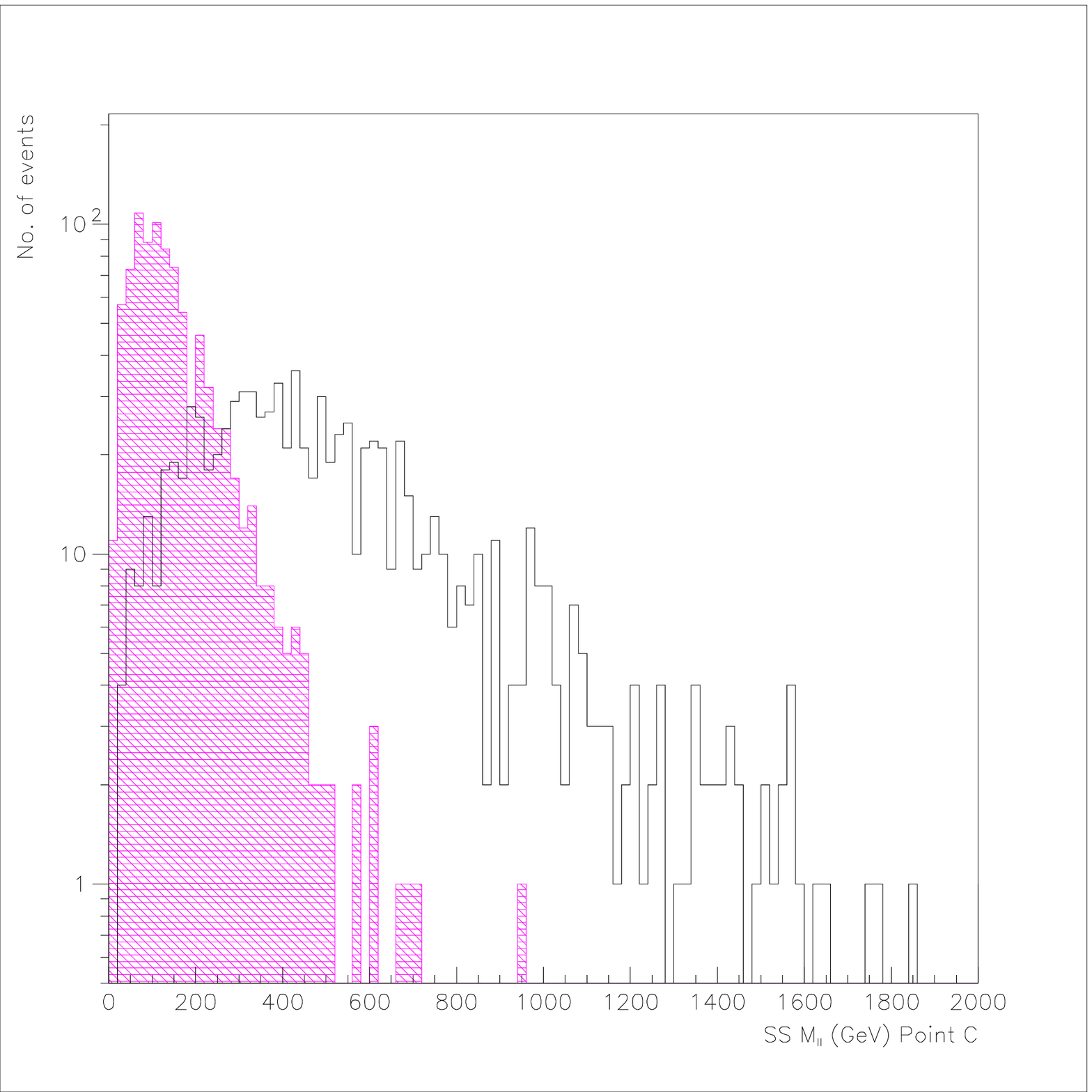}\\
    \includegraphics*[width=0.5\textwidth,totalheight=0.25\textheight]{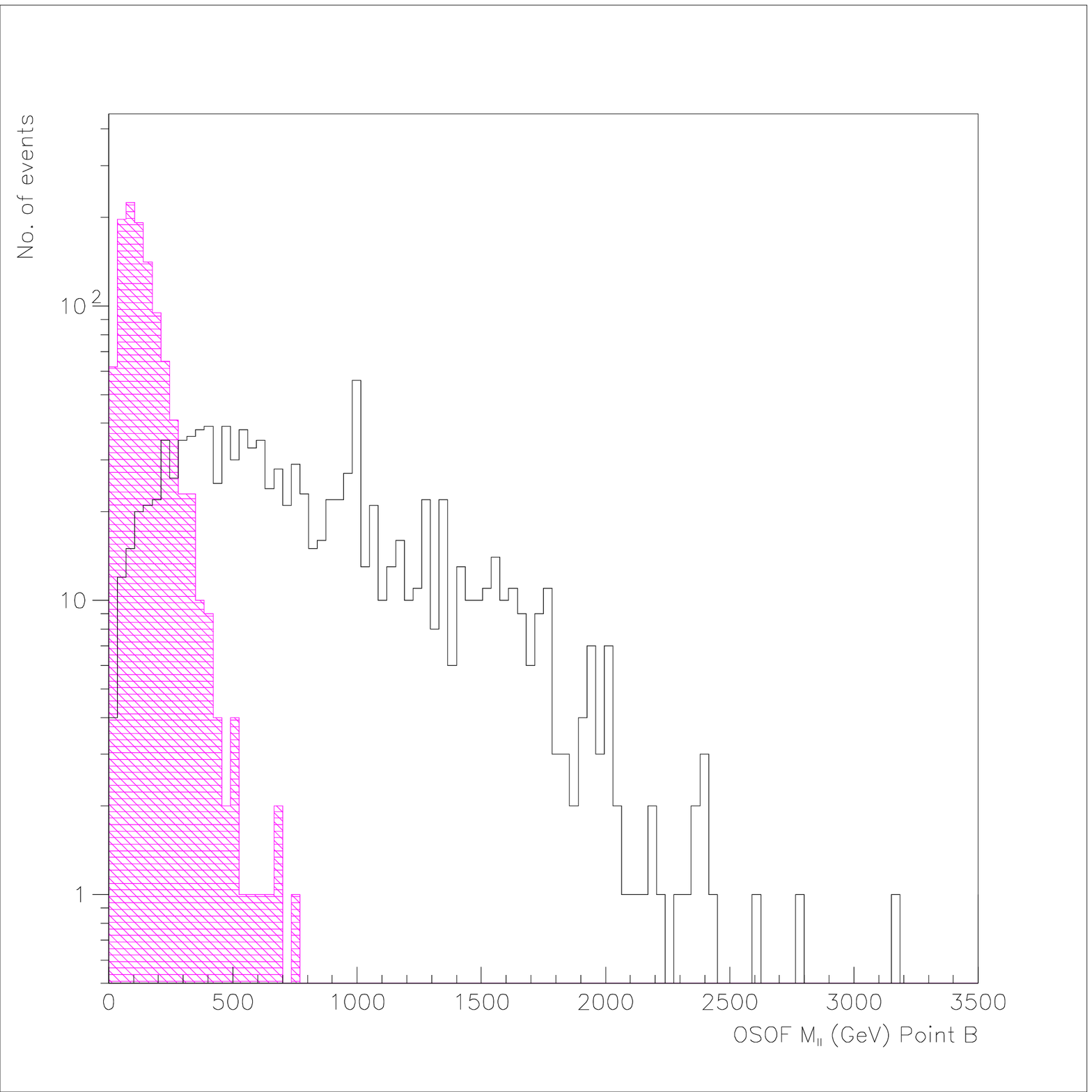}&
    \includegraphics*[width=0.5\textwidth,totalheight=0.25\textheight]{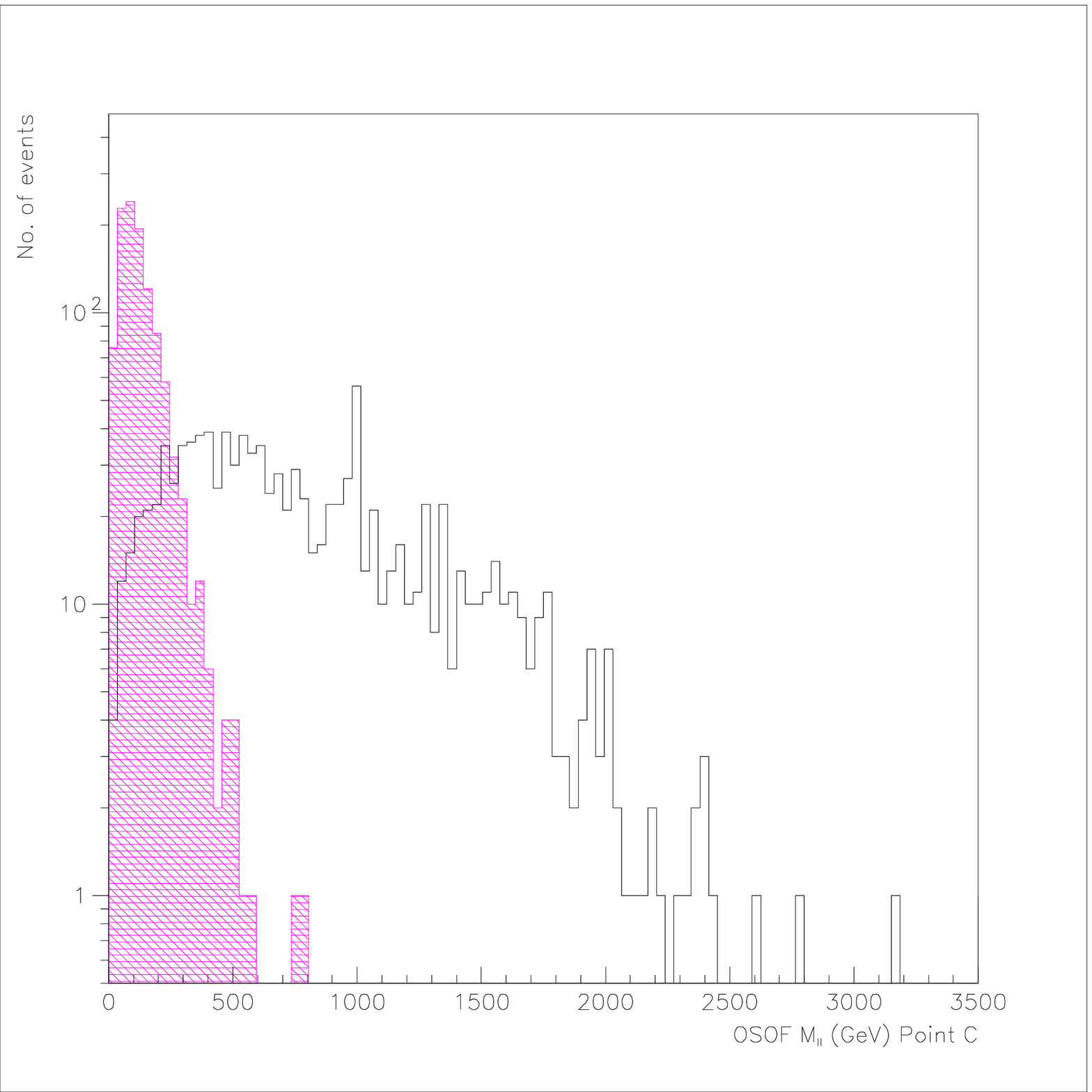}
    \end{array}$
    \end{center}
\caption{Invariant mass distributions at point B (left panels) and point C (right panels). OSSF (top panels), SS (middle
panels) and OSOF (bottom panels) distributions clearly discriminate SUSY and BH events.} 
\label{fig_susy_pt_bc}
\end{figure*}

Figure~\ref{fig_susy_pt_bc} shows the dilepton invariant mass distributions for points B and C. The background cuts
remove the SM contribution. Unless otherwise stated all dilepton histograms in this section show 2000 OSSF events, 900 SS events and
1100 OSOF events for both SUSY and SM. The SUSY invariant mass distribution is represented as the shaded (pink) histogram.
\begin{table}[htbp]
\caption{BRs of high-$P_T$ isolated dileptons for SUSY points B and C for 50,000 events. The high value of
$\tan\beta$=10 for point C manifests itself as a greater production of OSSF dileptons.}
\begin{center}
\begin{tabular*}{0.70\textwidth}%
{@{\extracolsep{\fill}}c|ccc}
\hline
\hline
~High $P_T$ isolated dileptons~ & $OSSF$ & $OSOF$ &\\
\hline
Point B                   & 12\% & 11.5\% &\\
\hline
Point C                   & 16\% & 11.6\% &\\
\hline  
\end{tabular*}
\end{center}
\label{table:table2}
\end{table}

\begin{figure*}[thbp]
\begin{center}$
\begin{array}{cc}
    \includegraphics*[width=0.5\textwidth,totalheight=0.25\textheight]{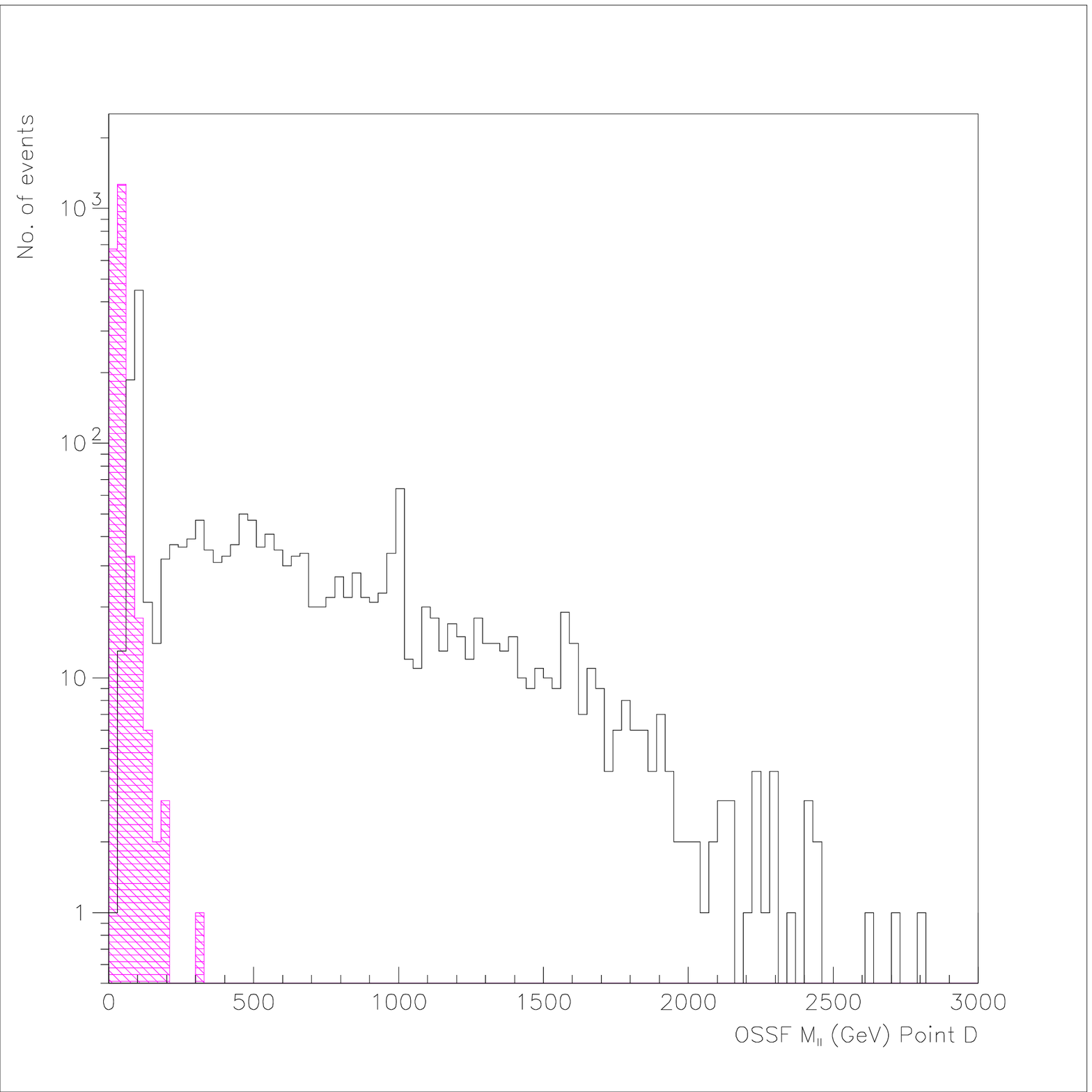}&
    \includegraphics*[width=0.5\textwidth,totalheight=0.25\textheight]{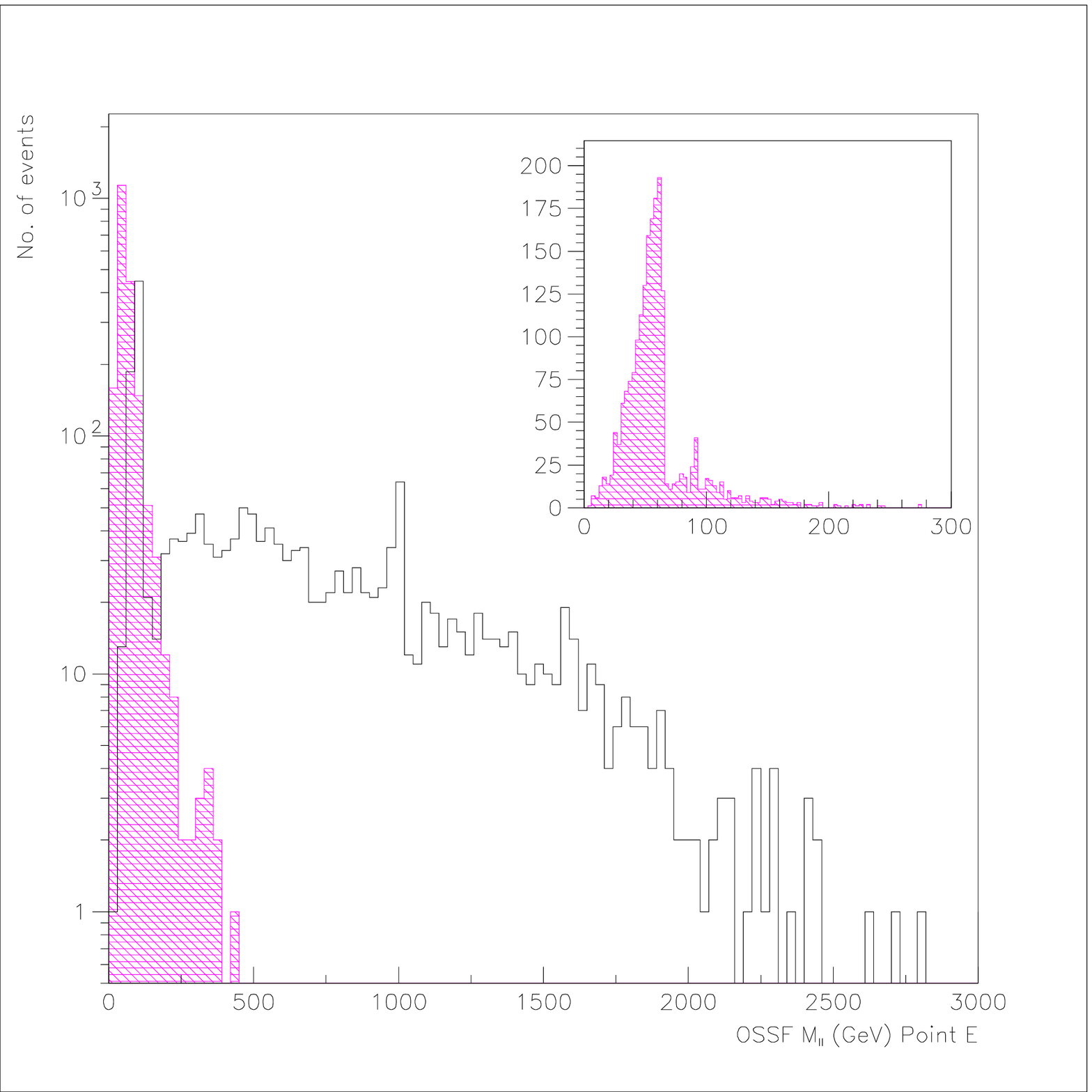}\\
    \includegraphics*[width=0.5\textwidth,totalheight=0.25\textheight]{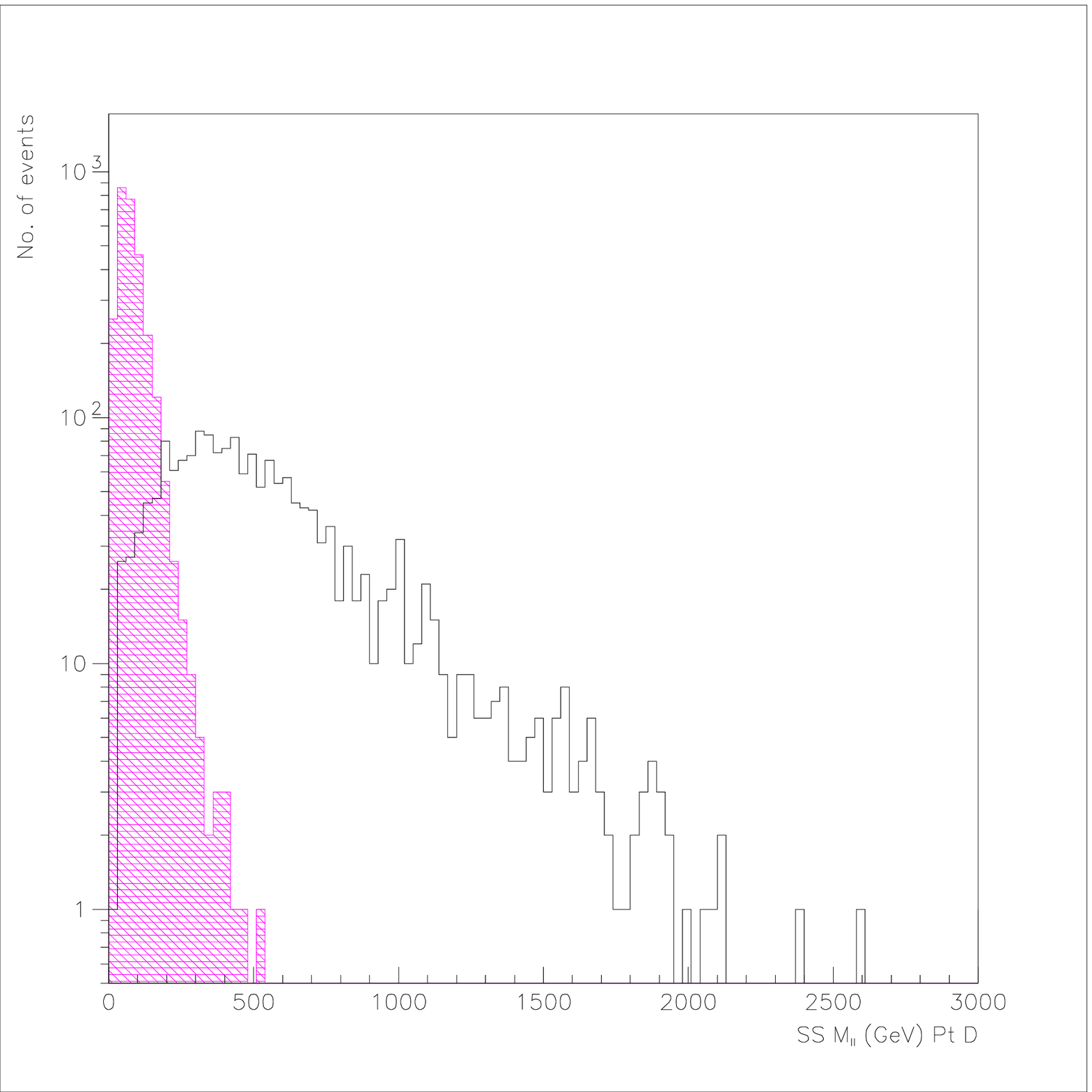}&
    \includegraphics*[width=0.5\textwidth,totalheight=0.25\textheight]{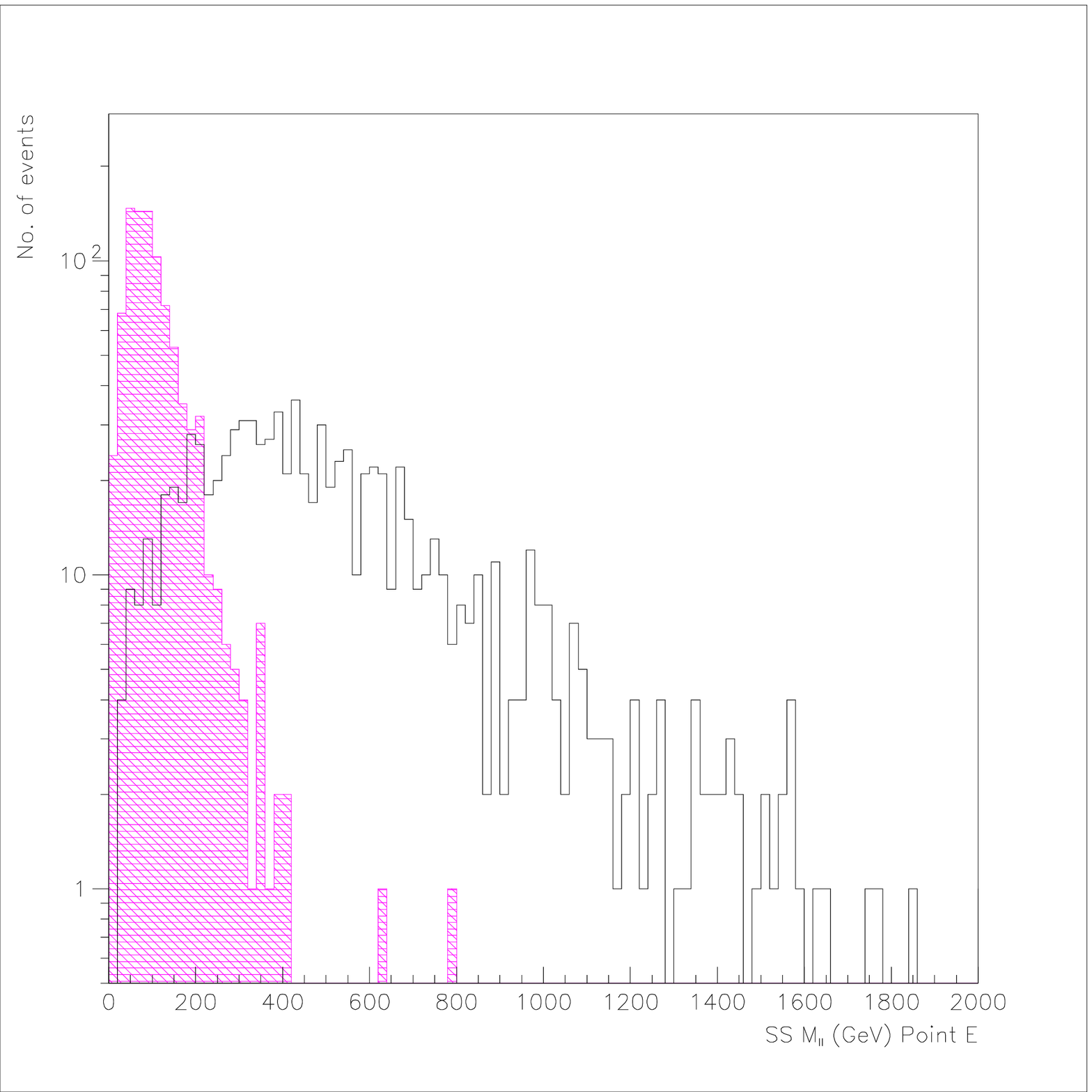}\\
    \includegraphics*[width=0.5\textwidth,totalheight=0.25\textheight]{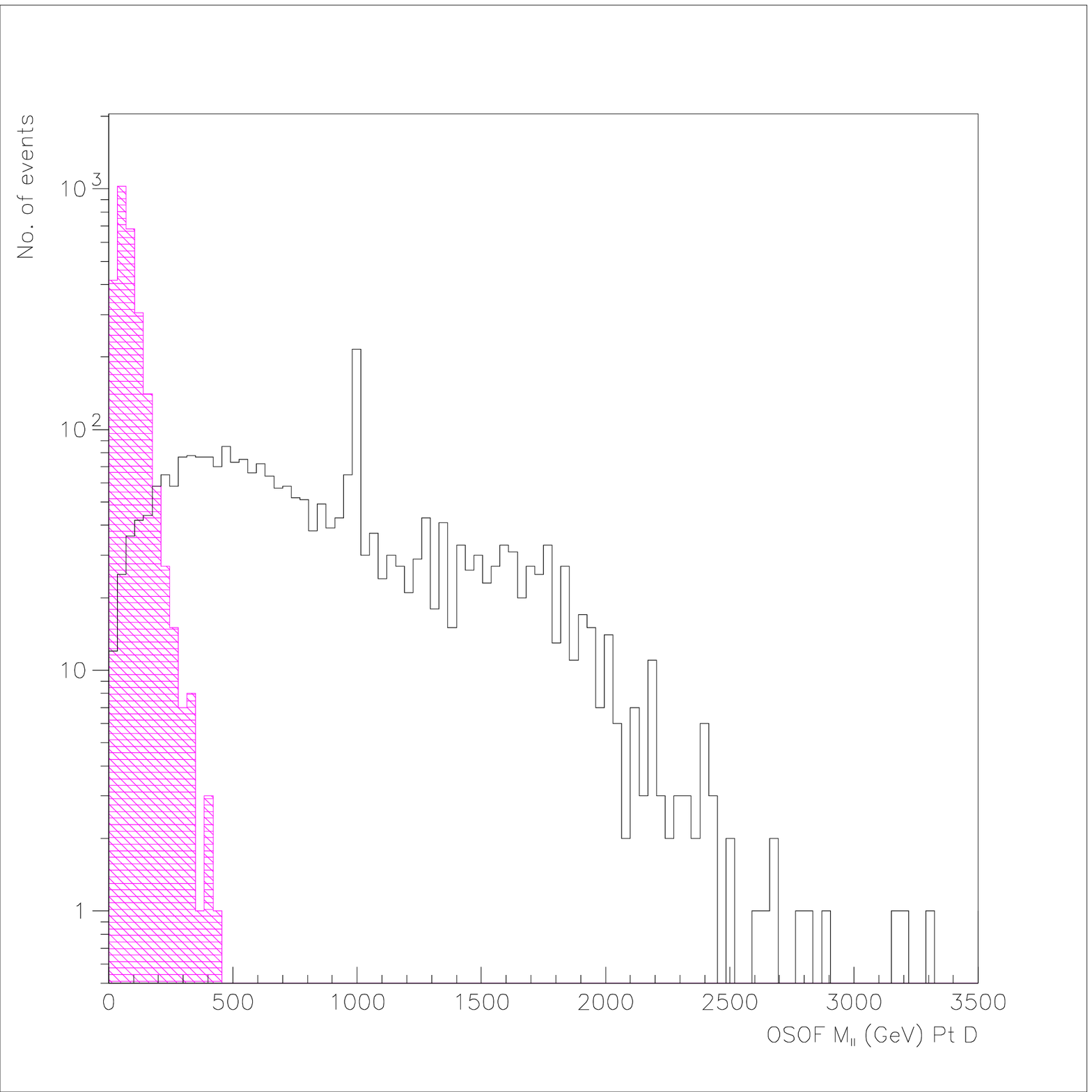}&
    \includegraphics*[width=0.5\textwidth,totalheight=0.25\textheight]{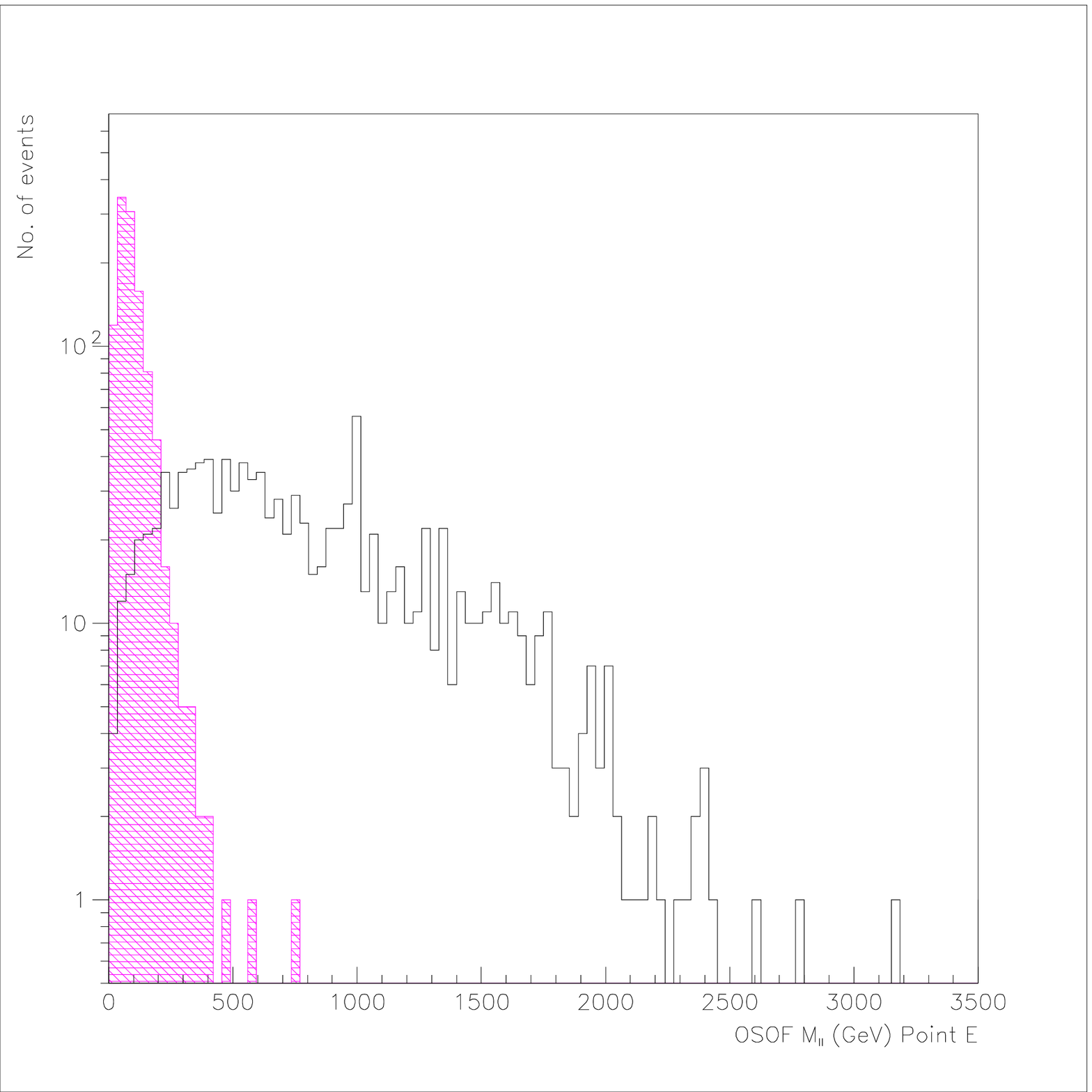}
    \end{array}$
    \end{center}
\caption{Invariant mass distributions at point D (left panels) and point E (right panels). Top Panel: Distribution of 2000 SUSY and BH OSSF dilepton events. At point E there is enhanced $Z^0$ production because of a high value of $\tan\beta$. Middle Panel: Event distribution for $\sim$ 1700 SUSY and BH SS dilepton events. The SM background is negligible. Bottom Panel: SUSY and BH distribution for $\sim$ 2600 OSOF events. The major SM background due to $\bar{t}t$ is negligible.} 
%\ContinuedFloat
\label{fig_susy_pt_de}
\end{figure*}

The dilepton invariant mass distributions for point D (left panel) and E (right panel) are shown in Fig.~\ref{fig_susy_pt_de}. From the edge in the OSSF invariant mass distribution (top panel of Fig.~\ref{fig_susy_pt_de}) the difference in mass of the two lightest
neutralinos can be obtained: $M_{\tilde{\chi}_2^0}-M_{\tilde{\chi}_1^0}$  $\sim$ 50 GeV. \missPT\ calculations are generally not
performed at point D \cite{Baer:1995nq,Baer:1995va,Paige:1997xb}. This is because lepton isolation requirements and the presence of $b$ jets are effective in reducing the SM background~\cite{ATLAS:1999,Drozdetsky:2007zza,Kcira:2007ty}. The bottom panel of Fig.~\ref{fig_susy_pt_de} shows the OSOF dilepton distribution for $\sim$ 2600 SUSY and BH events. More than half of the background contribution is from $\bar{t}t$ events \cite{Andreev:2007fy}. Simulations show that this background is  negligible due to the low BR of $W$ bosons into leptons.  The OSOF dilepton invariant mass distribution at point E is characterized by the $Z^0$ peak and an undistinct invariant mass edge.
%\clearpage
%
%
\begin{figure*}[t!]
\begin{center}$
\begin{array}{cc}
    \includegraphics*[width=0.5\textwidth]{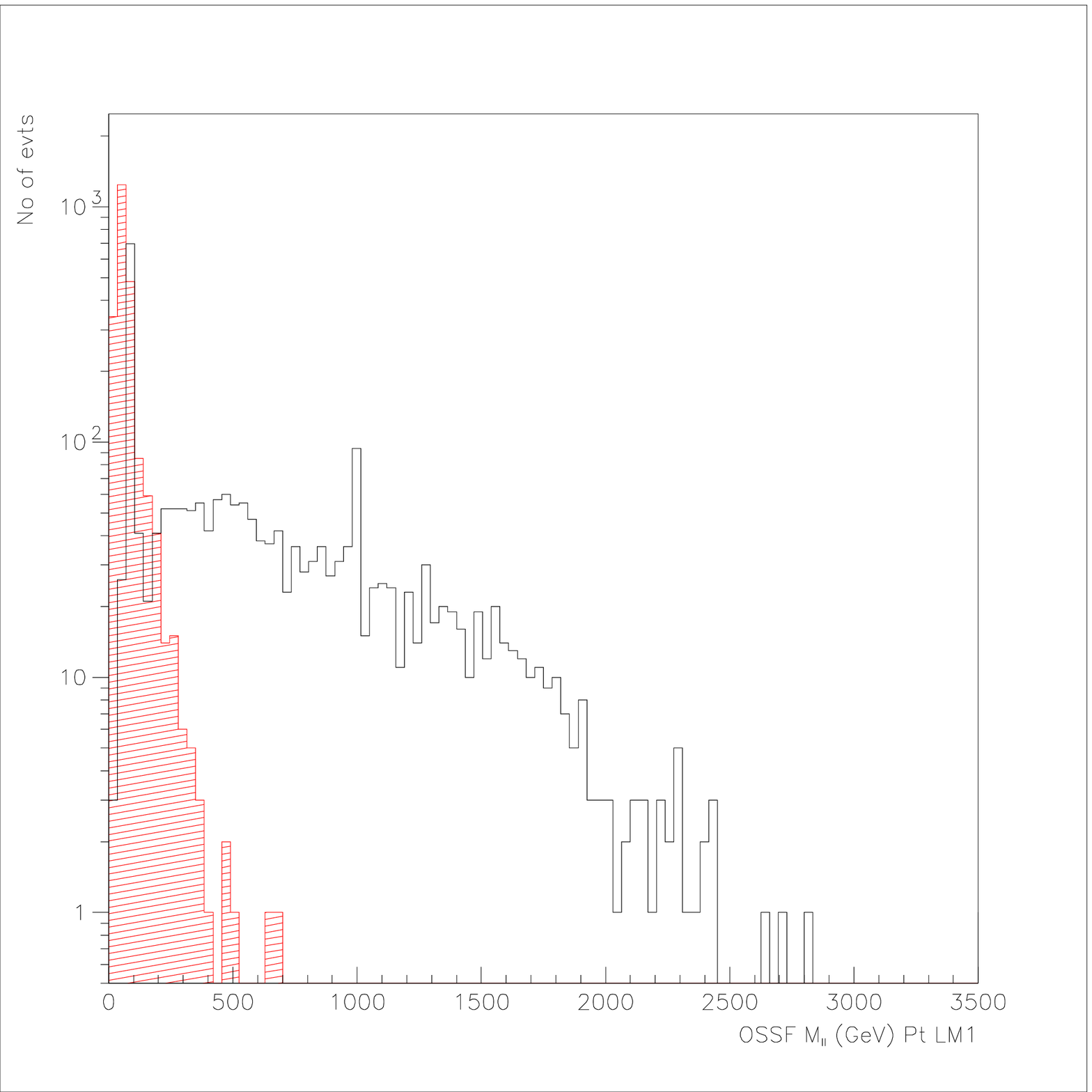}&
    \includegraphics*[width=0.5\textwidth]{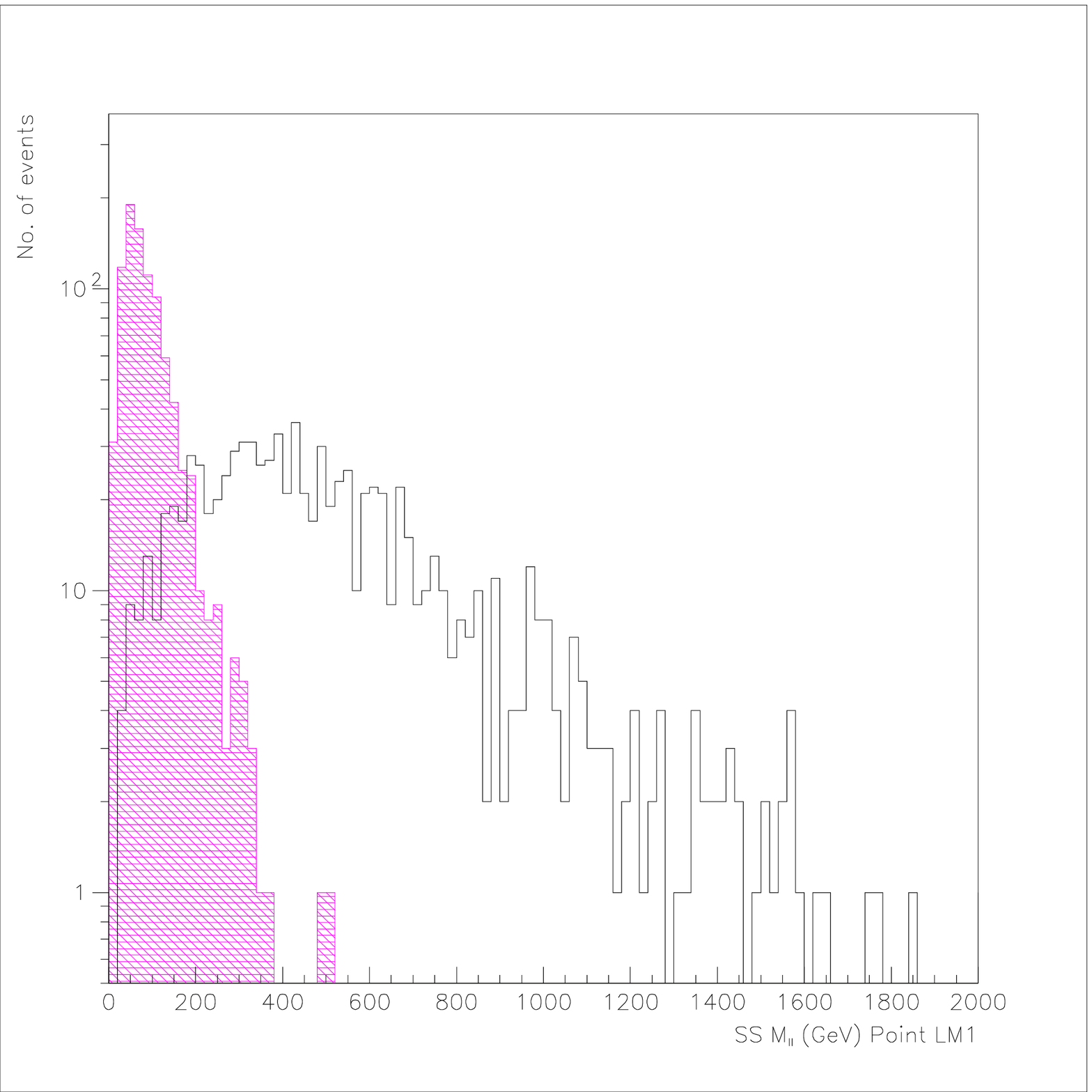}\\
    \multicolumn{2}{c}{\includegraphics*[width=0.5\textwidth]{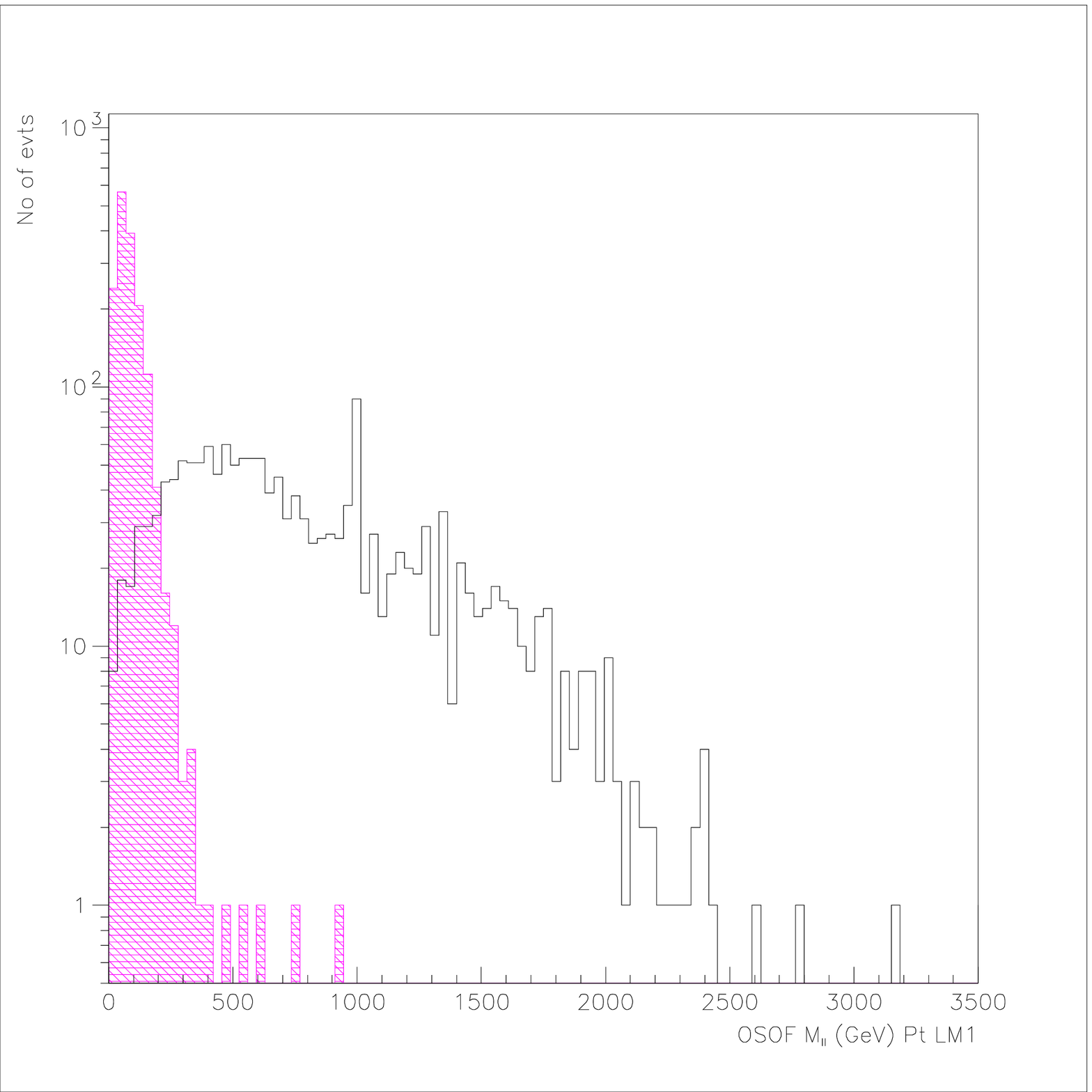}}
    \end{array}$
    \end{center}
\caption{Top Left Panel: Invariant mass distribution for SUSY and BH OSSF dilepton events
for the CMS benchmark point LM1. Likewise to point A, the SUSY distribution shows the endpoint due to
the presence of the LSP. Top Right Panel: invariant mass distribution for same sign dileptons. Bottom
Panel: OSOF invariant mass distribution. A high value of $\tan\beta$ enhances the production of OSOF
dileptons as explained in the text.} 
%\ContinuedFloat
\label{fig_susy_pt_lm1}
\end{figure*}
The dilepton distribution for point LM1 is shown in Fig.~\ref{fig_susy_pt_lm1}. The edge in the OSSF dilepton invariant
mass distribution is not as evident as in point A because of the large amount of missing momentum which is carried away
by the neutrinos in the $\tau$ decay \cite{Andreev:2007fy}.

Therefore, in conclusion, it is observed that irrespective of the LHC point the dilepton invariant mass always provides effective discrimination between SUSY and BHs.
\clearpage
%
%\clearpage
%
%\newpage
\section[ED and BH Event Analysis]{ED and BH Event Analysis\label{evt_analysis_edbh}}
In this section we present our original analysis of BH and graviton events at the LHC. Simulation results using energy, momentum and  event shape variables as discriminators are first presented followed by an analysis of dileptons.
\subsection{Energy, Momentum and  Event shape variables\label{e_m_shape_kk}}
\begin{figure*}[ht] 
\centerline{\includegraphics*[width=0.75\textwidth]{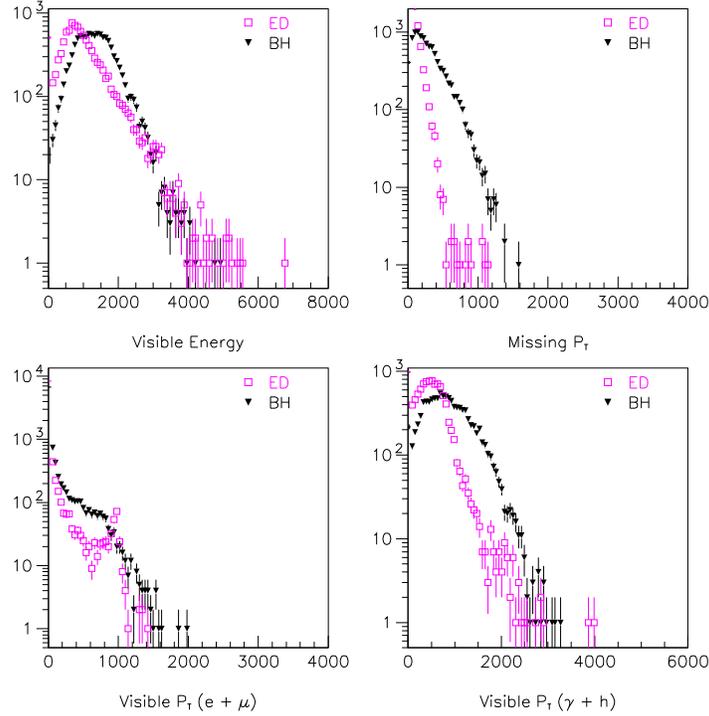}} 
\caption{Comparison of
10,000 RS graviton and BH benchmark events. Visible energy and \missPT\ (top
panels) are comparable due to the presence of invisible channels in both models. Bottom panels show visible $P_T$ due to leptons (left panel) and photons and hadrons (right panel).}
\label{FIG2_kk}
\end{figure*}
Figure~\ref{FIG2_kk} shows visible energy, \missPT\, and $P_T$ of leptons and
hadrons \& photons for 10,000  RS graviton emission and BH benchmark events. The amount of visible energy is comparable
for the two scenarios, even in the absence of a BH remnant. The \missPT\ is due to the presence of  neutrinos for the RS graviton model. Graviton decays are also capable of producing isolated leptons with high $P_T$. This explains the peak at 1 TeV in the visible $P_T$ distribution of leptons. As with SUSY and BH, isolated leptons may provide a discriminating signature between BH and graviton decays (See Sect.~\ref{leptons_kk}.). 
%
%\newline
%
\begin{figure*}[ht]
\centerline{\null\hfill
    \includegraphics*[width=0.75\textwidth]{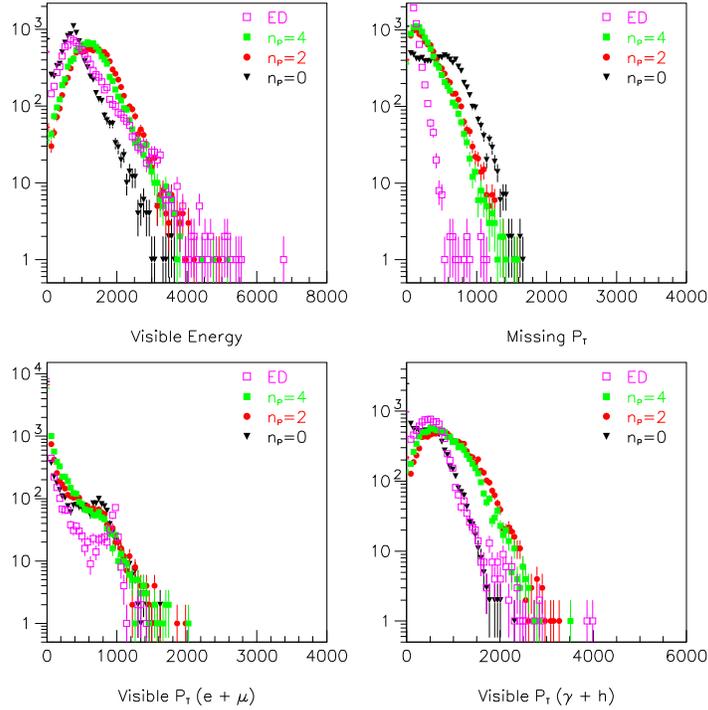}
    \null\hfill}
\caption{Distribution of visible energy, \missPT\ and transverse momenta of leptons and hadrons \& photons. RS graviton
plots are shown as pink open squares. The four plots show the effect of different decay modes in the
Planck phase of ten-dimensional BHs: remnant formation ($n_p=0$, black filled triangles), two-body decay ($n_p=2$, red
filled circles) and four-body decay ($n_p=4$, green filled squares). The fundamental Planck scale is $M_\star=1$ TeV.}     
\label{FIG3_kk1}
\end{figure*}
%
%\newline

\begin{figure*}[th]
\centerline{\null\hfill
    \includegraphics*[width=0.75\textwidth]{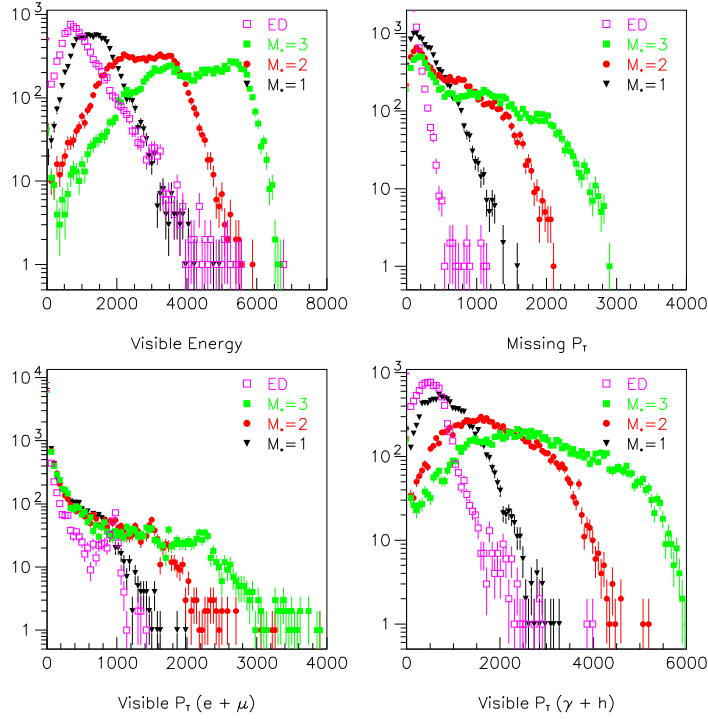}
    \null\hfill}
\caption{Distribution of visible energy, \missPT\ and transverse momenta of leptons and hadrons \& photons. RS graviton
results are shown as pink open squares. 
The four plots show the effect of varying the fundamental Planck scale: $M_\star=1$ TeV (black
filled triangles), $M_\star=2$ TeV (red filled circles) and $M_\star=3$ TeV (green filled squares). The
ten-dimensional BHs decay in two hard quanta at the end of the evaporation phase.}     
\label{FIG3_kk2}
\end{figure*}

Fig.\ \ref{FIG3_kk1} compares BH events with zero-, two- and four-body final decay to graviton events. The variation of visible energy, \missPT, and visible transverse momenta of leptons and hadrons \& photons for different values of the fundamental Planck scale is shown in Fig.\ \ref{FIG3_kk2}.
%\clearpage
%
%\newline

\begin{figure*}[ht]
\begin{center}$
\begin{array}{cc}
    \includegraphics*[width=0.46\textwidth,totalheight=0.15\textheight]{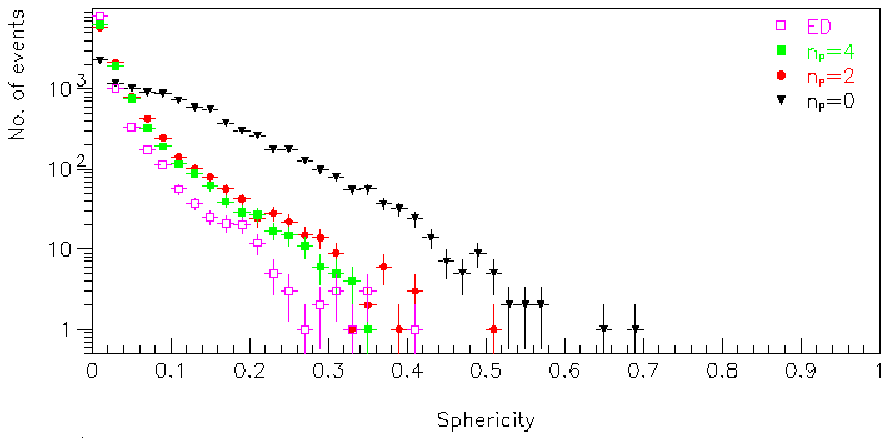}&
    \includegraphics*[width=0.46\textwidth,totalheight=0.15\textheight]{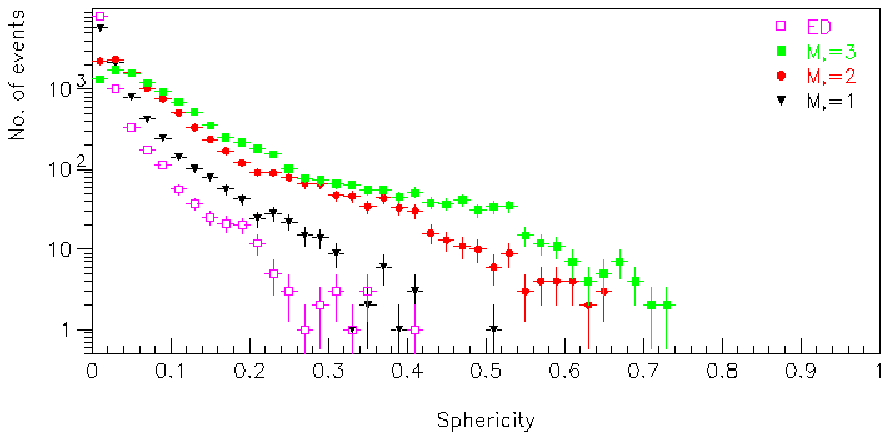}\\
    \includegraphics*[width=0.46\textwidth,totalheight=0.15\textheight]{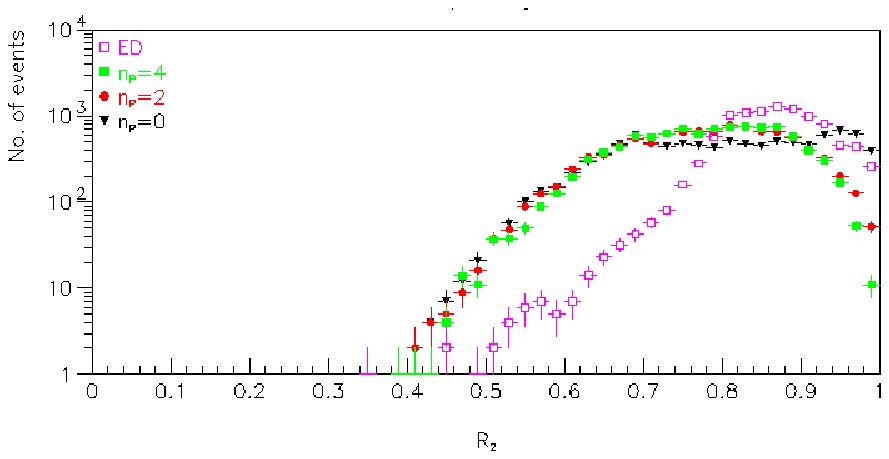}&
    \includegraphics*[width=0.46\textwidth,totalheight=0.15\textheight]{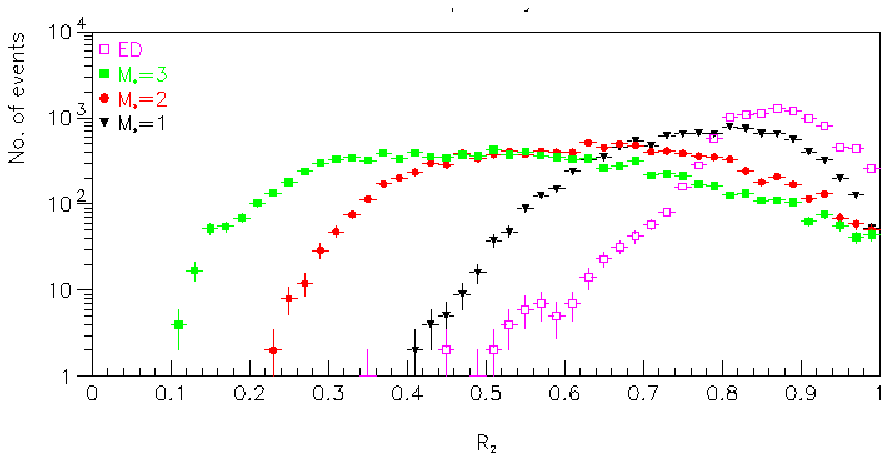}\\
    \includegraphics*[width=0.46\textwidth,totalheight=0.15\textheight]{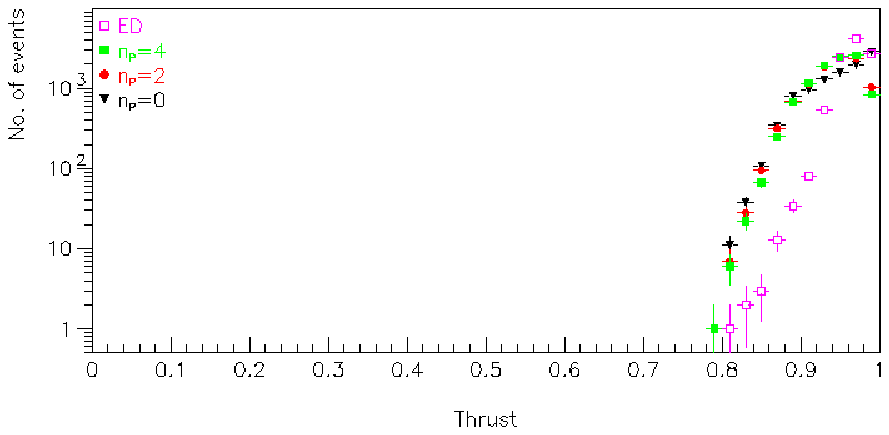}&
    \includegraphics*[width=0.46\textwidth,totalheight=0.15\textheight]{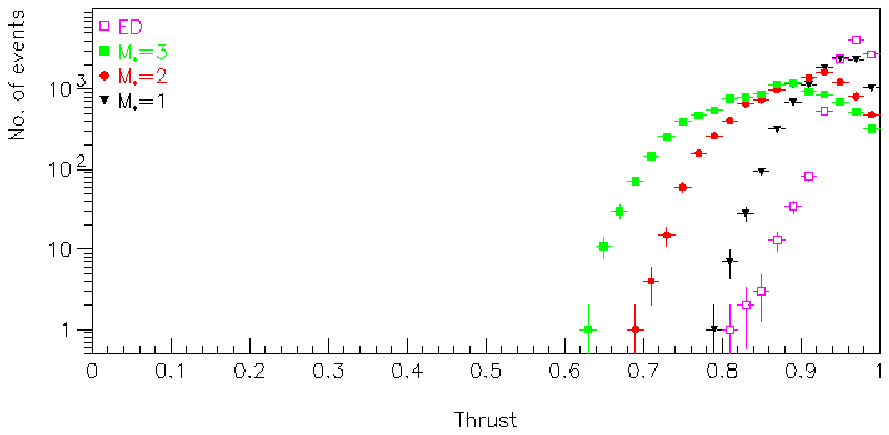}
    \end{array}$
    \end{center}
\caption{Sphericity (top panels) and $2^{\rm nd}$ Fox-Wolfram moment (middle panels) and
thrust (bottom panels) for 10,000 BH and graviton events (pink open squares). The left panels show the effect of
different Planckian decay modes: BH remnant (black filled triangles), two-body decay (red filled circles) and
four-body decay (green filled squares). The fundamental scale is $M_\star=1$ TeV and the number of EDs
is six. The right panels show the effect of different fundamental scales: $M_\star=1$ TeV (black filled triangles),
2 TeV (red filled circles) and 3 TeV (green filled squares). Here, the ten-dimensional BHs decay in two quanta at the end
of the Hawking phase. } 
\label{FIG4_kk}
\end{figure*}

The above analysis is complemented by studying event shape variables. Graviton decays are dominated by jets. Formation of a
BH remnant and high values of the fundamental scale lead to significant higher sphericity than graviton decays (top panels of Fig.\ \ref{FIG4_kk}). The $2^{\rm nd}$ Fox-Wolfram moment (middle panels of Fig.\ \ref{FIG4_kk}) is stable versus changes in the BH Planck phase and provides a good graviton/BH discriminator. BH models with higher $M_\star$ can be differentiated more easily from events with graviton decays. Thrust (bottom panels of Fig.\ \ref{FIG4_kk}) conveys a similar information.

\begin{figure*}[ht]
\begin{center}$
\begin{array}{cc}
    \includegraphics*[width=0.46\textwidth,totalheight=0.15\textheight]{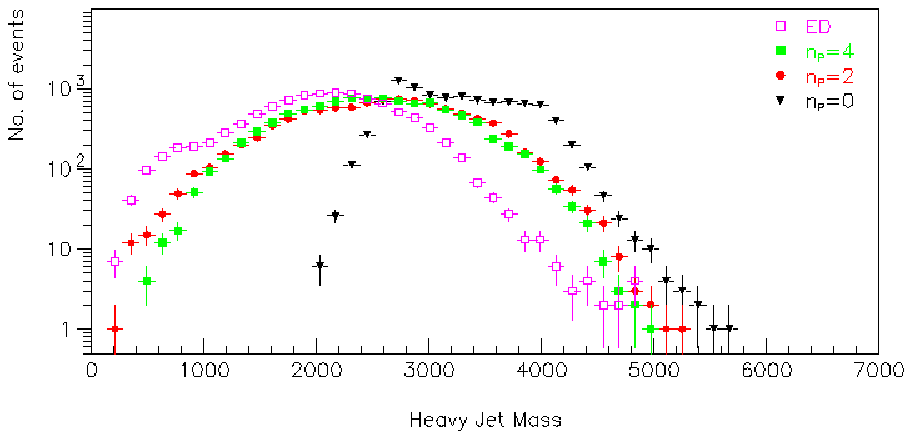}&
    \includegraphics*[width=0.46\textwidth,totalheight=0.15\textheight]{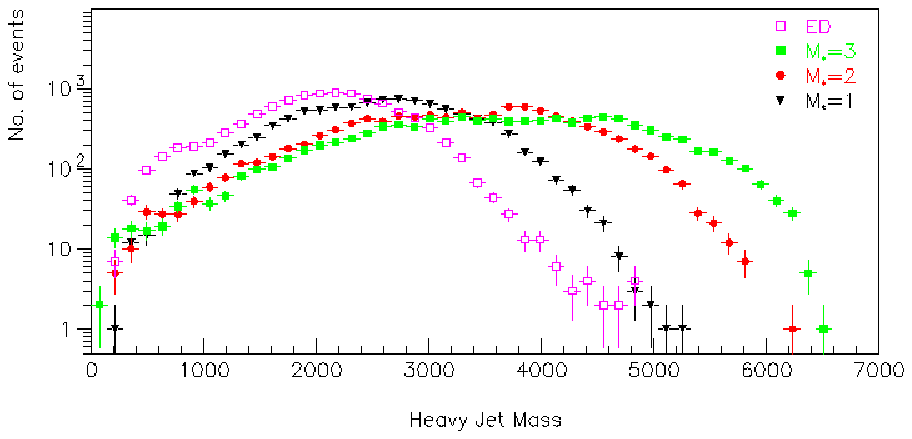}\\
      \includegraphics*[width=0.47\textwidth,totalheight=0.15\textheight]{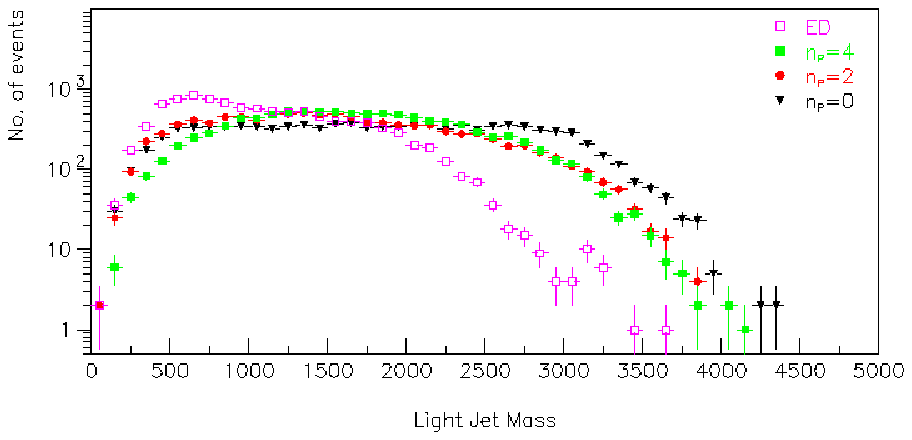}&
    \includegraphics*[width=0.47\textwidth,totalheight=0.15\textheight]{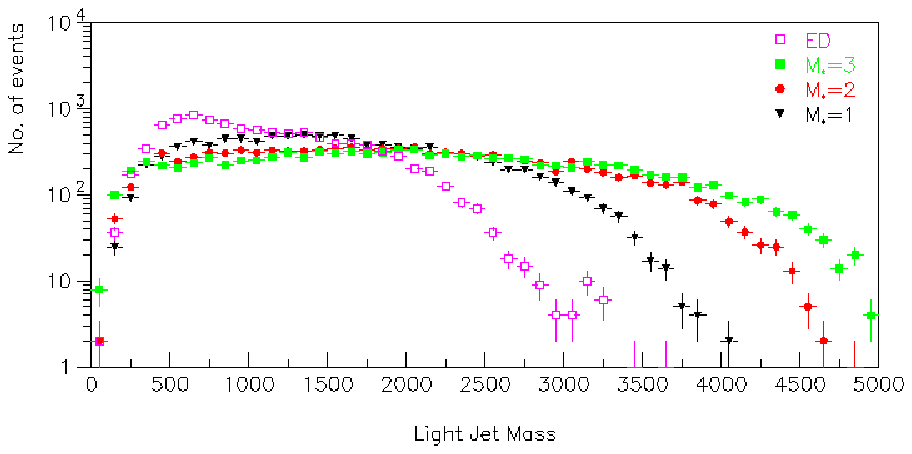}\\
      \includegraphics*[width=0.46\textwidth,totalheight=0.15\textheight]{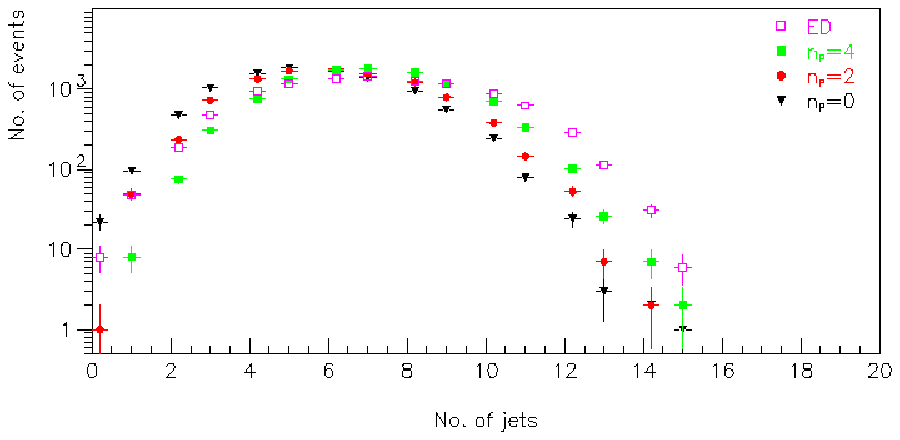}&
    \includegraphics*[width=0.46\textwidth,totalheight=0.15\textheight]{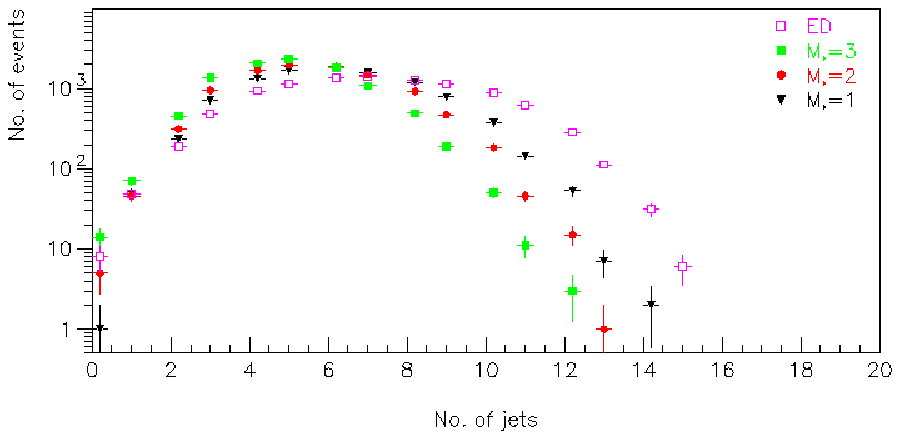}
    \end{array}$
    \end{center}
\caption{Heavy and light jet masses (top and middle panels) and number of jets (bottom panels) for 10,000 BH and graviton events.
Symbols are like in previous figures.} 
\label{FIG5_kk}
\end{figure*}

Similar conclusions can be reached by looking at jet the masses and at the number of jets. Graviton events generate more
lighter jets than the BH model due to the mass of the graviton being lower than the mass of the BHs (Fig.\ \ref{FIG5_kk}). The difference is specially significant for high values of $M_\star$ and in the presence of BH remnants. 
\subsection{High-$P_T$ leptons\label{leptons_kk}}
Graviton decay into dileptons provides a cleaner signature. The ATLAS and CMS detectors are expected to detect gravitons with mass as high as 3.5 and 2.08 TeV for c=0.1 and 0.01, respectively \cite{Allanach:2000nr, Allanach:2002gn, Collard:2002}. The discovery reach is better with electrons than with muons because of a better energy resolution of electrons. 

The SM background is removed by imposing the following cuts on leptons to remove the SM background: $P_{Tl} \ge 15$ GeV, $|\eta_l| < 2.5$, isolation cut~$\sum_i P_{T_i} < 7$ GeV in a cone of $R=0.2$ and \missPT $\ge 100$ GeV where the symbols have their usual meanings.

In the RS scenario, the majority of high $P_T$ leptons originate from the decay of the graviton and the $Z^0$ boson. The neutral
graviton decays into a pair of OS dileptons with a OF-to-SF ratio of $\sim$ 1:14. OSOF lepton pairs are originated
either from the graviton decay into $W$ bosons or from $t$ quark pairs, where the top quark decays into leptons.
Production of OSSF dileptons is less frequent in the BH model than in the RS graviton model. This is because of graviton decays in the $Z^0$ channel with the $Z^0$ decaying into leptons. Our analysis shows a 1:7 ratio of BH-to-ED dilepton events at fixed luminosity.  

\begin{figure*}[t]
\centerline{\null\hfill
    \includegraphics*[width=0.65\textwidth]{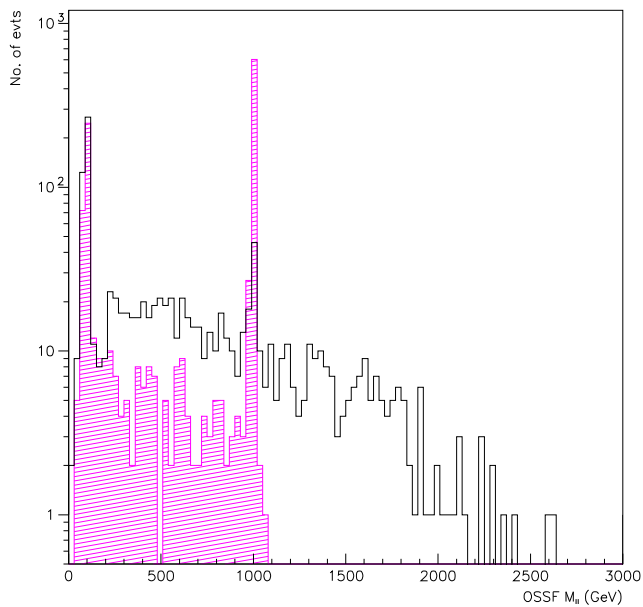}
    \null\hfill
    }
\caption{Invariant mass distribution of 1100 OSOF dilepton events. Most high-$P_T$ leptons in the RS model come
from the decay of the graviton and $Z^0$ bosons. This is indicated by the two peaks in the invariant mass
distribution. As discussed before in Sect.~\ref{susybhevt}, the BH distribution is also characterized by two peaks. The discriminating
signature is the high-$P_T$ tail in the BH invariant mass distribution.}
\label{fig1kk}
\end{figure*}

Figure \ref{fig1kk} shows the dilepton invariant mass distribution for RS graviton production (shaded pink plot) and
the BH model with final two-body decay. The RS graviton scenario is characterized by two peaks in the invariant mass
distribution, at $\sim$ 90 GeV and a second peak at 1 TeV. The first peak is due to dilepton events produced from the
decay of $Z^0$ bosons. The second peak at $\sim$ 1 TeV is due to graviton decay into a pair of leptons. The tail of
the BH distribution is due to uncorrelated leptons from the decay phase. Since there is no similar process in the RS scenario to generate leptons, the invariant mass distribution of the graviton does not have a tail at high energies.

Table~\ref{table:tablekk}  shows the BR's of 1000 high-$P_T$ isolated dileptons for the RS graviton model and the BH model. OSSF
dileptons are dominant in both cases. However, due to the flavor conserving nature of the graviton decay, the number of SS
dileptons produced in the RS scenario is virtually zero. This is not the case for BH events, where SS dileptons account for 18\% of the total dileptons produced. Therefore, the SS dilepton invariant mass would be a promising discriminating signature for RS graviton events and BH events. 
\begin{table}[htbp]
\caption{BRs of high-$P_T$ isolated dileptons for the RS and the BH models. Respectively $2.8\times10^4$ and $10^6$ events were simulated in the two cases, yielding approximately 1000 dilepton events.}
\begin{center}
\begin{tabular*}{0.90\textwidth}%
{@{\extracolsep{\fill}}c|ccccc}
\hline\hline
~High $P_T$ isolated dileptons~ & $Graviton$ & \% & $BH$ & \% &\\
\hline
OSSF                   & 987 & 99 & 523   & 50  &\\
\hline
OSOF                   & 13  & 1  & 341   & 32  &\\
\hline
SS                     & 0  & 0  & 190   & 18  &\\
\hline
\end{tabular*}
\end{center}
\label{table:tablekk}
\end{table}

\begin{figure*}[htbp]
\centerline{\null\hfill
    \includegraphics*[width=0.65\textwidth]{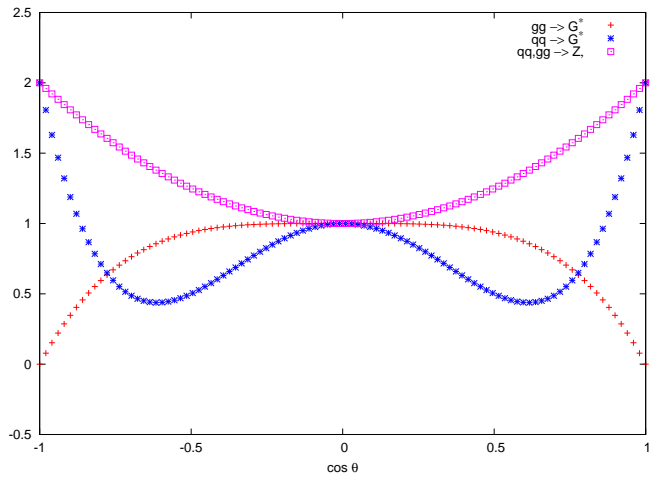}
    \hfill}
\caption{Angular distributions of electron from $Z'$ and graviton decays.} 
\label{angdistfig}
\end{figure*}

It is worthwhile to study the decay of the $Z'$ boson which is predicted by many extensions of the SM \cite{PDG, Leike:1998wr, Rizzo:2006nw}. The $Z'$ boson is a heavy, (mass $>$ 500 GeV) \cite{PDG}, colorless, neutral, spin 1 particle. As a result, $Z'$ decay into leptons could compete with graviton decay in the same channel\footnote{However, the decay of $Z'$ into photons is suppressed.}. This is especially significant for the LHC which could probe $Z'$ masses in the TeV range. The observation of a TeV resonance in the dilepton and diphoton channels could provide a strong evidence of a RS graviton, distinguishing it from a $Z'$ boson. An additional discriminating property between  the graviton and the $Z'$ is the angular distribution of the emitted particles, which is dependent on the spin of the resonance \cite{Rizzo:2006nw}. The angular distribution is strongly dependent on the production mechanism. The angular distributions of the $G^{\star}$ and  $Z'$ are shown in Table~\ref{angdist} \cite{Allanach:2000nr} and the corresponding distributions are shown in Fig.~\ref{angdistfig}. Gluon fusion predominates at the LHC. For graviton production, the quark-antiquark fusion contribution tends to flatten the angular distribution\cite{Allanach:2002gn}. 
\begin{table}[htbp]
\caption{Angular distributions of $Z'$  boson and graviton decay into dielectrons. $\theta$ is the angle between the beam direction and the electron in the rest frame of the resonance.}
\begin{center}
\begin{tabular*}{0.6\textwidth}{@{\extracolsep{\fill}}c|cc}
\hline\hline
Process & Distribution &\\
\hline
$gg\rightarrow G^{\star}\rightarrow e^{+}e^{-}$	& 1 - $\cos^4\theta$ &\\
\hline
$q\bar{q}\rightarrow G^{\star}\rightarrow e^{+}e^{-}$ 	& 1 - 3 $\cos^2\theta + 4 \cos^4\theta$ &\\
\hline
$q\bar{q},gg\rightarrow Z'\rightarrow e^{+}e^{-}$ 	& 1 + $\cos^2\theta$ &\\
\hline 
\end{tabular*}
\end{center}
\label{angdist}
\end{table}

\newpage
\section[SR and SM Event Analysis]{SR and SM Event Analysis\label{evt_analysis_string}}
Events with high-$P_T$ photons are a powerful discriminator of SR events from the SM background. SR events are characterized by high values of visible energy and \missPT\ as the photon and the jet originate
directly from the $2\times2$ interaction. Isolated photons provide a further means to extract string signals: being directly produced from the SR, they are harder than SM photons. Analysis of event shape variables enhance the effectiveness of the above method. String simulations use the following benchmark: $M_s$=1 TeV and $P_{Tmin}$=50 GeV. Our analysis uses $\sim$ 10000 and 200 string events for $M_s$=1 and 2 TeV produced from a sample run of $10^7$ events at an integrated LHC luminosity of $1~fb^{-1}$, respectively.

\begin{figure*}[t!]
\begin{center}$
\begin{array}{c}
    \includegraphics*[width=0.65\textwidth]{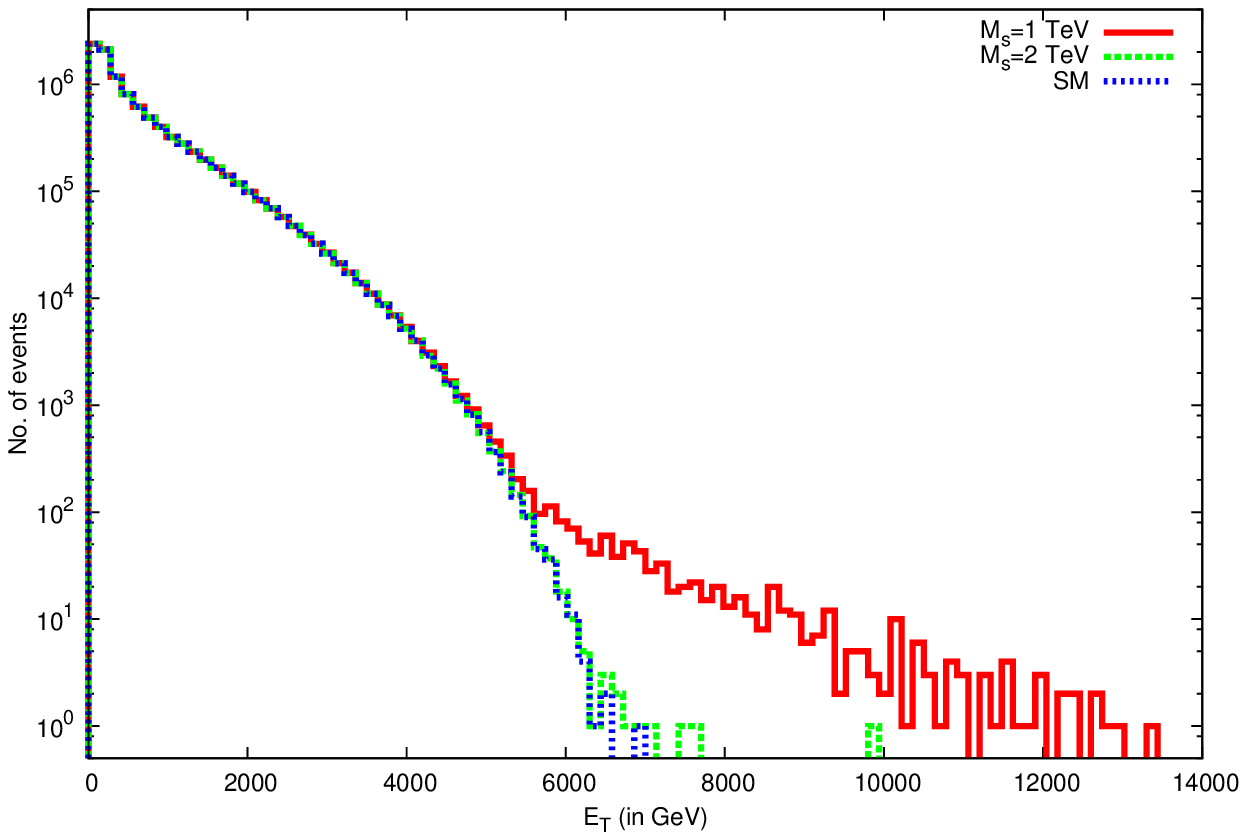}\\
    \includegraphics*[width=0.65\textwidth]{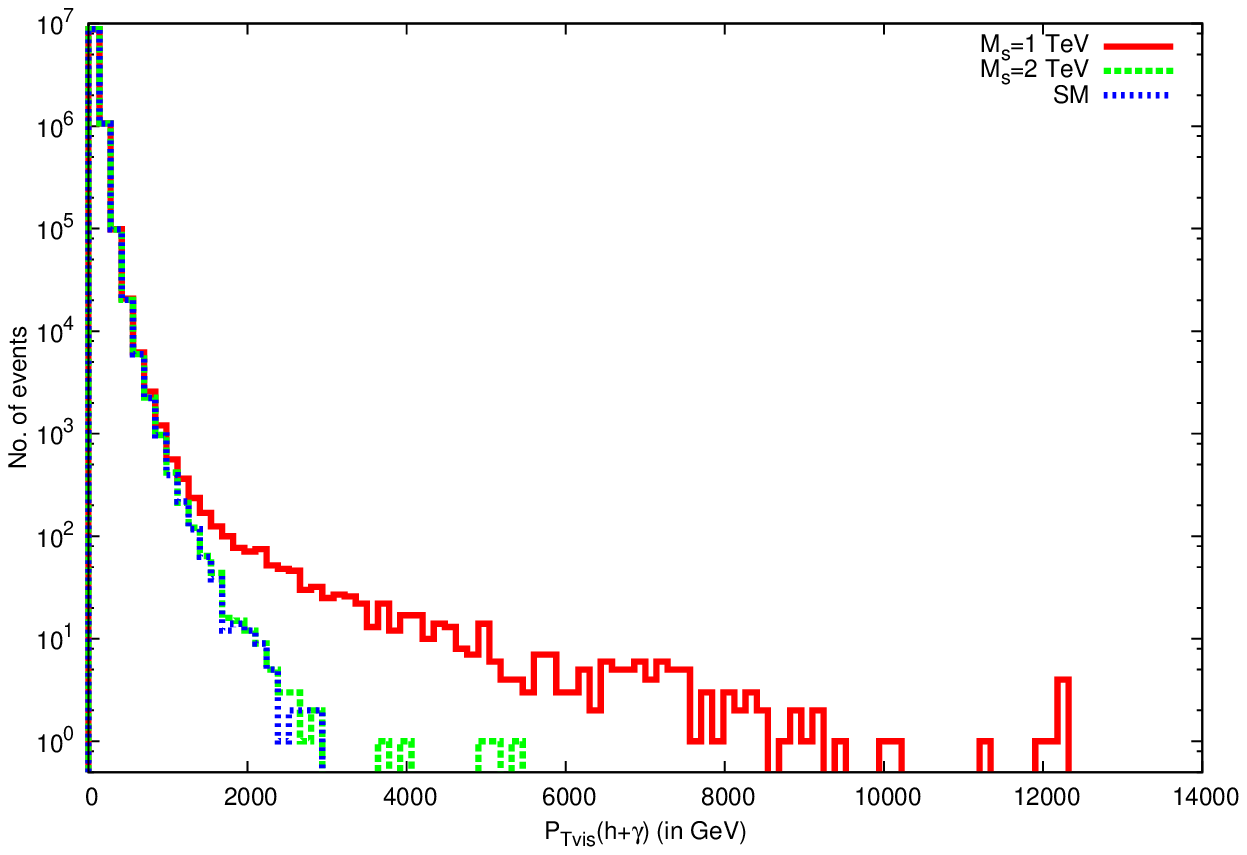}\\
    \end{array}$
    \end{center}
\caption{Top panel: Visible energy distribution for string+SM and SM only events. The result for string events is shown
by the solid red histogram ($M_s$=1 TeV) and the dashed green histogram ($M_s$=2 TeV). String events can be identified
from the high-$E_T$ tail for $M_s$=1 TeV. Bottom panel: Distribution of visible $P_T$ for $\gamma$+hadrons. The
high-$P_T$ tail is a strong indicator of the presence of SR.} 
\label{figs1}
\end{figure*}

Figure~\ref{figs1} shows the visible energy (top panel) and the $P_T$ of hadrons \& photons (bottom panel)
for 10 million string+SM and SM only events. The visible energy and the $P_T$ are produced by the hard
photons and the jets of the string decay. Their distributions are characterized by a long tail. The observation of
events with $P_T$ (visible energy) in excess of 3~(6) TeV would provide strong evidence of the existence
of a SR. 

In our analysis of the photon momentum, we select the hadronic jets according to the following criteria. The
detector is assumed to have an absolute value of pseudorapidity of 2.6. This ensures
that the jets are originated in the hard $2\times2$ scattering rather than in multiple interactions or from the beam
remnants. The contribution of jets which do not originate in the hard scattering are minimized by fixing the
minimum transverse energy of all particles comprising the jet ($\Sigma_i E_{T_i}$) to  40 GeV \cite{Gupta:2007cy}. The
particles of the jet must be within a cone of $R=\sqrt{(\Delta\eta^2+\Delta\phi^2)}$~=~0.2 from the jet initiator,
where $\theta$ and $\phi$ are the azimuthal and polar angles of the particle w.r.t. the beam axis, respectively.
Following Ref.~\cite{Gupta:2007cy}, the cuts on the photon jets are $P_{T_{\gamma}} \ge$ 80 GeV, $\eta < $ 2.6 and an
isolation cut $\Sigma_n P_T <$ 7 GeV in a cone of $R$~=~0.4. 

\begin{figure*}[t!]
\begin{center}$
\begin{array}{cc}
    \includegraphics*[width=0.65\textwidth]{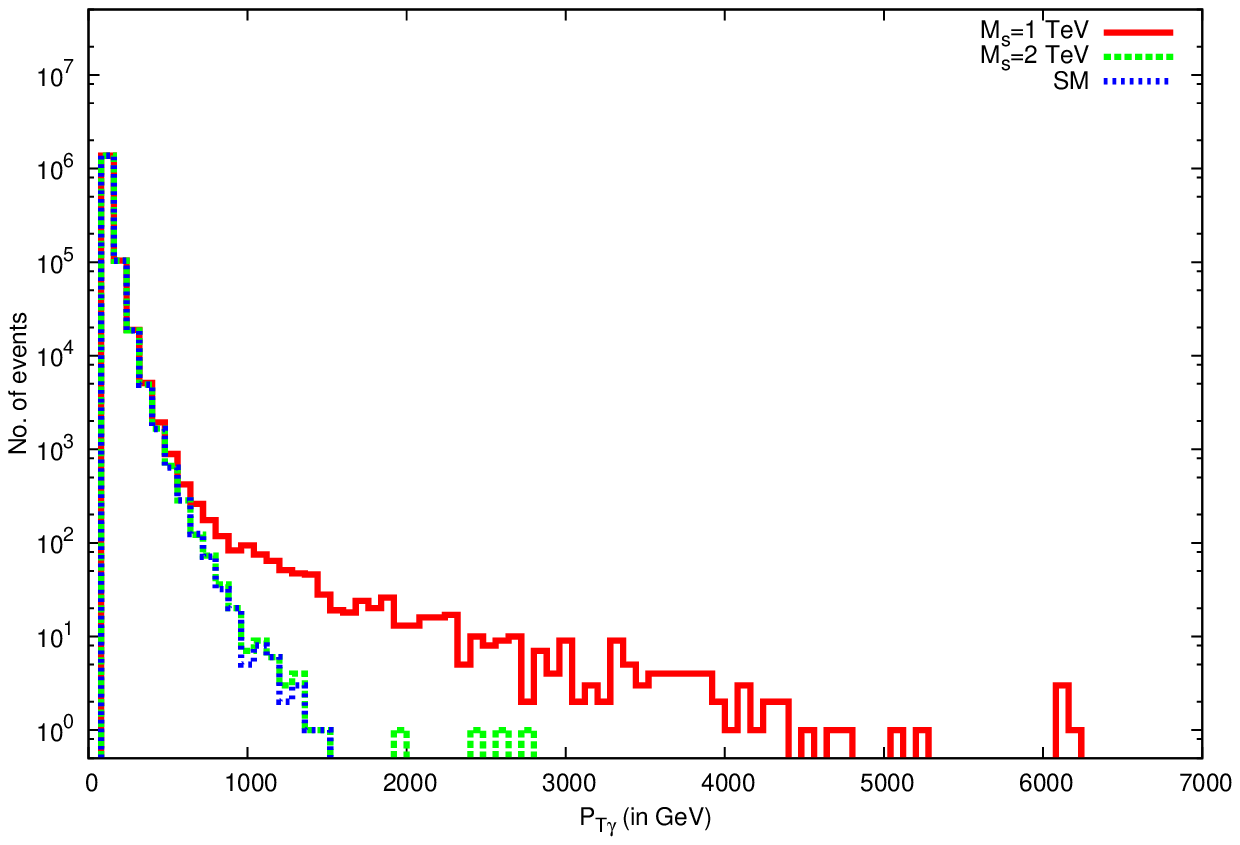}\\
    \includegraphics*[width=0.65\textwidth]{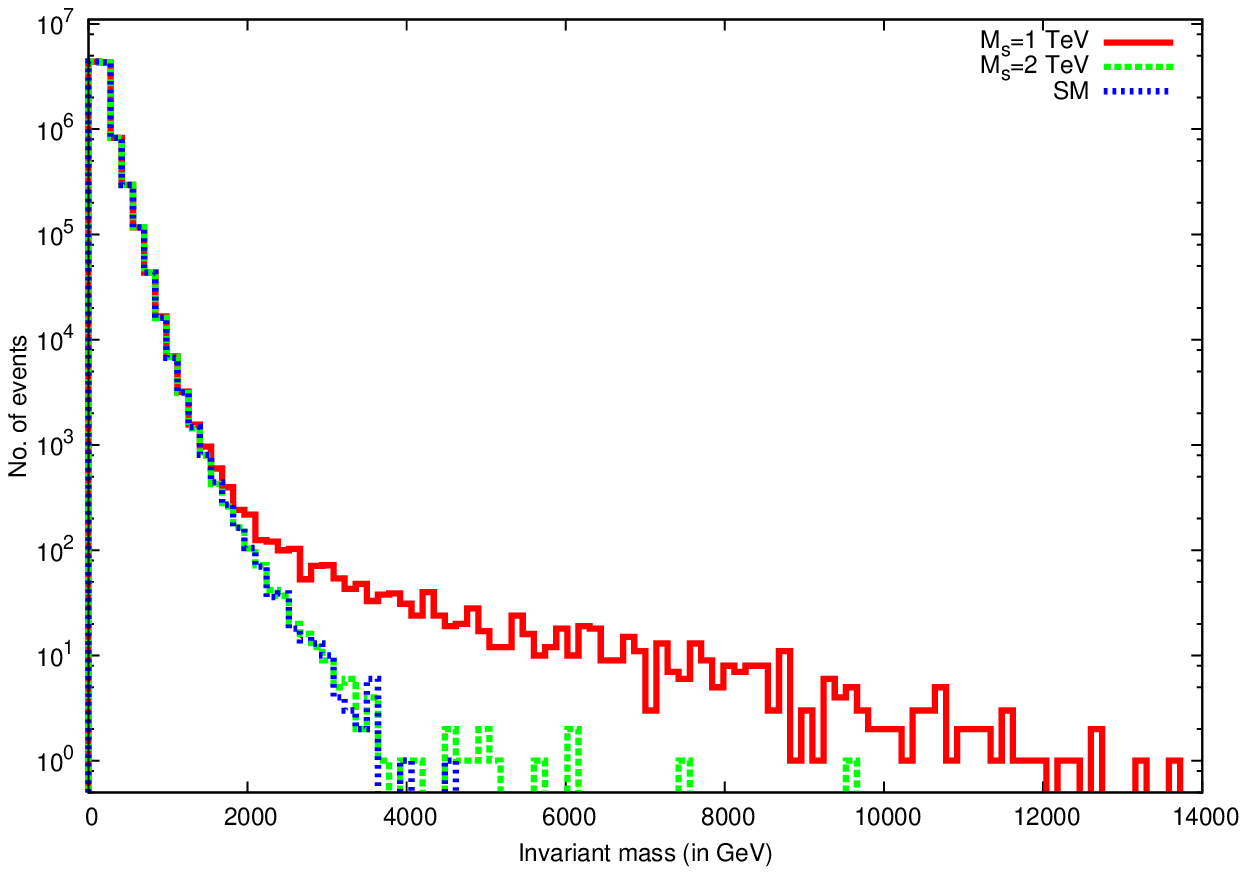}\\
    \end{array}$
    \end{center}
\caption{Top panel: Distribution of the highest $P_{T_{\gamma}}$ for each event. $\gamma$'s with high $P_T$
created in the string decay are the source of the long tail. Bottom panel: Dijet invariant mass distribution. String
decays may result is large invariant mass.} 
\label{figs3}
\end{figure*}
The top panel of Fig.~\ref{figs3} shows the distribution of the highest $P_{T_{\gamma}}$  of isolated photons for
string+SM and SM only events. Single photons from SR are expected to have a higher $P_{T_{\gamma}}$ than
SM photons because they are the direct products of the string decay. The main source of background for direct photons
are jet fluctuations  and photons originating from the initial and final state radiation \cite{Gupta:2007cy}. In the
former case, a jet consists of a few particles including high-$P_T$ $\pi_0$'s which decay into a pair of photons with a
$\sim$ 99\% BR \cite{Gupta:2007cy}. Due to the high boost, the photons have a relatively small angular
separation and therefore ``fake" a single photon in the electromagnetic calorimeter. The rate of this process is 1 out
of $\sim 10^3~\hbox{to}~10^4$ events \cite{Gupta:2007cy}. Other sources of fake photons are $H\rightarrow\gamma\gamma$ \cite{Pieri:2006bm} or processes from other exotic phenomena, e.g. SUSY \cite{Abulencia:2007ut} or LEDs \cite{Arkani-Hamed:1998rs,Antoniadis:1998ig,Arkani-Hamed:1998nn}. Isolation cuts on the photon can effectively reduce the number of fake photons. 

The invariant mass plot of the two highest $P_T$ jets for each event is shown in the bottom panel of
Fig.~\ref{figs3}. Due to the nature of the interaction under consideration, the bulk of the events for both string+SM
and SM only events are comprised of dijets. These were selected using the same cuts as before. 
As is expected, the SM invariant mass distribution is negligible beyond a certain threshold ($\sim$ 4 TeV). This is due to the production of direct soft photons and jets from the SM interaction. The string+SM distribution is characterized by a
long tail of energy up to several TeV (three times more than the SM). This tail is originated from the decay of string
resonances into hard jets and photons. Therefore, the measure of a large invariant mass would provide strong evidence of a string mediated interaction.

\begin{figure*}[t!]
\begin{center}$
\begin{array}{cc}
    \includegraphics*[width=0.65\textwidth,totalheight=0.25\textheight]{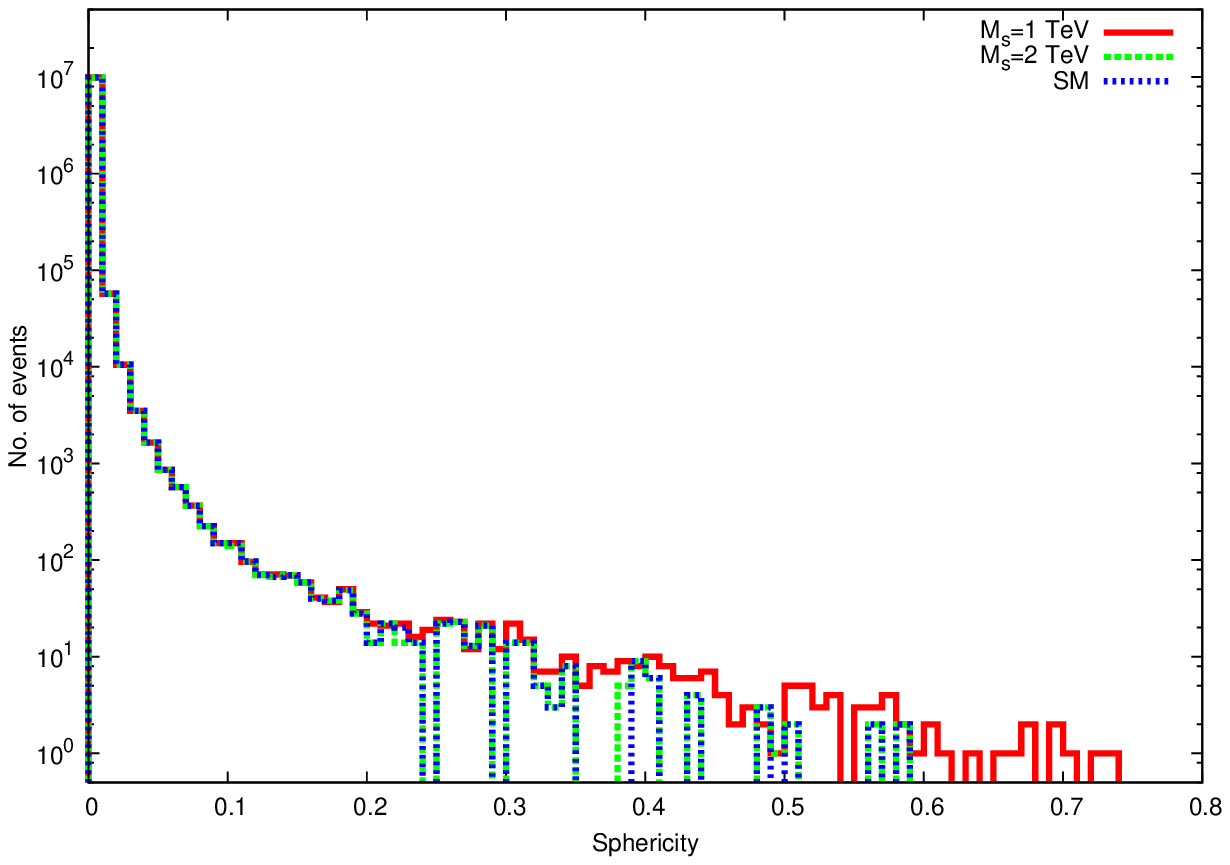}\\
    \includegraphics*[width=0.65\textwidth,totalheight=0.25\textheight]{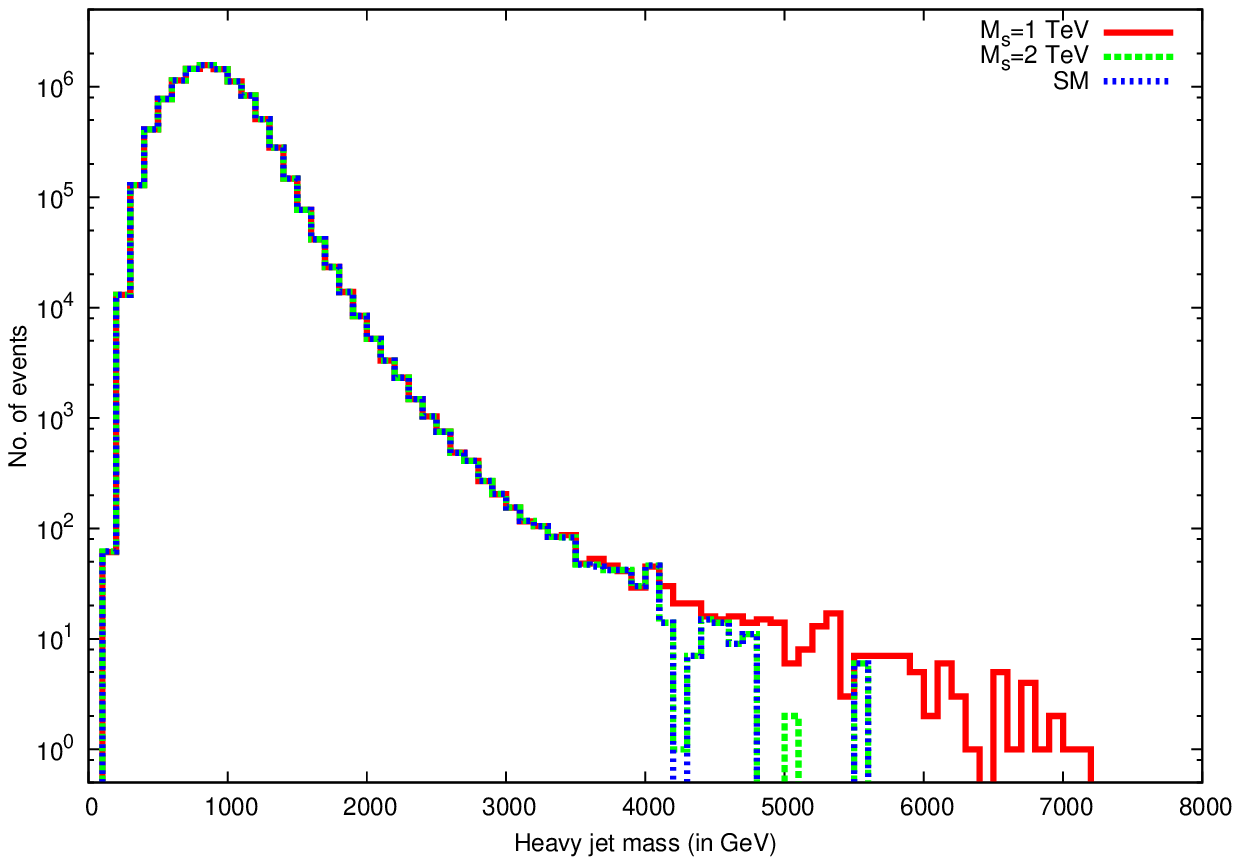}\\
    \includegraphics*[width=0.65\textwidth,totalheight=0.25\textheight]{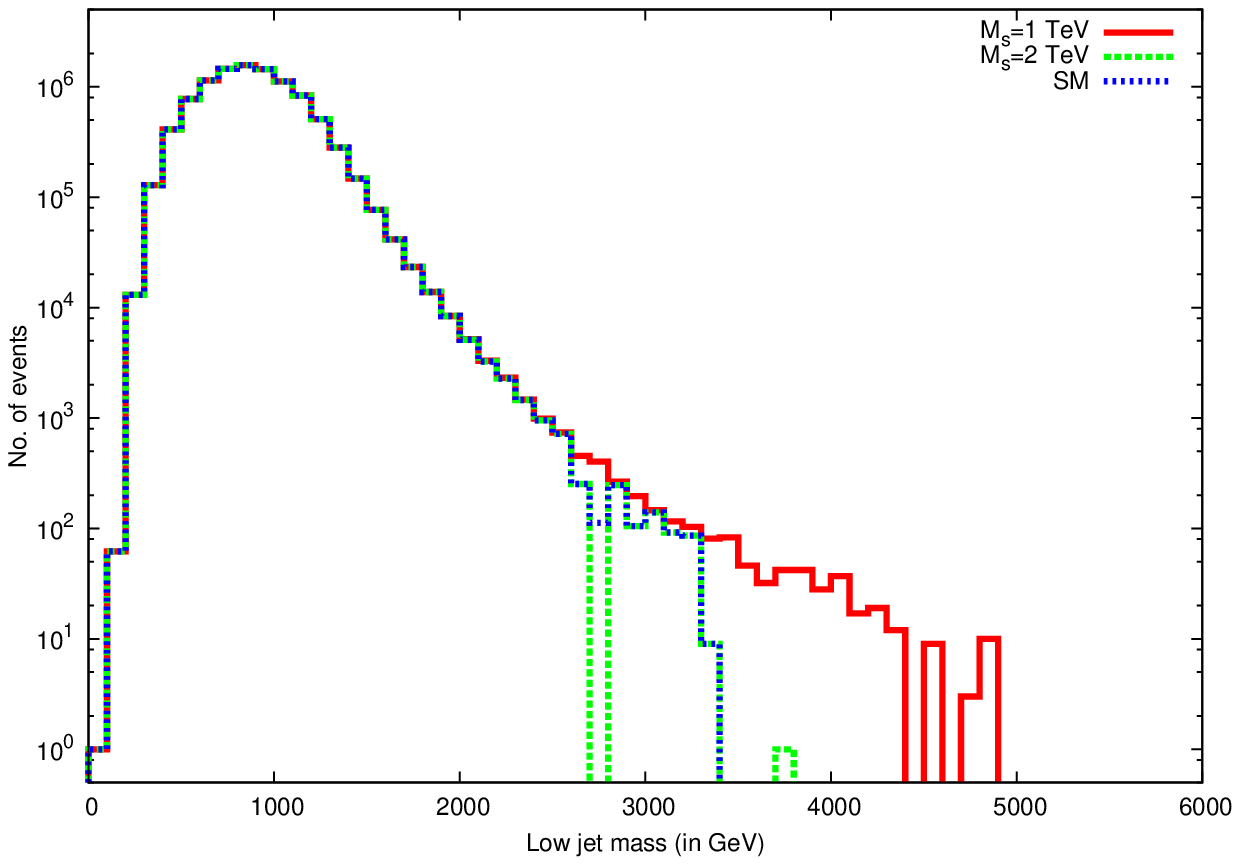}
    \end{array}$
    \end{center}
\caption{Histograms of event shape variables for 10 million string+SM and SM only events. String events are shown in
solid red ($M_s$=1 TeV) and dashed green ($M_s$=2 TeV). SM events are shown in dotted blue. String events have on the
average a slightly higher sphericity than SM events due to the slight increase in the number of jets (top panel).
Similar conclusions are reached from the heavy and low jet mass distributions (middle and bottom panels, respectively).} 
\label{figs2}
\end{figure*}
Histograms for different event shape variables are shown in Fig.~\ref{figs2}. String events tend to be more
spherical than the background: string+SM decays generally produce a distribution of high $P_T$ jets at slightly higher
values than the SM background.  The jets originate from the decay of SRs into photons and hadrons. The SM
generates less heavier jets than its string counterpart as shown in the middle and bottom panels of Fig.~\ref{figs2}.

\begin{figure}[htbp]
\centerline{\null\hfill
    \includegraphics*[width=0.65\textwidth]{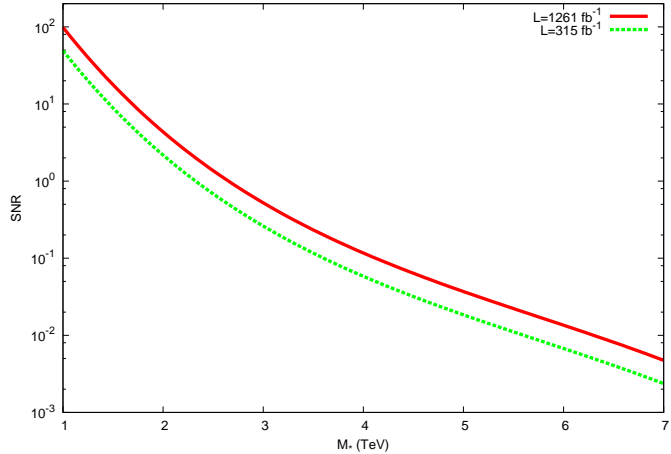}
    \null\hfill}
\caption{SNR for a integrated luminosity of 1300 $fb^{-1}$ (solid red line) and 300 $fb^{-1}$ (dashed green line), corresponding to 4 year and 1 year at an instantaneous luminosity of $10^{34} cm^{-2} s^{-1}$, respectively.}
\label{snr}
\end{figure}

Before ending this section, let us briefly discuss the discovery reach of SRs at the LHC. Fig.~\ref{snr} shows
the signal-to-noise (SNR) ratio for two values of the integrated luminosity as a function of $M_{\star}$ when
$P_{Tmin}$=50 GeV. The instantaneous luminosity is assumed to be the LHC design luminosity of $10^{-34} cm^{-2}
s^{-1}$ corresponding to an integrated luminosity of $\sim$ 1300 (300) $fb^{-1}$ in 4 (1) years. The SNR ratio for an integrated luminosity of 300 $fb^{-1}$ is $\sim$ 10 times less than the value obtained at 1300 $fb^{-1}$.
In an optimistic scenario, SRs at the TeV scale could be discovered after a successful year of running the LHC, assuming a signal-to noise ratio of 10.

\section{SR and BH Event Analysis\label{evt_analysis_string_bh}}
A powerful way of discriminating BH and string events would be searching for a $Z^0$ mass peak in the
invariant mass of high-$P_T$ leptons. $Z^0$ production is highly suppressed in case of string events
\cite{Anchordoqui:2008ac}. On the contrary BH decay is characterized by the production of a variety of particles with
high $P_T$. A rough counting of the number of degrees of freedom of these particles shows that the
estimated rate of hadron-to-lepton production is 5:1 and the rate of $Z^0$ and $\gamma$ production is comparable
($\sim$ 2\% to 3\%) with the $Z^0$ bosons decaying into opposite-sign leptons with a 3.4\% BR. The
invariant mass distribution of BH events peaks at $\sim$ 92 GeV, confirming the production of a $Z^0$ boson
\cite{Roy:2008we}. Therefore, the presence of a peak at $\sim$ 92 GeV in the invariant mass of leptons would
effectively rule out formation of SRs in favor of BH production. Due to the nature of the
Hawking radiation, BH events are also expected to have on average higher sphericity and a larger number of jets than string
events. This is because in the $pp\rightarrow \gamma+jet$ channel string events are dominated by dijets, whereas BH
decay produces multijets.

\chapter*{Conclusions\label{conclusions}}
\addcontentsline{toc}{chapter}{Conclusions}
In this thesis we studied different scenarios for physics beyond the SM: SUSY, EDs and SRs.

Signatures of SUSY and TeV-scale BH events at the LHC have been discussed in Sect.~\ref{susybhevt}. A combined analysis of event-shape variables and dilepton events has shown that it is possible to distinguish BHs and SUSY at the ATLAS and
CMS detectors. Event shape variables alone cannot unequivocally discriminate between SUSY and BHs. For example, the thrust distribution for BH events with no remnant production is indistinguishable from the thrust distribution for SUSY events.
However event  shape variables may prove useful when combined with the analysis of the leptonic channel. Isolated dileptons could provide the ``smoking gun'' for detecting BHs at the LHC. The BH dilepton invariant mass shows a tail at high energy which is
absent in the SM or MSSM. This analysis can be further strengthened by looking at the number and flavor of isolated leptons. 

The possibility of discriminating BH and graviton events was discussed in Sect.~\ref{evt_analysis_edbh}. Our simulations showed that by analyzing the dilepton channel, BH and graviton events can be clearly discriminated at the LHC. As with SUSY and BHs, event shape variables provide additional means of discrimination.

Creation of SRs at the LHC was investigated in Sect.~\ref{evt_analysis_string}. If the string scale is $\sim$1 TeV, our analysis has shown that SRs can be detected. With the LHC operating at a luminosity of $2\times10^{33} cm^{-2} s^{-1}$ SRs could be detected within a year of the start of the LHC. String events show higher sphericity and higher visible energy than the SM background. These quantities provide an effective means of detection when combined with the measure of the $P_T$ of isolated photons and the dijet invariant mass. Since the final products of the SR are directly produced from the string decay, the dijet invariant mass is characterized by a tail at high energies which is absent in the SM. A powerful way of discriminating between BH and string events would be searching for a $Z^0$ mass peak in the invariant mass of high-$P_T$ leptons. SR decay into leptons and photons may be another effective method for distinguishing SRs and BHs. 

The investigations in this thesis provide a starting point for a quantitative study of discrimination of BHs from SUSY, graviton events and SRs at the LHC. However, many open questions remain unanswered. For example, our work has not dealt in depth with sources of background at the LHC. The actual sensitivity of the method for specific experiments (CMS or ATLAS) is also not discussed here. Moreover, the kinematical cuts implemented in Chapter~\ref{simanaly} although commonly used in the literature, lack a study at the generator level of their efficiency. Finally, it would also be worthwhile to study the detector response with \texttt{CMSSW}, the
software framework which is used for simulation and analysis of high energy collisions at the CMS detector.
%
%\begin{center}\chapter*{}\end{center}
\begin{comment}
\newpage
\begin{center}
\vspace*{\fill}
 BIBLIOGRAPHY 
 \vspace*{\fill}
\end{center}
\end{comment}
\nocite{*}
\bibliographystyle{utphys}
\addcontentsline{toc}{chapter}{Bibliography}
\bibliography{mybib1}
%\begin{center}\chapter*{}\end{center}
\begin{comment}
\newpage
\begin{center}
\vspace*{\fill}
 APPENDICES
 \vspace*{\fill}
\end{center}
\end{comment}
%
\chapter*{Appendix A: Units\label{appendix}}
\addcontentsline{toc}{chapter}{Appendix A: Units}
In this thesis we use 3 different system of units: high energy, SI and natural. Table~\ref{units1} shows the value of the fundamental constants in the 3 systems of units. Conversion factors between these 3 systems are summarised in Table~\ref{units2}.
\counterwithout{table}{chapter} 
\setcounter{table}{0}
\begin{table}[htbp]
\caption{Fundamental constants.}
\begin{center}
\begin{tabular*}{1\textwidth}{@{\extracolsep{\fill}}c|cccc}
\hline\hline
Quantity & High energy units & SI units & Natural units\\
\hline
$\hbar$ & 6.588$\times10^{-28}$ GeV s  &  1.055$\times10^{-34}$ J s & 1&\\
\hline
$c$  & 2.998$\times10^{23}$ fm/s  &   2.998$\times10^{8}$ m/s& 1 &\\
\hline
$G_4$ & 6.707$\times10^{-39}$ GeV  &  6.674$\times10^{-11} {\rm m}^3 {\rm kg}^{-1} {\rm s}^{-2}$ & $M_{PL}^{-2}$&\\
\hline\hline
\end{tabular*}
\label{units1}
\end{center}
\label{units}
\end{table}
\begin{table}[htbp]
\caption{Conversion factors.}
\begin{center}
\begin{tabular*}{1\textwidth}{@{\extracolsep{\fill}}c|cccc}
\hline\hline
Quantity & High energy units & SI units & Natural units\\
\hline
Length &1.97$\times10^{-20}$ fm  &  1.97$\times10^{-35}$ m & 1 $M_{PL}^{-1}$ &\\
\hline
Mass & 1.22$\times10^{19}$ GeV  &  1.78$\times10^{-8}$ kg & 1 $M_{PL}$ &\\
\hline
Energy & 1.22$\times10^{19}$ GeV  &  1.95$\times10^{9}$ J & 1 $M_{PL}$ &\\
\hline
Time & 8.2$\times10^{-20}~{\rm GeV}^{-1}$  &  6.58$\times10^{-44}$ s & 1 $M_{PL}^{-1}$ &\\
\hline
Temperature & 1.22$\times10^{19}$ GeV  &  1.42$\times10^{32}$ K & 1 $M_{PL}$ &\\
\hline\hline
\end{tabular*}
\label{units2}
\end{center}
\label{units}
\end{table}
\chapter*{Appendix B: Acronyms\label{appendix}}
\addcontentsline{toc}{chapter}{Appendix B: Acronyms}
\counterwithout{table}{chapter} 
\setcounter{table}{0}
\begin{center}
\begin{longtable}{ll}
\hline\hline
ADD & ArkaniHamed-Dimopoulos-Dvali\\
BD & Black Disk \\
BH & BlackHole \\
BR & Branching Ratio \\
CERN & European Organization for Nuclear Research \\
CL & Confidence Level \\
CM & Center-of-Mass \\
CP & Charge-Parity \\
dof & degrees of freedom \\
ED & Extra Dimension \\
GUT & Grand Unified Theory \\
KK & Kaluza-Klein\\
LED & Large Extra Dimension \\
LEP & Large Electron-Positron \\
LHC & Large Hadron Collider \\
LSP & Lightest Supersymmetric Particle \\
MC & Monte Carlo \\
MSSM &Minimal Supersymmetric extension of the SM\\
mSUGRA &minimum SUperGRAvity\\
OSOF &Opposite Sign Opposite Flavor\\
OSSF &Opposite Sign Same Flavor\\
PDF &Parton Distribution Function\\
QCD &Quantum ChromoDynamics\\
QED &Quantum ElectroDynamics\\
RS &Randall-Sundrum\\
SM&Standard Model\\
SR &String Resonance\\
SSOF &Same Sign Opposite Flavor\\
SSSF &Same Sign Same Flavor\\
ST &String Theory\\
SUSY &SUperSYmmetry\\
TS &Trapped Surface\\
\hline\hline
\end{longtable}
\end{center}
\label{acro}
\chapter*{Appendix C: Symbols\label{appendix}}
\addcontentsline{toc}{chapter}{Appendix C: Symbols}
\counterwithout{table}{chapter} 
\setcounter{table}{0}
\begin{center}
\begin{longtable}{ll}
\hline\hline
$m_{bare}$ & Bare Higgs mass \\
$m_{Higgs}$ & Physical Higgs mass\\
$\Lambda$ (Sect.~\ref{BSM})& Higgs mass cutoff \\
$X$ and $\mu$ (Sect.~\ref{BSM}) & Momentum transfer scale \\
$b_i$ & Numerical prefactors \\
$m_0$ & common mass of scalar particles at $M_{GUT}$ \\
$m_{\frac{1}{2}}$ & common mass of gauginos and Higgsino at $M_{GUT}$ \\
$ A_0$ & trilinear coupling at $M_{GUT}$ \\
$\tan\beta$ & ratio of vacum expectation values of two Higgs fields \\
$\mu$ & sign of the Higgsino mass parameter \\
L(B) & Lepton (baryon) number \\
$\textbf{s}$ & Particle spin \\
$n$ &Number of extra-spatial dimensions \\
$R$ & Size of ED in ADD model \\
$\psi$ & Scalar field \\
$m_{\psi}$ &Mass of the scalar field\\
$x_i$ & 3 spatial coordinates\\
$G_4$ & 4-dimensional Newton's constant\\
$G_{n+4}$ & (n+4)-dimensional Newton's constant\\
$S_d$ &Surface area of unit sphere in $d$ dimensions\\
$F_4$ &Force between two test masses in 4 dimensions\\
$F_{n+4}$ &Force between two test masses in (n+4) dimensions\\
$D$ &Total number of dimensions\\
$\phi_{g(4)}$ &4-dimensional gravitational potential\\
$\phi_{g(n+4)}$ &(n+4)-dimensional gravitational potential\\
$\rho_m$ &Mass density\\
$V_n$ &Volume of ED in ADD model\\
$x_n$ &Bessel function roots\\
$\Lambda$ (Sect.~\ref{rs})& Strength of graviton-matter coupling in RS model \\
$\sigma$ &cross section\\
$M_{PL}$ &Planck mass\\
$M$ &BH mass\\
$\hat{s}$ &Parton square CM energy\\
$s$ &Particle square CM energy\\
$Q$ (Sect.~\ref{bhc}) &Momentum transfer scale\\
$T_H$ &Hawking temperature\\
$\tau$ &BH lifetime\\
$\alpha^{'}$ &Slope parameter\\
$l_s$ &String length scale\\
$M_s$ &String mass scale\\
$g_s$ &String coupling\\
$\theta_W$ &Weinberg angle\\
$p_z$ &Momentum along $z$ axis\\
$L$ &Luminosity\\
$f$ &Proton bunch crossing frequency\\
$n_b$ &Number of protons in each bunch\\
$A$ &Bunch area\\
$S$ &Sphericity\\
$T$ &Thrust\\
$R_2$ &Second Fox-Wolfram moment\\
$E_{vis}$ &Visible energy\\
$P_{T}$ &Transverse momenta\\
$P_{Tmin}$ &Minimum transverse momenta\\
$P_{Tl}$ &Transverse momenta of leptons\\
$P_{T_{\gamma}}$ &Transverse momenta of photons\\
\missPT\ &Missing transvese momenta\\
\missET\ &Missing transvese energy\\
$\eta$ &Pseudorapidity\\
\hline\hline 
\end{longtable}
\end{center}
\label{symb}
\end{document}